\documentclass[acmsmall]{acmart}
\usepackage{makecell}
\usepackage{graphicx}
\usepackage{multirow}
\usepackage{color,soul}
\usepackage{subcaption}
\usepackage{pifont}
\usepackage{rotating}
\usepackage{makecell}

\newcommand\mypara[1]{\noindent\textbf{#1}}
\definecolor{darkgreen}{rgb}{0.0, 0.5, 0.0}
\definecolor{darkgrey}{rgb}{0.3, 0.3, 0.3}
\newcommand{\y}{\textcolor{darkgrey}{\ding{52}}}
\newcommand{\n}{\textcolor{darkgrey}{\ding{56}}}
\newcommand{\na}{\textcolor{orange}{}}
\newcounter{myblock}

\newcounter{myblockcounter}
\newcounter{issuecounter}

\newenvironment{myblock}{
    \par\noindent\ignorespaces
    \refstepcounter{myblockcounter}
}{
    \par\ignorespacesafterend
}
\newenvironment{issue}{
    \par\noindent\ignorespaces
    \refstepcounter{issuecounter}
}{
    \par\ignorespacesafterend
}

\def\RQone{\emph{What are the challenges of UXDs and SDEs collaboration?}}
\def\RQtwo{\emph{What are the potential best practices to overcome the challenge/to solve the problem?}}
\def\RQthree{\emph{What are the current practices and tooling support of the collaboration between UXDs and SDEs?}}

\newcommand{\thumbsup}{\begingroup
\setbox0=\hbox{\includegraphics[width=4mm, height=4mm]{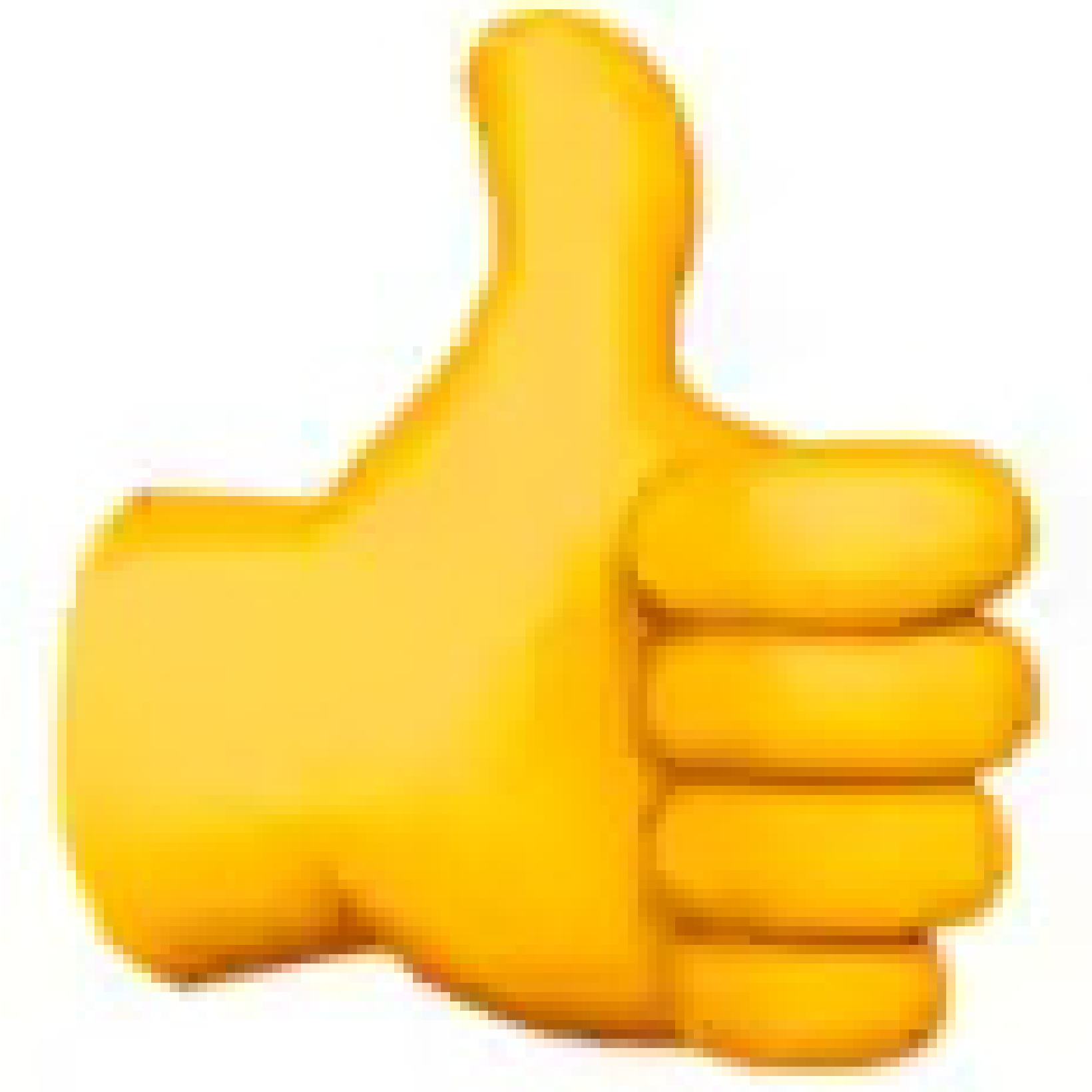}}\parbox{\wd0}{\box0}\endgroup}

\newcommand{\uxlabel}{\begingroup
\setbox0=\hbox{\includegraphics[width=10mm, height=4.5mm]{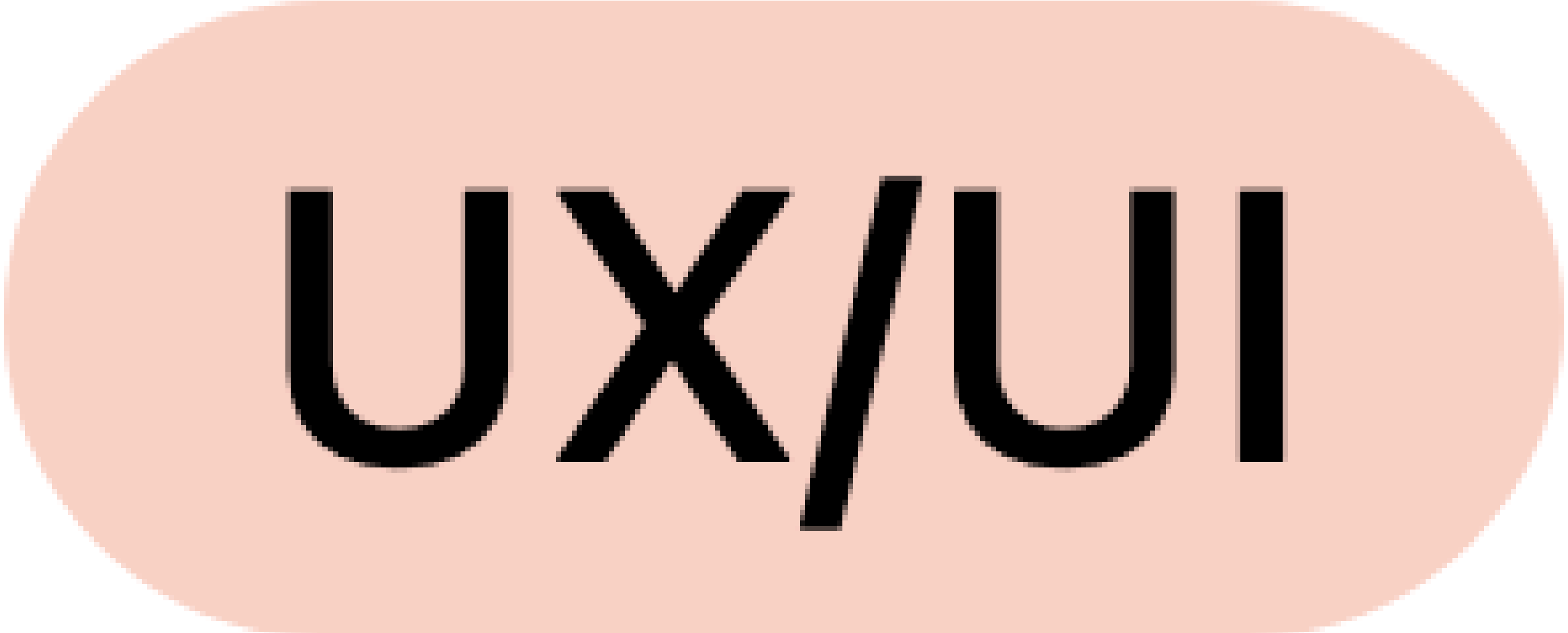}}\parbox{\wd0}{\box0}\endgroup}

\AtBeginDocument{%
  }

\setcopyright{acmlicensed}
\acmJournal{PACMHCI}
\acmYear{2025} \acmVolume{9} \acmNumber{2} \acmArticle{CSCW207}
\acmMonth{4}\acmDOI{10.1145/3711105}

\begin{document}

\title[A Comprehensive Review on Designer-Developer Collaboration]{Who is to Blame: A Comprehensive Review of Challenges and Opportunities in Designer-Developer Collaboration}

\author{Shutong Zhang}
\email{shutong.zhang@alumni.utoronto.ca}
\orcid{0009-0000-7011-5530}
\affiliation{%
  \institution{University of Toronto}
  \city{Toronto}
  \state{ON}
  \country{Canada}
}

\author{Tianyu Zhang}
\email{tianyutheodosia.zhang@alumni.utoronto.ca}
\orcid{0009-0008-8035-8040}
\affiliation{%
  \institution{University of Toronto}
  \city{Toronto}
  \state{ON}
  \country{Canada}
}

\author{Jinghui Cheng}
\email{jinghui.cheng@polymtl.ca}
\orcid{0000-0002-8474-5290}
\affiliation{%
  \institution{Polytechnique Montreal}
  \city{Montreal}
  \state{QC}
  \country{Canada}
}

\author{Shurui Zhou}
\email{shuruiz@ece.utoronto.ca}
\orcid{0000-0002-6346-6073}
\affiliation{%
  \institution{University of Toronto}
  \city{Toronto}
  \state{ON}
  \country{Canada}
}

\begin{abstract}
  Software development relies on effective collaboration between Software Development Engineers (SDEs) and User eXperience Designers (UXDs) to create software products of high quality and usability. While this collaboration issue has been explored over the past decades, anecdotal evidence continues to indicate the existence of challenges in their collaborative efforts. To understand this gap, we first conducted a systematic literature review (SLR) of 45 papers published since 2004, uncovering three key collaboration challenges and two main categories of potential best practices. We then analyzed designer and developer forums and discussions from one open-source software repository to assess how the challenges and practices manifest in the status quo. Our findings have broad applicability for collaboration in software development, extending beyond the partnership between SDEs and UXDs. The suggested best practices and interventions also act as a reference for future research, assisting in the development of dedicated collaboration tools for SDEs and UXDs.
\end{abstract}



\received{16 January 2024}
\received[revised]{11 July 2024}
\received[accepted]{09 December 2024}

\maketitle

\section{Introduction}
In the collaborative software development process, multiple groups of experts with diverse backgrounds need to collaborate to achieve a common goal. 
With the increasing demand for human-centred design and software usability, the collaboration between designers and developers has become a key factor in the achievement of success of software development. 
Numerous widely recognized software development models, such as waterfall~\cite{Boehm_waterfall, 2012_waterfall_lifecycle} and agile~\cite{2001_agilemanifesto}, encompass multiple phases that engage both software development engineers (SDEs) and user experience designers (UXDs), such as requirement elicitation, design exploration and specification, and usability verification and validation~\cite{2007_gather_requirements, paper_42}. 
Although there are a lot of discussions in the industry about how to achieve an effective UXD-SDE collaboration, these discussions are scattered and not synthesized.  
It remains uncertain how the two groups should coordinate their efforts at specific stages of the development process and how they can collaborate efficiently and communicate effectively.

UX design, first introduced in the human-computer interaction (HCI) community, aims to enhance user satisfaction when interacting with systems or products~\cite{2006_uxlifecycle}. 
After more than a decade of development and evolution, UX design has cultivated its own methodologies and procedures, many of which also follow a central lifecycle comprising stages such as analysis, design, prototyping, and evaluation~\cite{2006_uxlifecycle, 2019_UXbook}. 
When comparing the UX lifecycles with those specified in classic SE literature, several points coincide or exhibit similarities.
Nevertheless, the disparity between these two disciplines, potentially introduces complexities and challenges that often hinder effective collaboration. 

In recent years, there has been an increasing number of literature exploring how to better integrate software engineering (SE) and UX design~\cite{ paper_105, 2010_integration, 2019_integration}.
Several systematic literature reviews (SLRs) have focused on the idea of integrating UX design into agile model~\cite{2018_integration_slr, 2016_slr_integration_3, 2010_integration_slr, 2016_slr_integration}, summarizing the distinction between agile and UX design~\cite{2010_integration_slr}, with a focus on the integration~\cite{2011_integration_slr}, new frameworks that combine common practices and processes~\cite{2016_slr_integration}, and UX management approaches~\cite{2022_slr_management}.
Other research has examined the integration of SE and UX through various angles, such as processes, artifacts, and training. For example, a literature review explored the training of software development engineers in usability engineering~\cite{2010_slr_training}, while two mapping studies concentrated on usability evaluation~\cite{2015_slr_evaluation} and the artifacts used for communication~\cite{2017_slr_artifacts}. 
However, most of these studies have only concentrated on single aspects of the problem, resulting in a fragmented understanding (see detailed summary in Section.~\ref{related}). 
Additionally, it is crucial to refresh the collection of published papers up to date in order to comprehend current practices and concerns and augment academic research findings with the insights and perspectives of practitioners.

Therefore, in this research, we undertake a study that emphasizes a holistic understanding of the interaction between the two groups of experts. We aimed to achieve this through a broad systematic literature review, supplemented by the analysis of discussions in forums and open-source issue trackers. 
We aim to gain a qualitative understanding of how these two fields overlap in theory, the collaboration challenges, the current status of collaboration, and how we could promote the two groups to work together better.
With this goal in mind, we pose the following research questions: 
\begin{itemize}
\item \textbf{RQ1:} \RQone ~To answer this question, we aim to investigate the obstacles and barriers that hinder collaboration between UX designers and SE developers.
\item \textbf{RQ2:} \RQtwo  ~For this question, We aim to identify strategies that could be used to overcome the challenges summarized in RQ1.
\item \textbf{RQ3:} \RQthree  ~Lastly, we examine how the results obtained from RQ1\&2 have manifested in real-world scenarios. Specifically, we evaluate the existing tooling support to determine whether the challenges identified in RQ1 have been tackled, if the best practices outlined in RQ2 have been effectively put into practice, and if there are any remaining areas where enhancements are required.
\end{itemize}
In this study, we reviewed 45 papers spanning from 2004 to 2023 to address RQ1 and RQ2. 
In order to explore the present methods and practices (as per RQ3), we conducted searches on posts and issues within four online forums that hosted a substantial community of SDEs and UXDs, as well as investigated the issue discussions in one open-source project that involved both UXDs and SDEs. 
Our examination reveals three key collaboration challenges: (1) Separate decision-making procedures; (2) Different professional behaviors; (3) Lack of mutual understanding, and two main categories of potential best practices: (1) Improve development workflow; (2) Enhance communication. 
Additionally, by examining the search results from forums and the issue discussion in one popular open-source project, we uncovered a disparity between the existing tooling support for development work involving both UXDs and SDEs and the recommended best practices. 
Together, our findings provided insights related to collaboration suggestions and future research directions.

In summary, we make the following contributions: (1) We present a summary of three key challenges, further categorized into eleven detailed classifications, and connect them with two main best practices, which are divided into six sub-practices to improve the efficiency of collaboration between UXDs and SDEs; (2) we highlight the gaps between current practices and tooling support and the proposed best practices; (3) we identify recommendations for improvement and multiple future research directions.

\section{Related Work} \label{related}
In this section, we summarize the prior studies focusing on collaboration challenges during the software development procedure, and then justify the importance of conducting an SLR on UXD-SDE collaboration within the HCI, CSCW, and SE communities.

\subsection{Collaboration Scenarios During Software Development Process}
The majority of software projects go beyond the capacity of a single SDE, necessitating collaboration among multiple SDEs either within the same team or across different teams. 
Researchers have explored collaboration practices and challenges among SDEs in diverse contexts, including co-located environments~\cite{strode2012coordination,bjarnason2022inter,anslow2014reflections}, distributed environments~\cite{noll2011global,herbsleb2007global,gutwin2004group}, as well as the transition~\cite{boland2004transitioning,klein2012scaling} and comparison~\cite{jolak2020design,dekel2005supporting} between the two scenarios. 
Moreover, collaboration scenarios can also be distinguished between asynchronous ~\cite{hattori2010enhancing} and asynchrous~\cite{sangwan2020asynchronous}, each presenting distinct tradeoffs that require careful consideration~\cite{green2010understanding}.
Researchers have examined various forms of awareness in collaborative work and have developed tools~\cite{dourish1992awareness} to consolidate team members' activities, enhancing awareness throughout the collaboration process~\cite{dabbish2012social,gutwin2004importance,sarma2011palantir,treude2010awareness}.
Recently, with the incorporation of machine learning (ML) techniques has resulted in an increasing number of software systems integrating ML components. This development has necessitated SDEs to engage in collaboration with ML experts, giving rise to interdisciplinary collaborative scenarios and the associated challenges covering different aspects (i.e., Communication, Documentation, Engineering, and Process)~\cite{mailach2023socio,2022_collaboration_ml}. 

Likewise, the HCI community has explored the collaboration among UXDs in the development of interactive systems~\cite{feng2023understanding} to improve communication and coordination efficiencies by analyzing collaborative UX tools~\cite{feng2022handoffs}. 
Along the line of designing ML-based systems, researchers looked into how UXDs work with data scientists to build high-quality human-centred AI products~\cite{zdanowska2022study}.

\subsection{Related SLRs on SE and UX Integration}
\label{ref:RelatedSLR}
The significance of software system usability in recent times has led to a growing body of literature that delves into the integration of SE and UX methodologies and processes~\cite{paper_105, 2010_integration, 2019_integration}.
Accordingly, SLR research~\cite{2016_slr_integration_3, 2014_slr_integration,2011_integration_slr} and systematic mapping studies~\cite{2016_slr_integration} have been conducted to summarize the overview of how user-centered design is introduced into the Agile development process. 

In particular, in 2010, Sohaib et al. conducted a literature review on difficulties and suggested approaches for usability engineering and agile software development integration~\cite{2010_integration_slr}. 
However, the study was not conducted systematically and it was unclear about paper selection, and screening procedure, leaving the quality of the study results questionable. 
Later in 2011, Silva et al. reviewed the integration of user-centred design and agile methods, with a focus on the merged process model and the artifacts that were used to support the collaboration and did not include human factors~\cite{2011_integration_slr}. 
Similarly, in 2016, Caballero et al. conducted a systematic mapping study, covering papers from 2001 to 2013, to understand how agile teams integrate user-centred design tools and work with UXDs to better understand user requirements~\cite{2016_slr_integration}. 
The research outlined various team structures that were reported from the paper corpus, such as cross-functional teams comprising both UXDs and SDEs,  as well as independent teams of SDEs and UXDs that collaborate on a project basis. 
Additionally, the paper listed several strategies that prior work reported to ensure collaboration efficiency, such as continuous communication and close collaboration between co-located UX and SE experts. 
However, the paper only listed the summaries from prior work and did not synthesize the findings into categories and themes. 

\looseness = -1
The closest SLR study to our work was conducted by Salah et al. which summarised 71 research papers published between 2000 and 2012, primarily examining the interaction between agile developers and user-centred designers~\cite{2014_slr_integration}. They identified several collaboration challenges, including difficulties in performing UCD activities, UX testing, and optimizing work dynamics. 
However, the primary emphasis of their study was on the challenges of process integration, leaving notable gaps in understanding the distinct professional behaviours and other social aspects during the collaboration process. 

\looseness = -1
Moreover, several SLRs focused on specific activities between SDEs and UXDs, such as the evaluation methods in Agile-UX~\cite{2015_slr_evaluation}, the training condition of SDEs in UX domain~\cite{2010_slr_training}, and the artifacts in agile user-centered design~\cite{2017_slr_artifacts}. The most recent SLR was conducted by Hinderks et al. in 2022, primarily focusing on the approaches that manage the UX process in the agile development model~\cite{2022_slr_management}. 
This review analyzed 49 relevant studies and identified several approaches for UX management.
The study treated ``collaboration and communication'' as a simple and successful approach to manage UX-related tasks.
Unlike previous reviews, our study focuses primarily on the collaboration aspects throughout the whole development process, covering various development models and also open-source context. 
Additionally, we conducted qualitative analysis on four popular forums that SDEs and UXDs tend to use and one popular open-source project in order to complement the findings from academic papers.

\section{Review Protocol}
\begin{figure}[h!]
    \centering
    \begin{minipage}{0.49\textwidth}
        \centering
        \includegraphics[width=1.0\linewidth]{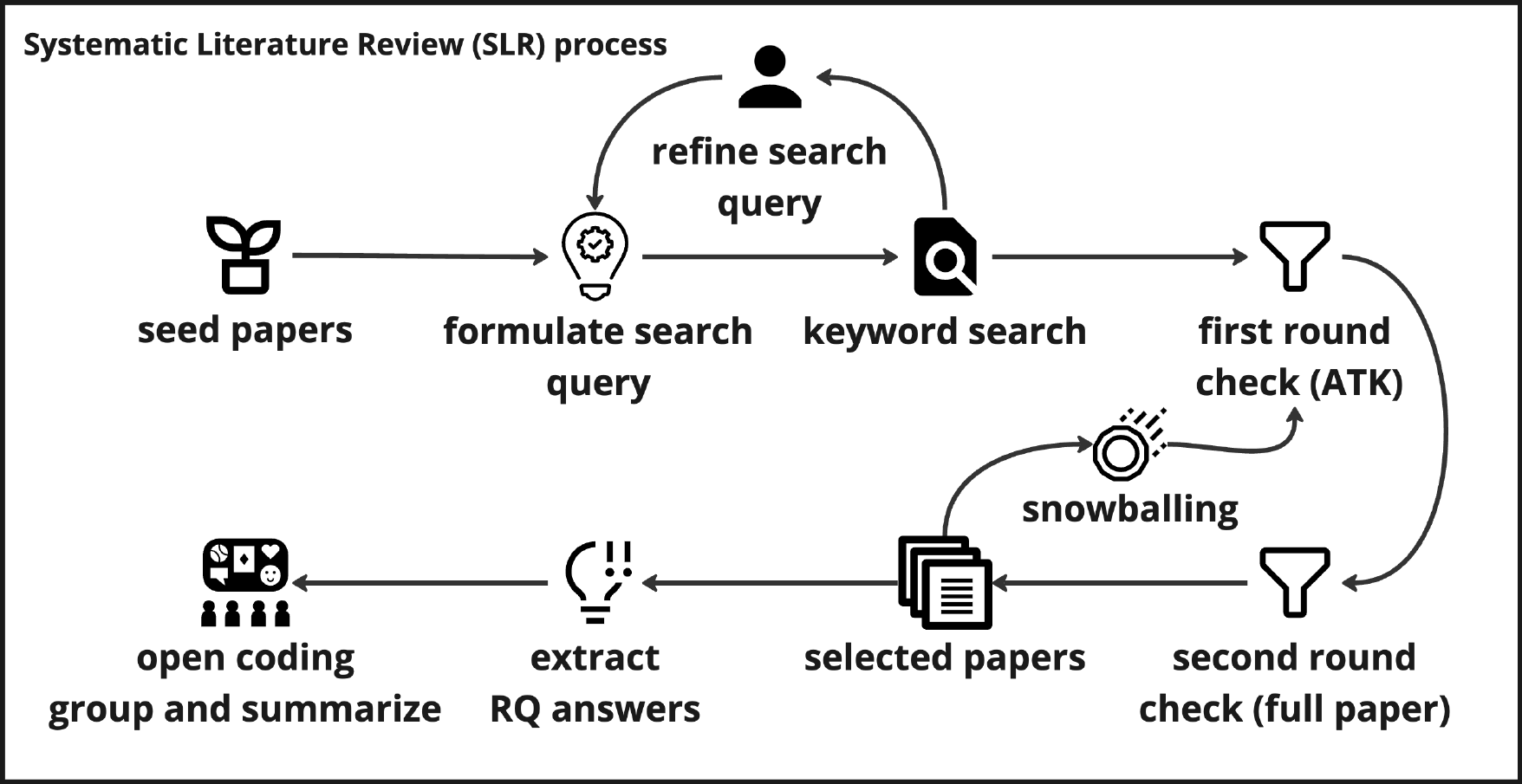} 
        \subcaption{Paper selection process}
    \end{minipage}%
    \hfill
    \begin{minipage}{0.49\textwidth}
    \centering
    \footnotesize
    \begin{tabular}{lrrr}
        \toprule
        \textbf{Database} & \textbf{\makecell{Initial \\ Papers}} & \textbf{\makecell{First \\ Round}} & \textbf{\makecell{Final \\ Papers}} \\
        \hline
        IEEE & 1855  & 37 & 13 \\
        ACM & 1430  & 25 & 11 \\
        SpringerLink & 173  & 13 & 5 \\
        Weily & 106 & 8 & 3 \\
        Engineering Village & 2172  & 37 & 9 \\
        Snowballing & -  & 12 & 4 \\
        \textbf{Total} & \textbf{5736}  & \textbf{132}  & \textbf{45} \\
        \bottomrule
    \end{tabular}
    \subcaption{Paper selection statistics}
\end{minipage}
    \caption{Paper Selection Process and Statistics. We first applied the keyword search and then conducted two rounds of filtering. Sub-figure (a) shows the overview of the selection process and sub-figure (b) shows the number of papers in each stage.}
    \label{fig:paper_selection}
\end{figure}

Our study relied on two different sources: (1) a systematic literature review  (SLR), and (2) a qualitative analysis of the practitioners' discussion in forums and issue trackers. 
Specifically, the SLR study covering five online databases aims to answer RQ1 (collaboration challenges) and RQ2 (best practices) following the guidelines established by Kitchenham et al.~\cite{2007_SLRguide,kitchenham2013systematic} and the PRISMA framework~\cite{moher2009preferred}. 
Figure~\ref{fig:paper_selection} shows an overview of the paper selection process.
Then to complement the SLR results, we analyzed posts and issues from four designer-developer forums and investigated the issue discussions from open-source projects hosted on GitHub, respectively. 
In the following, we detail the methods used to conduct the SLR 
(Section~\ref{ref:SLR}) and the qualitative analysis 
(Section~\ref{ref:GLR}).

\subsection{Systematic Literature Review}
\label{ref:SLR}

\subsubsection{Identification: Keyword Searching.}  Similar to prior SLRs on UX and SE focusing on process integration~\cite{2010_integration_slr, 2011_integration_slr, 2014_Integration_slr}, we divide our search string into SE and UX components. 
We first manually selected 14 papers as references and keywords from prior SLRs as initial keyword selections~\cite{2010_integration_slr, 2011_integration_slr, 2014_Integration_slr}, then iteratively refined the keyword list by using them to search papers in selected databases until all papers were covered. In specific, the keyword ``agile'' was selected because many related papers are from the Agile Conference.\footnote{\url{https://agile-gi.eu/}}
As a result, we defined two keyword lists used for querying papers (as shown in Table \ref{tab:keywords}). Combining all keywords, the final string we used for keyword search is presented as follows: \texttt{UX AND SE}.
\begin{table}[h]
\small
\centering
\caption{Keywords used to formulate searching string in the SLR. Combining all keywords, the final string we used for keyword search is presented as follows: \texttt{\textbf{UX} AND \textbf{SE}}.}
\begin{tabular}{>{\bfseries}c l}
\toprule
\textbf{Type} & \textbf{Search String} \\
\midrule
\texttt{\textbf{UX}} & \texttt{``designer'' OR ``usability'' OR ``ux'' OR ``user experience'' OR} \\
      & \texttt{``user-centered(centred)'' OR ``user interface''}  \\
\midrule
\texttt{\textbf{SE}} & \texttt{``developer'' OR ``software development'' OR ``software developing'' OR}\\
   &\texttt{``software process'' OR ``software engineer'' OR ``agile''} \\
\bottomrule
\end{tabular}
\label{tab:keywords}
\end{table}

We applied the above searching string in abstract, title, and keywords (ATK) following the guidelines~\cite{2007_SLRguide} in the IEEE Xplore digital database, the ACM Digital Library, the SpringerLink Digital Library, the Wiley Online Library, and the Engineering Village. Two co-authors iteratively examined papers randomly chosen from the search results to determine the relevant data source.  

\subsubsection{Screening, Eligibility, and Inclusion.} 
Following identification, we conducted screening and eligibility assessments, taking into account the specified inclusion criteria (IC) and exclusion criteria (EC) that were iteratively formulated.
In particular, an included paper must: 
\begin{itemize}
    \item \textbf{IC1}: be published in peer-reviewed journals or event proceedings;
    \item \textbf{IC2}: be written in English;
    \item \textbf{IC3}: be published between 2004 and 2023. Since the fields of SE and HCI are rapidly developing, we restrict the start year to 2004 to focus on papers published in the past 20 years, so that our study covers the most relevant information; 
    \item \textbf{IC4}: contains pertinent information related to either collaboration challenges between designers and developers, or best practices to facilitate collaboration between them.
\end{itemize}
At the same time, we exclude a paper if: 
\begin{itemize}
    \item \textbf{EC1}: The paper is a work-in-progress, poster, demo, or an extended abstract;
    \item \textbf{EC2}: The paper is a patent or standard;
    \item \textbf{EC3}: The paper is less than four pages;
    \item \textbf{EC4}: The paper is published in a local conference (Appendix \ref{apd:keywords} lists the collection of keywords);
    \item \textbf{EC5}: The main topic is not UXD and SDE collaboration but other fields (Appendix \ref{apd:keywords} lists the collection of these keywords);
    \item \textbf{EC6}: The paper constitutes secondary studies (i.e., an SLR or informal literature survey), to ensure concentration on primary studies~\cite{2007_SLRguide}.
\end{itemize}

\subsubsection{Paper Selection Process.}
The initial keyword search resulted in 5,736 papers from five databases published between January 2004 and August 2023. 
After removing duplicate papers, we have 5,112 papers in total. 
During our random checking, we observed that our initial result contained a large number of irrelevant papers. Thus, we implemented
a keyword filter to exclude papers irrelevant to software engineering or UX design. Specifically, we established a range of keywords including terms such as ``linear algebra,'' ``computer vision,'' ``gravity,'' and others. A full list of keywords is available in Appendix~\ref{apd:keywords}. 
We filtered out papers that contain at least one keyword from the list in their title, keywords, or abstract, resulting in 3,860 remaining papers.

\begin{figure}[h!]
    \centering
    \includegraphics[width=1.0\linewidth]{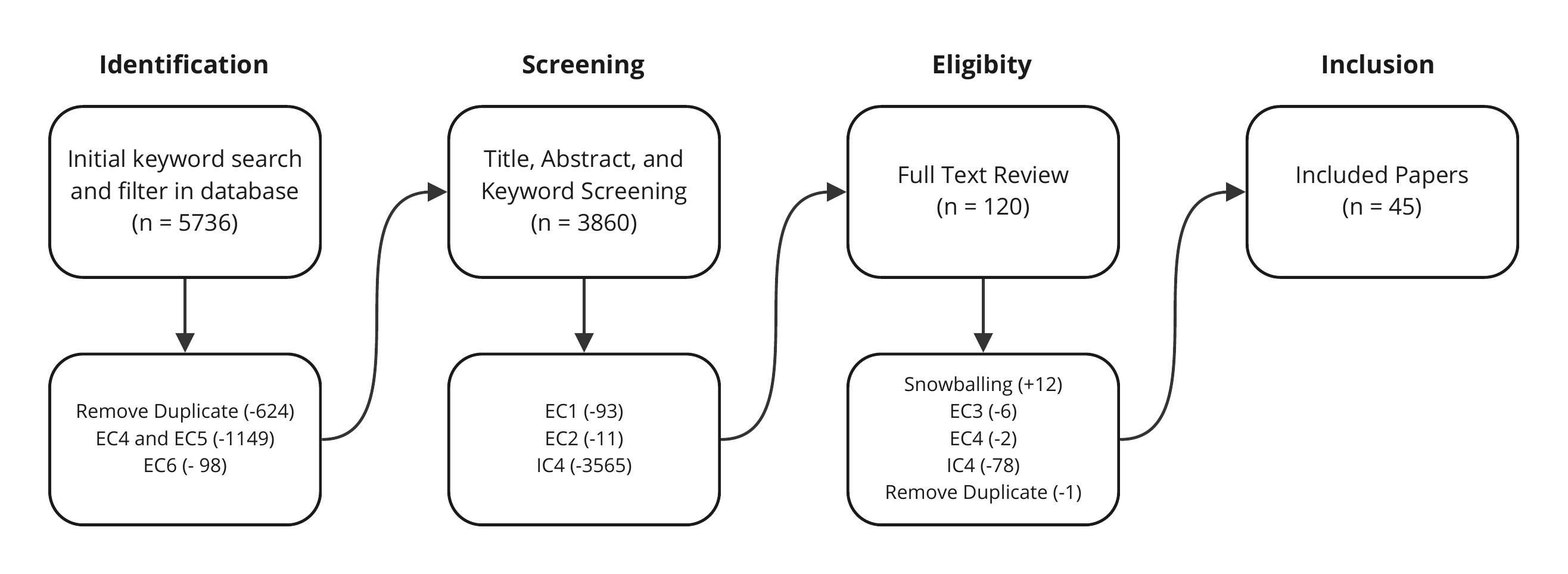} 
    \caption{Data collection process and statistics following Identification, Screening, Eligibility, and Inclusion.}
    \label{fig:process_stat}
\end{figure}

We then conducted the following three iterative processes of paper selection and information extraction as described below. To ensure accuracy and consistency, we iteratively evaluate and refine the review protocol and criteria before and after each process. For each paper, we define three types of labels, including \emph{Y} for include, \emph{N} for exclude, and \emph{NA} for not sure at each step. Figure.~\ref{fig:process_stat} shows the data collection process.

\begin{itemize}
    \item (\textbf{ATK-only}) We reviewed the abstracts, titles, and keywords (ATK) of all the papers, accessing them against the inclusion and exclusion criteria. In terms of eligibility, each author first reviewed 100 papers independently and then held meetings to address conflicts. After completing one round of labeling, we assigned papers with labels \emph{Y}, \emph{NA}, and $20\%$ papers with label \emph{N} to a second reviewer for cross-checking.  For papers with different opinions, we held meetings for discussion until the team reached an agreement. 
    This process led to the identification of 120 papers.
    \item  (\textbf{Full-text review})  We thoroughly examined the full text of each 
    paper against the criteria. Similar to the previous round, we carried out a pilot check to refine the review criteria and then applied the criteria for paper screening. This step led to 41 papers at the end.
    \item (\textbf{Snowballing}) While reading papers, we also applied backward snowballing to search for missing related papers in the initial keyword search~\cite{2014_snowballing}. Similar to the process described before, we filter out irrelevant papers and uncover responses to RQs. The initial snowballing process identified 12 papers; among them, 4 papers were added to the final review, resulting in 45 papers in our corpus.
\end{itemize}

\subsubsection{Data Analysis.}

\emph{\textbf{Extracting findings.}} Each author read a set of papers and then highlighted the quotes from the paper that are related to collaboration challenges (RQ1) and best practices (RQ2). Each paper was then assigned to a second author to cross-check and then discussed during meetings to resolve conflicts and uncertainty.
We extracted a total of 126 excerpts related to challenges and 66 excerpts related to best practices that could mitigate challenges. 

\emph{\textbf{Categorizing findings that share topical similarities using affinity diagramming.}}
Various papers organized their findings employing diverse terminologies and grouping approaches. 
In our project, we aim to present a comprehensive overview of both the challenges and corresponding solutions through a cohesive and consistent methodology. 
Therefore, we conducted affinity diagramming~\cite{affinity_diagrams} to group similar findings together. 
Adhering to the guidelines of the approach, we performed the activity on Miro.com by creating one sticky note for every highlighted sentence from our corpus. 
Two authors created the initial grouping and other authors validated the initial grouping and raised different opinions. Then all authors discussed disagreements during meetings, refined the groupings, and obtained the final themes reported in the results.
Regarding the sticky notes associated with challenges, we revisited the relevant sections to grasp the context, noting the possible causes and impacts of the challenges if they are discussed in the papers. 
For the sticky notes related to practices that mitigate challenges, we grouped them into themes and linked them to corresponding challenges that were identified in the first step. 

\subsection{Analyzing SE\&UX Discussions in Forums and Issue Trackers}
\label{ref:GLR}
To triangulate and complement the findings from SLR and further investigate the current state of SE\&UX collaboration in real-world settings, we qualitatively analyzed the relevant discussions targeting the SE-UX collaboration issues in four designer-developer forums. 
Furthermore, we perform a brief analysis of a widely used open-source project (i.e., VS Code) in anticipation of discovering firsthand instances of collaboration communication between UXDs and SDEs during the open-source development process.

\subsubsection{Data Selection From Designer-Developer Forums} \hspace*{\fill} \\
\mypara{Search Strategy.} 
We focused on four forums, including UX Stack Exchange~\cite{UX_stackexchange}, Reddit UX~\cite{Reddit_UX}, Reddit SE~\cite{Reddit_SE}, and Stack Overflow~\cite{stack_overflow}, that are renowned for hosting communities of professionals that provide experiences and discussions on SDEs\&UXDs collaboration.
Similar to SLR, we utilized the keyword search strategy in post titles to retrieve relevant posts. In addition to the initial keywords listed in Table~\ref{tab:keywords}, we added keywords related to collaboration and tooling to make the search more specific; Table~\ref{tab:glr_keywords} shows additional keywords we used in the post search.
Moreover, we did not use the UX keywords in UX forums and likewise for SE keywords, since by default, these forums will touch on the corresponding topics.
Combining all keywords, the final string we used for keyword search is presented as follows: \texttt{(forum\_type == software\_engineering ? UX : SE) OR Collaboration OR Tool}. 

\begin{table}[h]
\small
\centering
\caption{Additional keywords used to formulate searching string in the Post search. Combining keywords from Table~\ref{tab:keywords}, the final string we used for pose and issue search is presented as follows: \texttt{(forum\_type == software\_engineering ? \textbf{UX} : \textbf{SE}) OR \textbf{Collaboration} OR \textbf{Tool}}.}
\begin{tabular}{>{\bfseries}c l}
\toprule
\textbf{Type} & \textbf{Search String} \\
\midrule
Collaboration & \texttt{``collaboration'' OR ``collaborate'' OR ``cooperation'' OR } \\
      & \texttt{``cooperate'' OR ``challenge'' OR ``gap''}  \\
\midrule
Tool & \texttt{``tool'' OR ``platform''}\\
\bottomrule
\end{tabular}
\label{tab:glr_keywords}
\end{table}

\mypara{Selection Process.}
Similar to RQ1 and RQ2, we established inclusion and exclusion criteria to identify the relevant information to answer RQ3. In specific, an included post must:
\begin{itemize}
    \item \textbf{IC1}: be published in 2020-2023 to ensure the timeliness of our search results.
    \item \textbf{IC2}: contains pertinent information related to collaboration challenges, best practices, or collaboration tools.
\end{itemize}
Simultaneously, we consider the following exclusion criteria exclude a post if the post:
\begin{itemize}
    \item \textbf{EC1}: is not related to UXD-SDE collaboration.
    \item \textbf{EC2}: does not contain follow-up comments from either UXDs or SDEs.
\end{itemize}
Our initial keyword search resulted in 1,299 posts in four different online forums from 2020 to 2023. Two authors first checked the titles and main threads of all posts from the keyword search, ensuring that more than one author reviewed each post, then excluded posts following EC1 and EC2.
The initial review resulted in 65 posts. 
Next, two authors conducted a second round of filtering by reading the entire post and associated comments to check whether the post met IC2, a third author then validated the result.
This round of filtering identified 17 irrelevant posts, resulting in 48 posts from the four forums to address RQ3.
Table~\ref{tab:post_issue} shows the post and issue search statistics.

\begin{table}[h]
\small
\caption{Posts and Selection Statistics. This table shows the number of posts selected after each stage during the search process, the average number of responses, and their distribution. }
\begin{tabular}{lccccc}
        \toprule
        \textbf{Forum/Project} & \textbf{\makecell{Initial \\ Posts}} & \textbf{\makecell{After \\ Filter}} & \textbf{\makecell{Selected \\ Posts}} & \textbf{\makecell{Avg. \\\#Responses}} &
        \textbf{\makecell{\#Responses \\ Distribution}}\\
        \hline
        UX Stack Exchange & 275 & 14 & 13 & 7.3 & \includegraphics[width=0.1\linewidth]{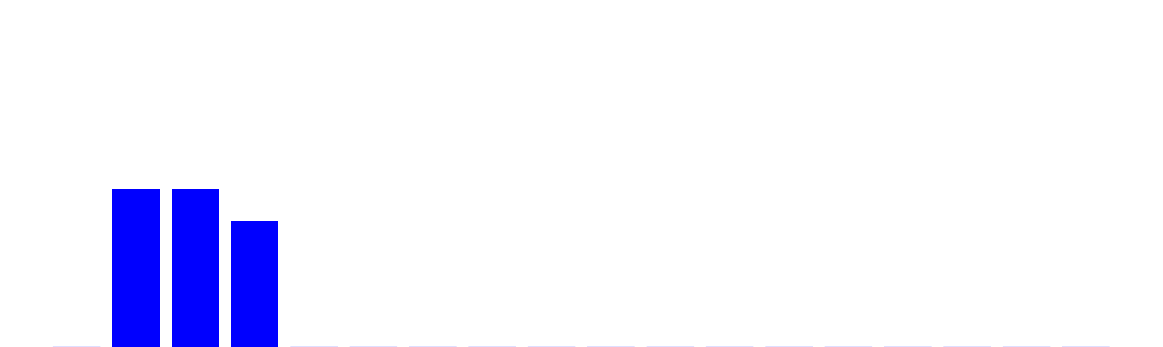} \\
        UX Reddit  & 742 & 47 & 32 & 23.3 & \includegraphics[width=0.1\linewidth]{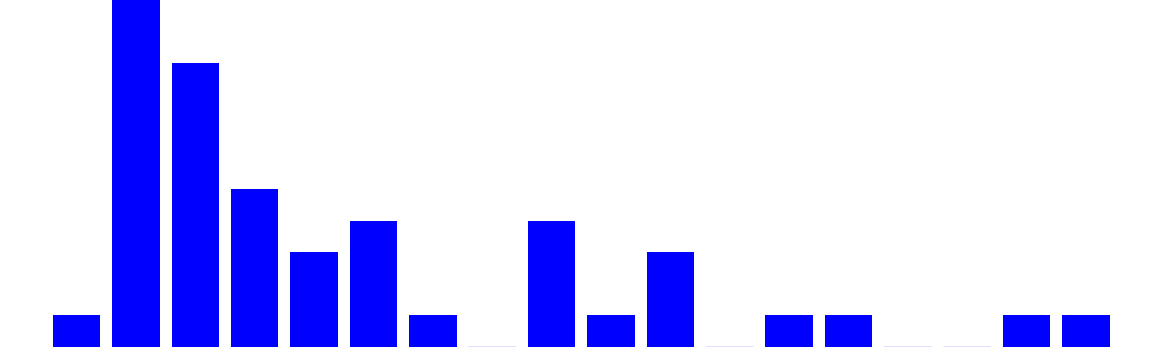}\\
        SE Reddit  & 201 & 4 & 3 & 6.3 & \includegraphics[width=0.1\linewidth]{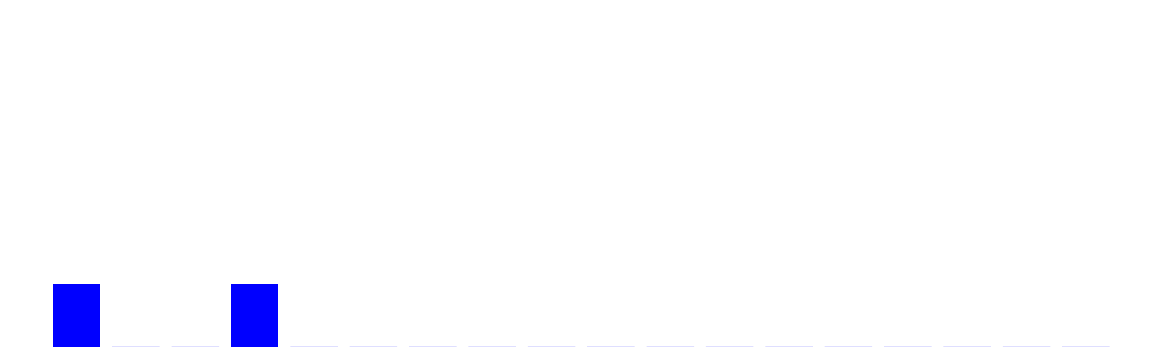}\\
        Stack Overflow  & 81 & - & - & - & -  \\
        \textbf{Total} & \textbf{1299} & \textbf{65} & \textbf{48} & \textbf{18.8} & \includegraphics[width=0.1\linewidth]{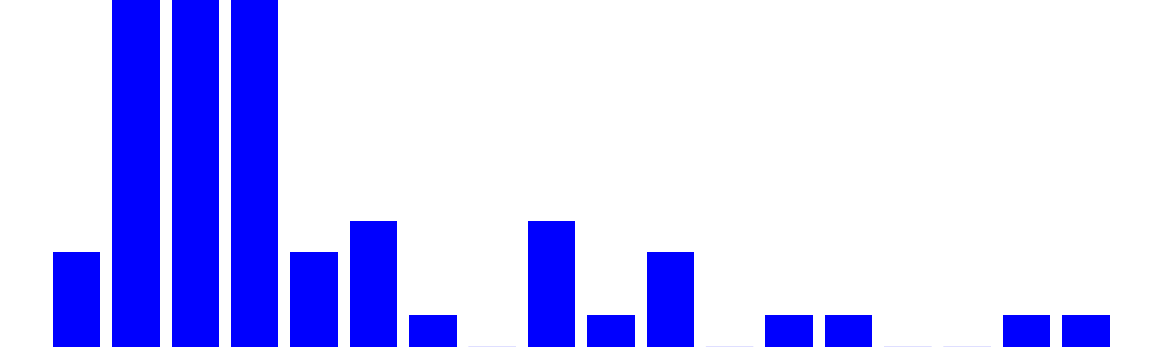} \\
        \bottomrule
    \end{tabular}
\label{tab:post_issue}
\end{table}

\subsubsection{Investigating Issue Discussions in Open-source Project} \hspace*{\fill} \\
Last but not least, we searched on GitHub for popular projects with a graphical user interface (GUI), which might contain instances of UXD-SDE interactions and collaboration. 
We defined the following inclusion criteria:
\begin{itemize}
    \item \textbf{IC1:} The project contains a GUI;
    \item \textbf{IC2:} The project contains issues labeled with UX.
\end{itemize}

Correspondingly, we also defined:
\begin{itemize}
    \item \textbf{EC1:} The issue participants 
 should include both UXDs and SDEs; 
    \item \textbf{EC2:} The project does not consistently utilize issue labels.
\end{itemize} 

We identify the background of each issue participant by examining their GitHub profile portfolio. Additionally, we conduct searches for their names to locate personal websites or LinkedIn accounts, if available.
We initially reviewed the top 20 GitHub repositories with the most stars from Gitstar Ranking\footnote{gitstar-ranking.com/} that meet IC1\&2.
The results show that the VS Code project\footnote{github.com/microsoft/vscode/} is the only project among the top 20 that meets our criteria. Since its creation on September 3rd, 2015,  the project has accumulated over 1.8k issues that have UX labels out of a total of 195k issues.  The earliest issue marked with a UX label dates back to late 2015.
To discover more appropriate projects, we conducted a manual review of the top 100 GitHub repositories with the highest number of stars that meet IC1\&2.
Subsequently, we identified additional four repositories, including \emph{Excalidraw},\footnote{github.com/excalidraw/excalidraw/}  \emph{Terminal},\footnote{github.com/microsoft/terminal/} \emph{Godot},\footnote{github.com/godotengine/godot/} and \emph{Storybook}.\footnote{github.com/storybookjs/storybook/}
However, we have discovered that none of the four projects are suitable for our investigation purpose. 
For example, although \emph{Excalidraw} used the UI/UX label to categorize the issues, it is not used consistently and the majority of the issues are not labeled, so we filtered it out by EC2. 
Moreover, we found it challenging to determine whether a participant is a developer or a designer for issues in \emph{Godot} and \emph{Storybook}.
Considering that both Terminal and VS Code are developed by Microsoft, it is plausible to assume they follow comparable development practices. Therefore, as an initial exploratory case study, our focus will be on VS Code, which was created in 2015, two years prior to the creation of Terminal in 2017. Future research could further distinguish project practices by examining individual repositories rather than by GitHub account.

We utilized the GitHub API to fetch those issues from the \emph{VS Code} project. Specifically, we initially chose issues labelled as ``ux'' and those containing video/image attachments, representing instances of UI/UX design artifacts and discussions~\cite{paper_41}. This process yielded a total of 1,019 issues. Then, we filtered out issues that violated the following criteria: the issue (1) has no follow-up comments; (2) is still open; (3) has ``bug,'' ``duplicate,'' ``out of scope,'' ``debt,'' or ``as-designed'' label; (4) involves no SDE or UXD, as they are not able to reflect sufficient communication and collaboration between SDEs and UXDs.
During this stage, 259 issues have been filtered out, leaving us with 760 issues needing review.

To filter out issues irrelevant to RQ3, we applied a second round of checks to eliminate issues that do not contain collaboration/discussion between UXDs and SDEs. 
The first two authors labeled the 760 issues with Y (include),  N (exclude), and NA (not sure). 
Another author reviewed issues labeled with NA to re-label them as Y or N.
Finally, 157 issues were selected. 
Lastly, the first two authors then read each issue separately to decide whether it complements collaboration challenges or good practices.

\subsubsection{Data Analysis} \hspace*{\fill} \\
For the selected posts and issues, two authors read them in detail and extracted information relevant to collaboration challenges and best practices. 
We then applied affinity diagramming to organize similar information into groups.
Following that, we held meetings to address conflicts that arose during grouping and finally used refined groups to answer RQ3. 
\renewcommand{\arraystretch}{1.5}

\subsubsection{Limitations and Threats to Validity}\hspace*{\fill} \\
Every research methodology possesses constraints that pose potential threats to the validity and credibility of its findings. 
Despite the fact that we carefully employ selection methods, it is possible that we may have overlooked certain papers.
Some reflections on collaboration between SDEs and UXDs might not be included only in peer-reviewed publications. To address this, we conducted one round of backward snowballing to include additional relevant studies. We also complemented our study by searching online forums and issue discussions of an open-source project on GitHub. In addition, our selection of included papers could potentially be biased because of the subjectivity of authors. We attempted to mitigate the potential bias by iteratively refining the selection criteria and cross-checking among the researchers. Each researcher indicates their decisions while reviewing the papers. Those results were cross-checked by another researcher, and discussion meetings were held to arrive at a consensus when there was a disagreement.

Another limitation to consider is that our data collection was primarily sourced from English-language forum sites. Consequently, the generalizability of our findings may be limited to the global communities of  SDEs and UXDs. This lack of diversity in representation could potentially influence our observations regarding prevailing practices and the adoption of tools, as discussed in Section~\ref{ref:GLR}. Similarly, we only searched for popular repositories on GitHub with GUI, and filtered out those lacking UX labels in their issue lists. This selection process resulted in the inclusion of just one project from GitHub, respectively. It is important to note that this aspect of our research aimed to establish an initial understanding of the collaboration between UXDs and SDEs, and a comprehensive comparison among various projects as well as an assessment of collaboration efficiency were beyond the scope of this study. The findings from our work serve as a promising first step in identifying best practices and challenges of collaboration between UXDs and SDEs during the development lifecycle.

\section{Results}
\looseness=-1
Below we report the analysis results based on the 45 papers selected from the SLR, as well as the 48 posts (`Px' to represent the ID) and 157 issues (`Ix' to represent the ID) from the post and issue search.  
\subsection{Quantitative Overview}
\begin{figure}[h]
\centering
\includegraphics[width=\linewidth]{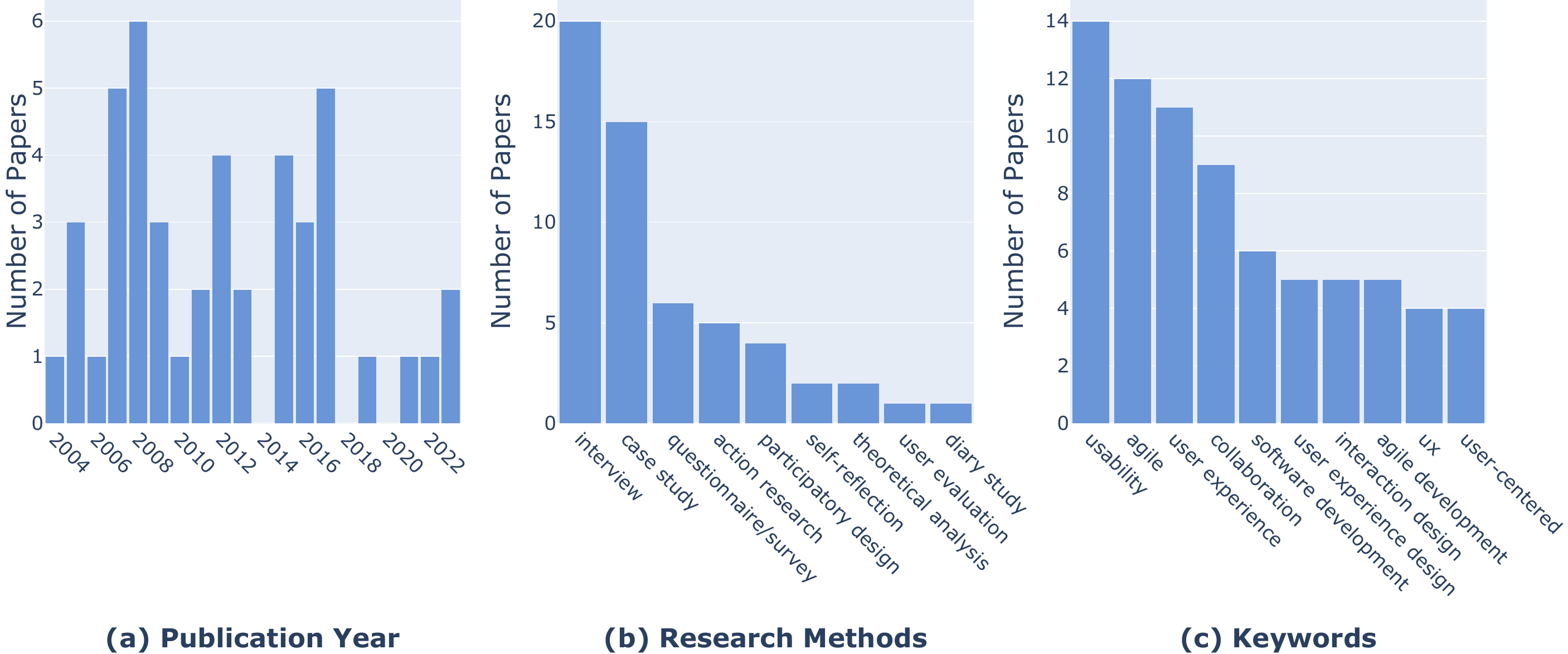}
    \caption{Statistics summary of the selected papers in terms of their (a) Publication Years, (b) Research Methods, and (c) Author Keywords.}
    \label{fig:method_keyword_year}
\end{figure}

\mypara{Publication Years.} Figure~\ref{fig:method_keyword_year}(a) shows the distribution of publication years of all selected papers in the SLR. 
It demonstrates a clear increasing trend in selected papers from 2004 to 2008. 
Despite fluctuations from 2004 to 2018, it is noteworthy that about $90\%$ of selected papers were published before 2018. 
This suggests a decline in the topic's popularity in recent years, indicating a waning interest among researchers.

\begin{figure}[h]
\centering
\begin{tabular}{@{}c@{\hspace{1mm}}c@{\hspace{1mm}}c@{}}
\includegraphics[width=0.325\linewidth]{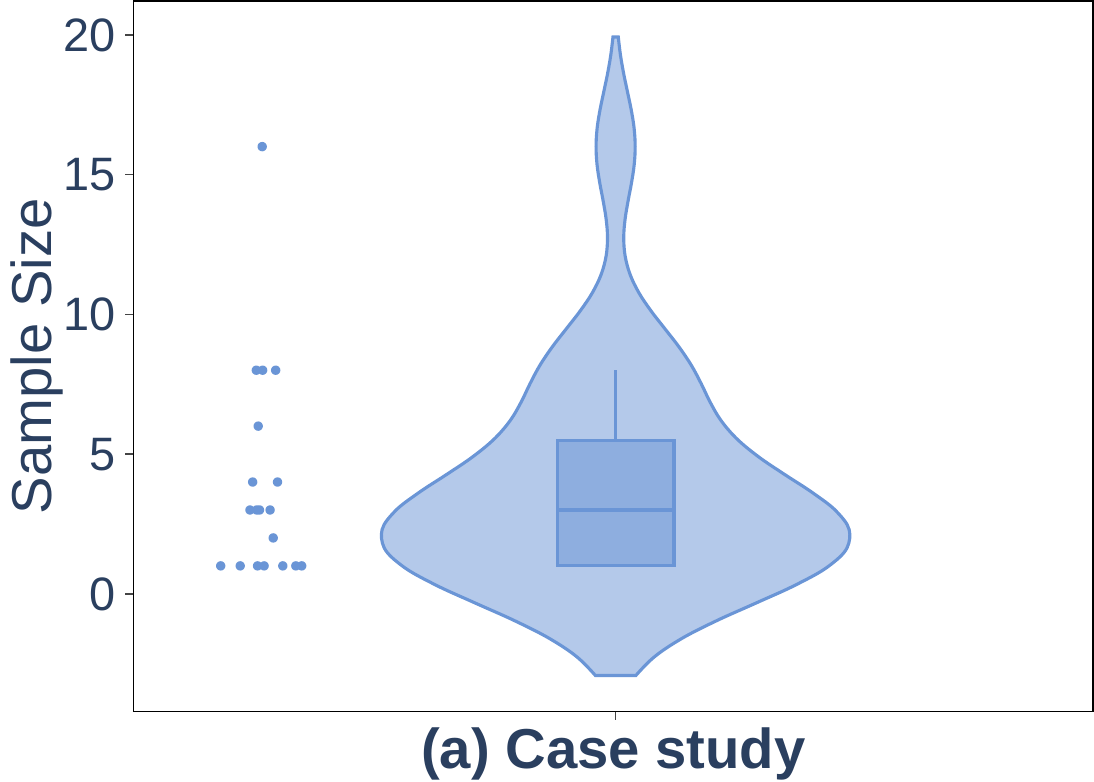} &
\includegraphics[width=0.325\linewidth]{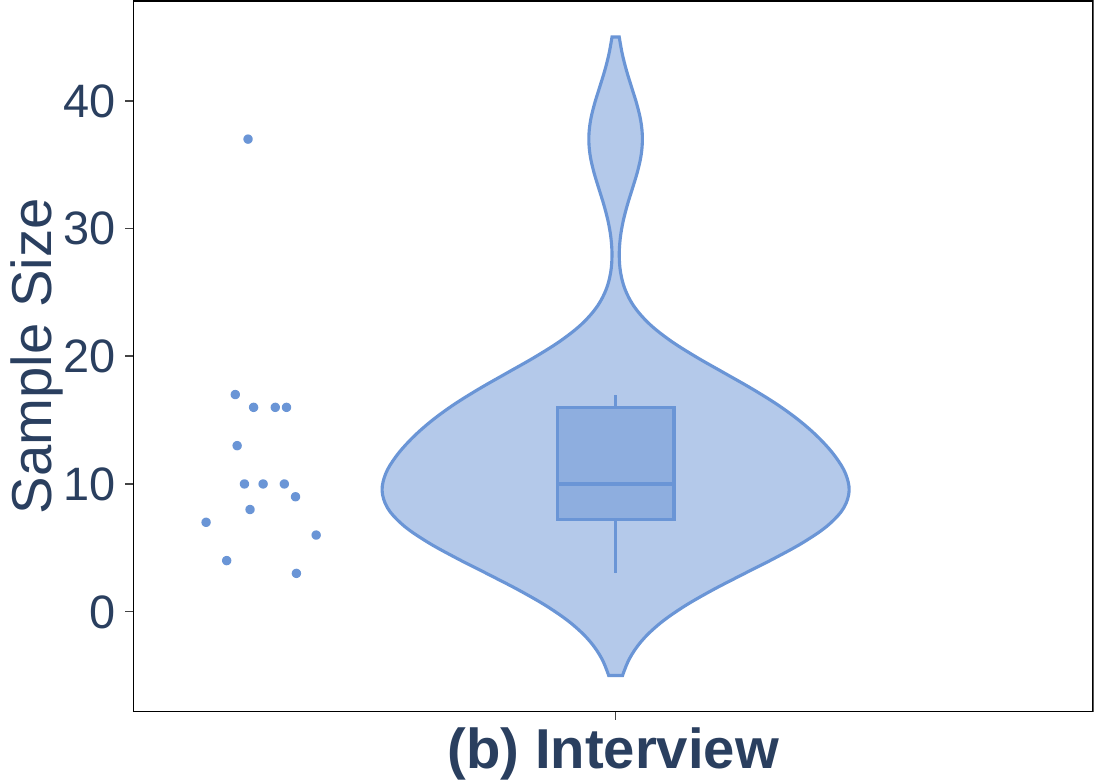}  &
\includegraphics[width=0.325\linewidth]{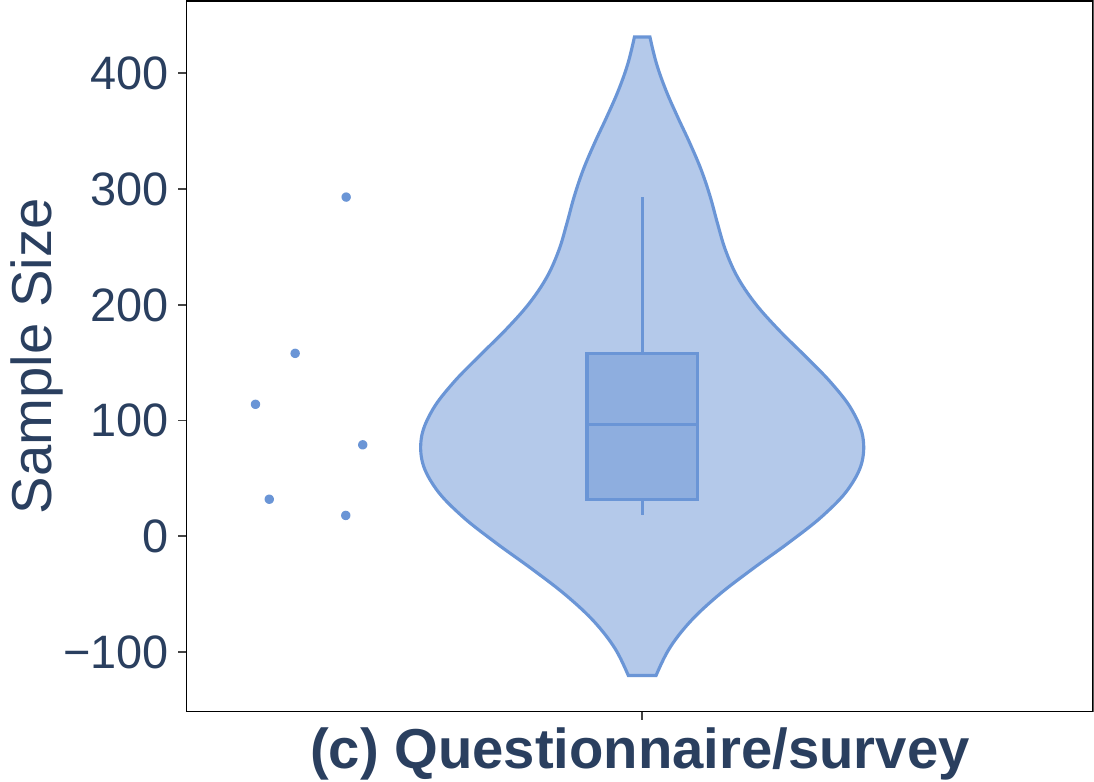}\\
\end{tabular}
    \caption{Research Methods Sample Size: (a) Case study, (b) Interview, (c) Questionnaire/survey.}
    \label{fig:sample_size}
\end{figure}

\mypara{Research Methods.} 
Figure~\ref{fig:method_keyword_year}(b) shows the frequencies of different types of research methods adopted by the reviewed papers. Noted that one paper may contain multiple research methods.
The three most widely utilized research methods are interview, ethnography/case study, and questionnaire/survey, used by 45\%, 34\%, and 14\% papers, respectively.
These three methods all involve human participants; the distributions of the participants' sample size in the papers that applied these three methods are presented in Figure~\ref{fig:sample_size}.

\looseness=-1
\mypara{Keyword Distribution.}
We also summarized author keywords of selected papers, presented in Figure~\ref{fig:method_keyword_year}(c). 
The most frequently appearing keywords are usability and agile, which appear in 14 and 12 papers, respectively. 
All popular keywords shown in the figure are relevant to UX, SE, or collaboration. 
This indicates that the papers we selected for the SLR match our intended research direction. 

\subsection{(RQ1) Challenges of UX and SE Collaboration} \label{RQ1}
\label{ref:RQ1}

In this research question, we aim to investigate the collaboration challenges between UX designers and SE developers. 
Among the 45 reviewed papers, 35 directly mentioned challenges regarding inefficient UXD-SDE collaboration. We classified these challenges into three main categories, each including several detailed sub-categories.
We compile a summary of the challenges and the relevant academic papers and online posts that discussed these challenges in Table~\ref{tab:challenges}.

\begin{table}[h]
\renewcommand{\arraystretch}{1}
\scriptsize
\centering
\caption{Collaboration challenges identified from related papers and online posts. We present three key challenges which are further categorized into eleven detailed classifications.}
\begin{tabular}{clcc}
\toprule
\textbf{Challenges} & \textbf{Detailed classification} & \textbf{Papers} & \textbf{Posts} \\
\midrule
\multirow{5}{*}{\makecell{Separate \\ decision-making \\ procedures}} & \makecell{Suboptimal team arrangement \\ due to geographical separation}  & \makecell{\cite{paper_7}, \cite{paper_s9}, \cite{paper_46}, \cite{paper_59}, \cite{paper_4}} & - \\
\cmidrule{2-4}
                            & Isolated decision making & \cite{paper_11}, \cite{paper_41}, \cite{paper_42}, \cite{paper_4}, \cite{paper_99}, \cite{paper_53}, \cite{paper_77} &-\\
\cmidrule{2-4}
                            & Unbalanced involvement   & \cite{paper_17}, \cite{paper_59}, \cite{paper_s12} & \makecell{[P\ref{UXR8}], [P\ref{UXR17}], [P\ref{SER3}], [P\ref{UX15}]}\\
\cmidrule{2-4}
                            & Blurry boundary          & \cite{paper_46}, \cite{paper_59} & [P\ref{UX2}] \\
\cmidrule{2-4}
&   Lack of group awareness & \cite{paper_94}, \cite{paper_45}, \cite{paper_95}, \cite{paper_46}, \cite{paper_11}, \cite{paper_103}, \cite{paper_105}, \cite{paper_50}    & [P\ref{UXR17}], [P\ref{UX3}], [P\ref{UXR18}], [P\ref{UX11}], [P\ref{UXR8}]      \\
\midrule

\multirow{3}{*}{\makecell{Different \\ professional \\ behaviors}}   & Educational disparities & \cite{paper_7}, \cite{paper_94}, \cite{new_2}, \cite{paper_50}, \cite{paper_105}, \cite{paper_41}, \cite{paper_42}, \cite{paper_99} & \makecell{[P\ref{SER3}], [P\ref{UX3}], [P\ref{UX4}], [P\ref{UX10}] [P\ref{UXR8}], \\ \ [P\ref{UXR44}], [P\ref{SER2}]}\\
\cmidrule{2-4}
                                            & Methodology detachment    & \makecell{\cite{paper_s9}, \cite{paper_95}, \cite{paper_22}, \cite{paper_46},  \cite{paper_17}, \cite{paper_59},
                                            \cite{2004_new}
                                            \\ \cite{paper_50}, \cite{paper_103}, \cite{paper_99}, \cite{paper_53}, \cite{paper_s1}, \cite{paper_104}, \cite{new_3}} & \makecell{[P\ref{UX1}], [P\ref{UX5}], [P\ref{UX13}], [P\ref{UXR8}], [P\ref{UXR4}], \\ \ [P\ref{UXR17}], [P\ref{UXR46}]} \\
\cmidrule{2-4}
                                            & Different artifacts    & \cite{paper_46}, \cite{paper_41}, \cite{paper_42}, \cite{paper_99}, \cite{paper_37} & - \\
\midrule

\multirow{4}{*}{\makecell{Lack of mutual \\ understanding}}  &  UX de-emphasized & \makecell{\cite{paper_94}, \cite{paper_45}, \cite{paper_95}, \cite{paper_57}, \cite{2004_new} \\ \cite{paper_59}, \cite{paper_105}, \cite{paper_99}, \cite{paper_104}, \cite{paper_77} } & [P\ref{UX11}] \\
\cmidrule{2-4}
&   Absence of trust    & \cite{paper_94}, \cite{paper_s9}, \cite{paper_57}, \cite{paper_105}, \cite{paper_48}, \cite{paper_99} &  [P\ref{UX3}], [P\ref{UXR17}]\\
\cmidrule{2-4}
                                            & Mismatch of expectations   & \cite{paper_11}, \cite{paper_103} & [P\ref{UX14}], [P\ref{UXR17}] \\
\bottomrule
\end{tabular}
\label{tab:challenges}
\end{table}

\subsubsection{\textbf{Separated Decision-Making Procedure}}\label{sec:sep_dec}
\hfill\\
\textbf{Suboptimal Team Arrangement Due to Geographical Separation.}
SDEs and UXDs are sometimes separated geographically, often with timezone differences~\cite {paper_46, paper_4, paper_s9}. This geographic separation adds to the discipline separation, contributing to difficulties in communication and further resulting in poor integration of UX design work and its implementation~\cite{paper_59}. 
When the geographic separation creates timezone differences, it could even make one team unable to attend meetings, resulting in a breakdown in communication between two teams~\cite{paper_7, paper_4}. This separation aggregates the isolated decision-making as we elaborate next.

\textbf{Isolated Decision-Making.}
Effective communication between two groups during decision-making is important, as exchanging information among knowledgeable individuals can alleviate uncertainties and address questions. Non-participation in such discussions might result in postponed deliverables~\cite{paper_11}.
A prominent theme discussed in the literature is related to design decision-making in isolation~\cite{paper_77, paper_41, paper_42}.
For example, some design teams make early design decisions without informing the development team, making SDEs unaware of the current state of the design and devoid of opportunities to observe design evolution~\cite{paper_41, paper_42, paper_77}. 
As a result, certain design choices might not be incorporated into the final implementation, as developers remain unaware of them~\cite{paper_42}. 
Moreover, upon review by developers, some other designs may be technically infeasible, wasting designers' effort to redesign~\cite{paper_94, paper_99}.

\looseness = -1
A cause for this isolation is the untimely communication between two groups, where UXDs are not informed about whether their design is implemented, resulting in no user research and testing being implemented~\cite{paper_4}.
Another reason for this separation is the underestimation of UXDs by developers or their organizations.
Some developers view UXDs as external entities and perceive UX work as insignificant external resources rather than part of the decision-making process, leaving the usability specialists outside the decision-making and planning process~\cite{paper_53}. 
Consequently, they occasionally modify designers' work after the designer-developer handoff without proper communication with designers.
Furthermore, situations arise where developers, facing unclear design documents from designers, rely on their past experience to implement the design rather than request communication with designers~\cite{paper_4}.
In some organizations, UXD's work is limited to the early design phase of the project. They lack the authority to participate in the development decision-making or even arrange meetings with SDEs~\cite{paper_99}.
All those challenges result in discrepancies between the initial design and the final implementation. 
The absence of designers' input during development meetings might potentially give rise to unidentified UX issues, subsequently requiring developers to invest additional time in correcting them afterwards, finally leading to delays in deliverables~\cite{paper_4}.

\textbf{Unbalanced Involvement.} In some cases, although collaboration teams are geographically close, interaction and communication gaps still exist due to unbalanced monetary compensation or uneven workload~\cite{paper_s12, paper_17, paper_59}. 
For example, several papers reported cases studied regarding the involvement of UX experts in the organization falling short~\cite{paper_s12, paper_59}. 
The difference in the number of UXDs and SDEs within the team increases the team's specialization. 
However, this specialization could result in challenges when it comes to distributing tasks equitably and flexibly, potentially leading to misunderstandings among team members during the handoff process~\cite{paper_17}.
Moreover, this deficiency in the number of UX Designers may also restrict testing and validation activities, making it challenging to sustain high-level collaboration with the development team~\cite{paper_59}.
This could finally lead to additional workload for UX designers and make them feel frustrated~\cite{paper_s12}.

Similar concerns are observed from the forum analysis, where
three discussions from two posts noted that developers often find themselves isolated from the design process with limited chance to contribute~[P\ref{UXR17}, P\ref{SER3}].
According to one designer, this issue arose due to the late involvement of developers, resulting in the inability to provide design feedback~[P\ref{UXR17}]. 
However, another developer argued that they could not provide valuable feedback or make technical contributions even if they were involved earlier~[P\ref{UXR17}].
Three other posts stated developers often lack clarity on when to consider UX, leaving designers minimal time to offer feedback and make contributions during the implementation phase~[P\ref{UXR8}, P\ref{SER3}, P\ref{UX15}].

\textbf{Blurry Boundary.}
Another factor that intensifies the challenge of separated decision-making is the unclear demarcation between designers' and developers' responsibility~\cite{paper_59, paper_46}. 
SDEs and UXDs often lack well-defined responsibilities concerning design or other associated activities within the organization; such absence usually leads to a deficiency of visibility of implementation activities and finally results in UX experts not having the opportunity to be involved in development-related decisions~\cite{paper_59}. Regarding developers, their decision on design issues during the implementation phase usually results in re-implementation after UX designers suggest changes. Thus, time is wasted due to redoing works~\cite{paper_46}.
In one post, a designer specified that the boundary between designers and developers and the responsibility for specific tasks is unclear;
he mentioned that it was unclear who should be responsible for the CSS colour variable in a previous project~[P\ref{UX2}].

\textbf{Lack of Group Awareness.}  During the collaboration procedure, it is also observed that the two groups of experts are likely to be unaware of each other's work content and timeline, causing the inability to inform joint decisions and finally leading to missing deadlines~\cite{paper_105, paper_46, paper_11, paper_94, paper_103}. 
For example, software developers and UX professionals often hold differing perspectives on the nature of UX tasks and the timeline of their integration~\cite{paper_105, paper_46}. 
Many developers view the work of UX designers as primarily focused on user research and completing user stories, which often happen before development. 
However, UX designers still play a vital role in identifying usability issues and requesting developers to implement necessary changes~\cite{paper_103, paper_94}.
To be more precise, this disparity can be viewed as a deficiency in the shared understanding of the project's goal~\cite{paper_105}.
Moreover, this lack of consensus may create communication barriers between UXDs and SDEs.
When knowledge is not shared among the team, team members will likely become isolated~\cite{paper_95, paper_50}.
Due to this lack of visibility and isolation, misunderstandings can arise between groups, exacerbating the communication gap and leading to isolated decision-making~\cite {paper_45, paper_11}.

It's important to highlight that the challenge of \emph{lack of group awareness} stands out as a prominent recurring theme in our analysis of grey literature, appearing nine times across five posts~[P\ref{UXR17}, P\ref{UX3}, P\ref{UXR18}, P\ref{UX11}, P\ref{UXR8}].
In two posts, four discussions addressed cases in which development teams made alterations to designers' work after the handoff, leading to a deviation between the final implementation and the original design~[P\ref{UXR17}, P\ref{UXR18}].

\subsubsection{\textbf{Different Professional Behaviors}}
\label{sec:diff_culture}
\hfill\\
\textbf{Educational Disparities.} 
SDEs and UXDs come from different educational and professional backgrounds and are equipped with distinct perspectives~\cite{paper_105,paper_50,paper_7,paper_42}. 
This divergence often results in increased or overlooked tasks within the workflow, miscommunication, and misunderstanding~\cite{paper_42,paper_41}. 
For instance, looking at it from the developers' point of view, numerous design concepts fail to take into account the technological limitations~\cite{paper_94, paper_105,paper_99,paper_41,paper_42, feng2023understanding, new_2}.  
This oversight may introduce divergence that could further result in developers having to invest additional efforts into modifying the design on their own and/or waste effort for both groups~\cite{paper_42}. 
From the viewpoint of UX designers, SDEs' lack of familiarity with usability testing led to a narrow focus on functionality testing rather than encompassing UX-related assessments~\cite{paper_105}. 

Similarly, the analysis of the grey literature review reveals the same challenges, with them appearing in element instances across seven posts out of a total of 48 posts~[P\ref{SER3}, P\ref{UX3}, P\ref{UX4}, P\ref{UX10}, P\ref{UXR8}, P\ref{UXR44}, P\ref{SER2}].
According to five posts, the main challenge in the collaboration process arises from designers' limited technical understanding, which makes them unable to understand the complexity of their design, consequently creating challenges for developers during the implementation process~[P\ref{SER3}, P\ref{UX4}, P\ref{UX10}, P\ref{UXR44}].
Three posts highlighted a different aspect that developers lack a comprehensive UX understanding~[P\ref{UX3}, P\ref{UXR44}, P\ref{SER2}]. 
As a result, they struggled to provide feasible design insights and faced difficulties integrating designers' work.
One participant, who used to be both a designer and a developer, noted that their divergent focuses caused conflicts during designer-developer collaboration: designers tend to prioritize usability aspects, while developers concentrate more on the functionality and coding aspects of the project~[P\ref{UXR8}].

\textbf{Methodology Detachment.} 
The reviewed papers have noted that a gap exists in the approaches to development and design due to the fundamental distinctions between these two fields, which often results in a lack of clear communication between the two groups~\cite{paper_104, paper_95, paper_s9}. 
More precisely, in terms of pace, the iterative cycle of software development moves more rapidly compared to UX design~\cite{paper_59, new_3}. 
Estimating the time required for both categories of work, particularly the exploratory nature of UX design, is demanding, making the alignment of these two lifecycles a daunting task~\cite{paper_17,paper_s9}. 
This could lead to the usability-related insights becoming outdated by the time the report is completed~\cite{paper_53,paper_s1}, or the product being released without the user interface having undergone comprehensive testing~\cite{paper_22,paper_103,paper_46,paper_99}. 
This discrepancy is often caused by the mismatch between the software development lifecycle (specifically, the agile approach) and the process of UX design~\cite{paper_s1, paper_103}.  
Furthermore, the documentation of UX design provided to SDE is sometimes identified as inadequate for effective and accurate implementation~\cite{paper_50}. 

The mismatch between the traditional software development cycle and the UX design process often poses a challenge when one team is transitioning their work to another. 
Although we do not identify these challenges from SLR, 
\emph{inefficient designer-developer handoff} is one of the most frequent challenges mentioned by practitioners in their discussions, appearing nine responses in seven posts~[P\ref{UX1}, P\ref{UX5}, P\ref{UX13}, P\ref{UXR8}, P\ref{UXR4}, P\ref{UXR17}, P\ref{UXR46}]. 
Designers mainly mentioned strained communication~[P\ref{UX1}], uncertain handoff process~[P\ref{UX13}, P\ref{UXR8}], and lack of interaction during the handoff stage~[P\ref{UXR17}, P\ref{UXR4}, P\ref{UXR46}].
Practitioners mentioned that the existing approach to the designer-developer handoff is only a transfer of responsibilities and materials.

\textbf{Different Artifacts.}
Another aspect that highlights the distinction between the two groups is the artifacts they generate -- `static and textual' versus `dynamic and visual' designs~\cite{paper_41, paper_42, paper_46, paper_37}.
For example, UX designers tend to utilize visual representations like images or videos of their screens to show their design~\cite{paper_41}. In contrast, developers primarily depend on text, including programming terminology, to elucidate information about their implementations~\cite{paper_42}.
When sharing artifacts with the counterpart group, essential information can become lost or create confusion, making it difficult for the receiving team to fully comprehend~\cite{paper_42}.
As a result, this discrepancy often demands extra effort or results in a misalignment between the initial design and the eventual deliverable~\cite{paper_99,paper_46}. 
Although this problem has been discussed over decades, it is still unclear what boundary objects for representing interaction can be shared by designers and developers~\cite{paper_42}. 
Thus, time is wasted and misunderstandings arise due to the need for designers to produce multiple files to illustrate their designs, developers having to recreate design documentation, and developers misinterpreting the designs, leading to redundancy and rework~\cite{paper_41}. 

\subsubsection{\textbf{Lack of Mutual Understanding}} 
\label{sec:lack_under}
\hfill\\
\textbf{UX De-emphasized.} With distinct professional behaviors and separate decision-making, SDEs and UXDs are likely to develop their own understanding of the collaborative project~\cite{paper_94, paper_99, paper_59}. 
This lack of mutual understanding usually results in an unwillingness to accept suggestions, criticism, or decisions from other groups~\cite{paper_94, paper_99}. 
For example, some UX designers feel they are treated as ``artists'' by developers; thus, their questions and suggestions are usually ignored~\cite{paper_99}. 
In the context of open-source software (OSS) projects, this situation could lead UX designers to have a shallow understanding of the project's goal and detailed procedures~\cite{paper_59}. 
As a result, they might tend to criticize excessively rather than contribute by enhancing the codebase or resolving issues. 
Ultimately, this could lead to their limited contribution to the project as a whole~\cite{paper_77}. 
From the developer's point of view, their limited understanding and knowledge in the UX field often leads to a lack of appreciation for UX work~\cite{paper_77}. 
They usually regard UX as trivial or lacking intellectual contribution and perceive it as the last phase of the project~\cite{paper_45, paper_94, paper_57}, thus reluctant to integrate UX work into their development and exclude UX designers out of the loop~\cite{paper_104}. 
The absence of consensus and the undervaluing of UX contribute to relegating UX work to a lower priority, and in some cases, even dismissing it altogether~\cite{paper_105, paper_99}.
In addition, according to [P\ref{UX11}], de-emphasizing UX not only happens between individual UX designers and developers but could also be an issue at the organizational level. When companies are development-oriented, they will ignore the importance of UX and focus more on functionality rather than usability. As a result, the management teams usually approve design changes made by development teams without adequate research or input from the design teams [P\ref{UX11}].

\textbf{Absence of Trust.} 
Given that UX work is usually de-emphasized and even dismissed, UX designers may perceive it as a lack of respect~\cite{paper_99}. 
Regarding communication or decision-making, this lack of mutual appreciation and trust between the two expert groups tends to make them defensive, avoiding addressing concerns from the other group~\cite{paper_s9, paper_48, paper_105}. 
The communication problem caused by lack of trust further isolated two groups of experts, making them unaware of the other group's efforts, resulting in an incomplete development process~\cite{paper_94, paper_105}. 
Specifically, certain UX-related aspects of development, such as user research, were conducted late or even missing~\cite{paper_105}. 
In addition, certain design works created by UX designers may not be presented in the final product~\cite{paper_57}, which leads to a waste of effort. 
P\ref{UX3} explicitly specifies that developers lack trust and are unwilling to engage in discussions with designers.

\textbf{Mismatch of Expectations.}
The different education and methodology between UXDs and SDEs usually result in mismatched expectations between two groups of experts~\cite{paper_103, paper_11}. Given that the agile model is technical and functional driven, it could be confusing for SDEs to know when a user story is ``done''~\cite{paper_103}.
 In particular, there is a lack of communication regarding how to support each other~\cite{paper_11}, which often leads to wasted planning effort and team members working overtime to meet deadlines. 
From our forum post analysis, we observed the same challenge where a designer stated that the expectations between the design and the development team lack clarification and mutual understanding, resulting in two groups of experts working towards different goals~[P\ref{UX14}].
Moreover, conflicting expectations could arise during the development phase, concerning the primary focus of the work. For instance, as cited in [P\ref{UXR17}], distinct priorities could emerge between design teams, who prioritize enhancing user experience (UX), and development teams, who prioritize optimizing code and functionality. These divergent emphases significantly contribute to conflicts encountered during the development process [P\ref{UXR17}].

\subsection{(RQ2) Best Practices to Overcome Collaboration Challenges} \label{RQ2}
\label{ref:RQ2}
\looseness = -1
In this section, we aim to identify strategies that could be used to overcome the challenges summarized in RQ1. 
Among all the 45 papers we collected, 25 proposed best practices that could be categorized into two high-level themes: (1) improving the processes and (2) enhancing communication, as presented in Table~\ref{tab:best_practice}. 
In the subsequent sections,  we elaborate on these best practices in detail.

\begin{table}[h]
\renewcommand{\arraystretch}{1}  
\centering
\caption{Best practices identified from related papers and the corresponding collaboration challenges they addressed. We present two main best practices, which are divided into six sub-practices.}
\scriptsize
\begin{tabular}{lccc}
\toprule
\textbf{Best Practice} & \textbf{Detailed Classification} & \textbf{Papers} & \textbf{Addressed Challenges}\\
\midrule
 \multirow{3}{*}{\makecell{\\\\Improve \\ development \\ workflow}} & \makecell{UX Design work should start \\ earlier than software development}  & \makecell{\cite{paper_s12}, \cite{paper_42}, \cite{paper_50}, \cite{paper_1}, \cite{paper_17}, \cite{paper_59}, \\ \cite{paper_46}, \cite{paper_s1}, \cite{paper_99}, \cite{paper_7}, \cite{paper_4}} & Methodology detachment \\
 \cmidrule{2-4}
 & \makecell{Enhance collaboration between UXD \\ and SDE throughout the process} & \makecell{\cite{paper_42}, \cite{paper_99}, \cite{paper_79}, \cite{paper_53}, \cite{paper_59}} 
 & \makecell{Isolated decision-making \\ UX de-emphasized \\ Absence of trust \\ Mismatch of expectations}\\
\cmidrule{2-4}
& \makecell{Establish more \\ effective boundary objects} & \makecell{\cite{paper_59}, \cite{paper_4}, \cite{paper_79}, \cite{paper_42}, \cite{paper_19}} & Different artifacts \\
\midrule
 \multirow{3}{*}{\makecell{\\Enhance \\ communication}} & \makecell{Ensure regular \\ and productive communication} & \makecell{\cite{paper_4}, \cite{paper_42}, \cite{paper_50}, \cite{paper_17}, \cite{paper_s9}, \cite{paper_95}, \\ \cite{paper_41}, \cite{paper_s12}, \cite{paper_53}, \cite{paper_22}, \cite{paper_102}, \cite{paper_11}, \cite{paper_45}} &
 \makecell{Lack of group awareness \\ Absence of trust }\\
\cmidrule{2-4}
& \makecell{Promote UX in \\ development-drive teams} & \makecell{\cite{paper_1}, \cite{paper_17}, \cite{paper_26}, \cite{2004_new}} & \makecell{UX de-emphasized } \\
\cmidrule{2-4}
& Establish shared vocabulary & \makecell{\cite{paper_41}, \cite{paper_42}, \cite{paper_46}, \cite{paper_99}} & \makecell{Educational disparities} \\
\bottomrule
\end{tabular}
\label{tab:best_practice}
\end{table}

\subsubsection{\textbf{Best Practice 1: Improving Development Workflow}} is essential for aligning timelines and enhancing collaboration between SDEs and UXDs. This is a team-level best practice that reduces isolation (Section~\ref{sec:sep_dec}), integrates different professional behaviors (Section~\ref{sec:diff_culture}), and fosters better mutual understanding (Section~\ref{sec:lack_under}).

\mypara{UX Design Work Should Start Earlier Than Software Development.} One major difference between SE and UX is their iteration speed (as described in Sec.~\ref{sec:diff_culture}), where software development iterates faster than UX~\cite{paper_22, paper_17, paper_53, paper_103}, which leads to insufficient communication, rework, and delays. 
To bridge the disparity in timing, nine papers propose that UX activities, particularly UX evaluation, should commence and be completed earlier to synchronize with the software development schedule during the project's initial phase~\cite{paper_s12, paper_42, paper_50, paper_1, paper_17, paper_59, paper_46, paper_s1, paper_99}.
The primary cause for this gap is the tendency of UX designers to employ a `big design upfront' approach during the early stages of the design process~\cite{paper_7}, which leads to too much design work done ahead and redesigning afterwards.
Instead of gathering all requirements and including all usability testing at the early phase of the project, UXDs should apply ``little design upfront'' to create only basic project structures in sprint 0 and increment them in future sprints.
Similarly, other studies have also noticed that current practice involves too much upfront design work and recommended designers take the ``sprint 0'' strategy: create UX vision among the team while preserving the flexibility of UX work~\cite{paper_1, paper_46, paper_42, paper_50, paper_4}.
Some papers even proposed that design work should always be one sprint ahead to ensure timeliness and flexibility~\cite{paper_59, paper_17}.

\looseness =-1
On the other hand, developers also prefer the early involvement of designers~\cite{paper_99, paper_s12}. 
Involving UXDs at the early stage can help SDEs avoid significant changes in later sprints~\cite{paper_99}. 
Moreover, it can also facilitate early collaboration to reveal different thinking habits and avoid future misunderstandings~\cite{paper_s12}.

\mypara{Enhance the Collaboration Between UXD and SDE Throughout the Process.}
As described before, the current workflow of software development and UX design suffers from a lack of seamless integration, resulting in isolated decision-making (described in Section~\ref{sec:sep_dec}) and communication breakdowns during collaboration (described in Section~\ref{sec:lack_under}), ultimately causing workflow inefficiencies. 
To address these issues, previous efforts suggested a more effective integration of both groups of stakeholders into each other's work processes. 
In our analysis of forum posts, we noticed consistent recommendations for this best practice: a designer emphasized that the handoff process should involve prolonged cooperation and ongoing communication (P\ref{UXR46}).

\emph{Involving SDEs in UX design.}
Involvement of SDEs during the initial design phase can facilitate the creation of complex interactions and enhance developers' understanding of design principles~\cite{paper_42}. Furthermore, the collaborative team will become better equipped to minimize information gaps, identify edge cases, and establish a well-defined technical scope for designers.
Similarly, developers should engage in the design process by collaborating with designers to create sketches and prototypes~\cite{paper_99}.
This collaborative approach enables SDEs to gain a more comprehensive grasp of the UX vision.
Another study also suggests that SDEs should participate in UX evaluation and testing~\cite{paper_79}.
In the study, SDEs engaged in the usability testing session, taking on the role of technical assistants to facilitate the testing process.
Their recorded notes and comments offered valuable technical insights to the designers.

\emph{Involving UXD in development.} 
Involve UXDs in SE activities is also recommended~\cite{paper_53, paper_42, paper_59}.
It is beneficial for UXDs to have technical skills and code access to contribute to development~\cite{paper_53}.
Similar findings were observed by~\citet{paper_42}, highlighting that allowing designers to modify code can avoid misunderstandings and reduce iterations~\cite{paper_42}.
Designers should also participate in development sprint planning meetings;
this will provide them with improved control and greater transparency throughout the UX process, enabling them to maintain effective collaboration and communication with the implementation team~\cite{paper_59}.

\mypara{Establishing More Effective Boundary Objects.}
One challenge mentioned in Section~\ref{sec:diff_culture} is that SDEs and UXDs employ distinct types of artifacts, resulting in miscommunication and isolation. 
To bridge this gap, it is crucial to establish a clear definition of the boundary objects that are mutually shared by both groups of experts to support subsequent interactions and establish shared understanding~\cite{paper_19}. 
Many studies emphasize the importance of sharing design artifacts as boundary objects to facilitate collaboration between SDEs and UXDs.
For example, sharing the \emph{UI proposal and UX design} can facilitate early evaluation through comments and discussions~\cite{paper_59}.
Similarly, sharing \emph{user research and testing results} can help to clarify misinterpretation and enhance designers' participation~\cite{paper_4}. 
Further, the incorporation of \emph{paper prototypes} — initially used for usability testing, into UX storyboards provides a comprehensive guide that aligns the entire team with the project's vision~\cite{paper_79}.
In addition to design deliverables created by UXDs, contributions from developers should also be valued. 
For instance, while designers can provide visual representations capturing interaction snapshots, developers offer \emph{component diagrams describing connections between primitive graphical elements and functions}. 
These artifacts can then be combined to help the team reach a complete and shared understanding~\cite{paper_42}.

\subsubsection{\textbf{Best Practice 2: Enhance Communication.}} 
In addition to efforts aimed at enhancing workflow, another aspect to consider is enabling improved communication. This can foster better mutual understanding (Section~\ref{sec:lack_under}) and eliminate the isolated decision-making (Section~\ref{sec:sep_dec}).

\mypara{Ensuring Regular and Productive Communication.}
Twelve papers proposed that communication between SDEs and UXDs should be explicit~\cite{paper_4, paper_42, paper_50, paper_17}, as well as active and frequent~\cite{paper_17, paper_s9, paper_95, paper_41, paper_s12, paper_53}. 
Developers are encouraged to ask explicitly for design specifications and ideas~\cite{paper_4, paper_42}, while designers should actively provide details to resolve misunderstandings and facilitate integration~\cite{paper_50, paper_17, paper_42}.
Similarly, designers should request clarification and confirmation on technical issues~\cite{paper_42}. 
When it comes to the development process, explicit communication between two groups of experts can also help both groups be aware of the project timeline, avoiding late or missing deliverables~\cite{paper_17}.
In terms of active and frequent communication, collaborative teams are encouraged to communicate daily~\cite{paper_17, paper_s9}, and face-to-face or live meetings are valued~\cite{paper_95, paper_41, paper_s12}. 
Engaging in discussions and face-to-face debate enables SDEs and UXDs to avoid missing information, address edge cases, and develop better products~\cite{paper_95, paper_42}.
Moreover, communication tools and platforms are another aspect of consideration~\cite{paper_53, paper_s12}. 
Studies also demonstrate that practitioners prefer flexible and lightweight tools, as they believe such tools can enhance communication by making it more accessible and convenient~\cite{paper_s12}.
Despite being explicit and active, the relationship between SEDs and UXDs is also a critical factor affecting communication~\cite{paper_22}.
The foundation of such communication lies in the willingness of both groups to collaborate~\cite{paper_102, paper_11}. 
According to interviews conducted by~\citet{paper_45}, UXDs should build trust by reputation, position and pointing to data, while SDEs should respect UXDs and make them feel valued~\cite{paper_45}.

\mypara{Promote UX in development-drive teams.}
Another challenge discussed in Section~\ref{sec:lack_under} is the lack of mutual understanding between SDEs and UXDs, where SDEs usually de-emphasize and overlook UX.
To overcome this challenge, several studies have proposed to promote the position of UX in the overall development process~\cite{paper_1, paper_17, paper_26, 2004_new}.
It is beneficial to integrate UX closer with the development team;
to achieve this, all participants should reach a consensus regarding the positive impacts of UX and comprehend the UX vision through training or synchronization with users~\cite {paper_1, 2004_new}.
An alternative way to promote UX is through workshops and presentations; 
UX workshops have the potential to cultivate a shared UX understanding among teams, while the presentations by the UXDs can widen the perspectives of developers~\cite{paper_17, paper_26}. 

\mypara{Establishing a Shared Vocabulary.}
Work culture and background differences are another designer-developer collaboration challenge (Section~\ref{sec:diff_culture}).
To overcome this challenge, several studies suggest that both groups of experts should become acquainted with each other's terminology to guarantee effective communication with minimal confusion~\cite{paper_41, paper_42}. 
An observation study conducted by~\citet{paper_42} found that designers started to incorporate mathematical concepts while developers began formulating ideas using UX widgets to bridge the vocabulary gap. 
They also explored specific examples to validate the consistency of terminologies.
Some other papers further suggest that SDEs and UXDs should consider each other's work tradition and knowledge domain~\cite{paper_46, paper_99}. 
Instead of solely depending on management efforts, they should clearly understand each other's roles and expectations.
\citet{paper_99} discovered from a case study that having practitioners with both domain expertise and decision powers is a critical factor of success~\cite{paper_99}.

\subsection{(RQ3) Current State of UX \& SE Collaboration} 

In this section, we investigate how the findings from RQ1 and RQ2 were reflected in real-world settings. In particular, we study the existing tooling support to assess whether the challenges identified in RQ1 have been addressed, if the best practices summarized in RQ2 have been fully implemented, and if there are still areas where improvements are needed. 
We examined four online forums frequented by UX designers and software developers, focusing on discussions relevant to our research. We identified a total of 14 collaboration tools and/or platforms and highlighted challenges identified in Section \ref{rq3-tools} for some of these platforms discussed online. Additionally, we performed an initial case study on the VS Code project to complement our analysis of forums (Section~\ref{vscode}), aiming to explore how UX designers and software developers collaborate within a prominent and well-established real-world open-source project, and to assess their adherence to recognized best practices described in Section~\ref{ref:RQ2}.

\subsubsection{\textbf{Collaboration Tools and Platforms}} \label{rq3-tools}\hspace*{\fill} \\

\begin{figure*}[h]
\tiny
\centering
    \noindent\includegraphics[width=0.5\textwidth]{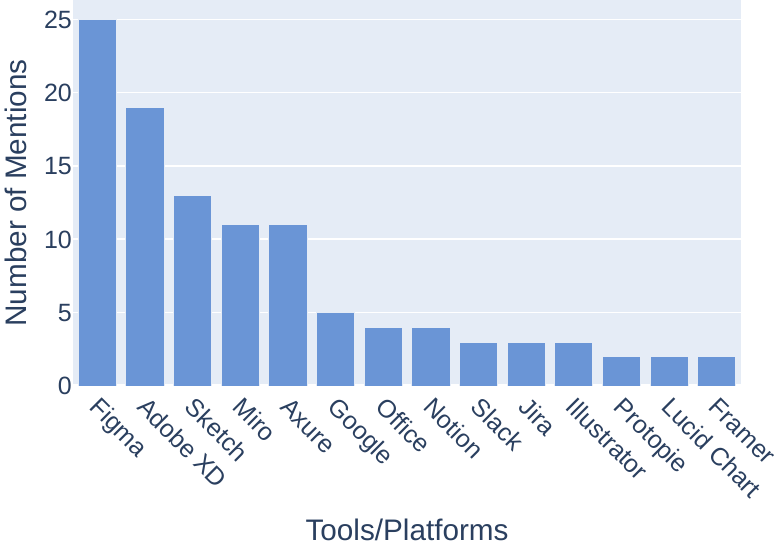}
    \caption{Popularity of collaboration tools and platforms. The Y-axis shows the frequency of the appearance in collected posts. This figure only presents tools and platforms that appear more than once.}
    \label{fig:platform_popularity}
\end{figure*}

Having tooling support is essential for the collaboration process to guarantee efficient sharing of information and communication~\cite{2006_collaboration_tool, 2004_collaboration_tool}. 
In the post-search phase across the four forums (i.e., UX Stack Exchange~\cite{UX_stackexchange}, Reddit UX~\cite{Reddit_UX}, Reddit SE~\cite{Reddit_SE}, and Stack Overflow~\cite{stack_overflow}), we specifically focused on posts that discussed tools and platforms. We aimed to gather insights and opinions from professionals and practitioners. 
We identified 26 posts that are relevant to our focus in total. 
Figure~\ref{fig:platform_popularity} displays the frequency of mentions for each tool/platform within these posts. 
Among these posts, we only identified eight posts focused explicitly on the pros and cons of five tools from the collaboration perspective, including Figma~\cite{figma}, Adobe XD~\cite{adobe_xd}, Axure~\cite{axure}, Protopie~\cite{protopie}, and Miro~\cite{miro}. 
Other posts primarily emphasize the technical functionality or accessibility of those tools. 
We carefully review the compiled pros and cons of each tool, comparing them with the collaboration challenges outlined in Section~\ref{ref:RQ1} to identify whether any of the tools effectively address specific challenges. 
Table~\ref{tab:tool_mapping} shows a mapping between the tools and corresponding challenges identified in RQ1. We use \y ~ to represent the challenge that can be alleviated by a tool and \n ~to show that the challenge can not be alleviated.
\renewcommand{\arraystretch}{1}

\begin{table}[h]
    \centering
    \caption{Matching the summarized collaboration challenges summarized in Section~\ref{ref:RQ1} with the relevant features provided by each tooling support in order to address and mitigate these challenges. \y ~represents the challenge can be alleviated and \n ~means the challenge can not be alleviated.}
    \footnotesize
    \begin{tabular}{llccccc}
        \Xhline{2\arrayrulewidth}
         &  & \multicolumn{5}{c}{\textbf{Tools}} \\
         &  & \textbf{Figma} & \textbf{Adobe XD} & \textbf{Miro} & \textbf{Axure} & \textbf{Protopie} \\ \cline{1-7} 
        \multirow{11}{*}{\begin{sideways}\textbf{Challenges}\end{sideways}} & \multicolumn{1}{r|}{Suboptimal team arrangement} & \multicolumn{1}{c|}{\na} & \multicolumn{1}{c|}{\na} & \multicolumn{1}{c|}{\na} & \multicolumn{1}{c|}{\na} & \multicolumn{1}{c|}{\na} \\ \cline{3-7} 
         & \multicolumn{1}{r|}{Isolated decision making} & \multicolumn{1}{c|}{\na} & \multicolumn{1}{c|}{\na} & \multicolumn{1}{c|}{\y} & \multicolumn{1}{c|}{\na} & \multicolumn{1}{c|}{\na} \\ \cline{3-7} 
         & \multicolumn{1}{r|}{Unbalanced involvement} & \multicolumn{1}{c|}{\na} & \multicolumn{1}{c|}{\na} & \multicolumn{1}{c|}{\na} & \multicolumn{1}{c|}{\na} & \multicolumn{1}{c|}{\na} \\ \cline{3-7} 
         & \multicolumn{1}{r|}{Blurry boundary} & \multicolumn{1}{c|}{\y} & \multicolumn{1}{c|}{\na} & \multicolumn{1}{c|}{\na} & \multicolumn{1}{c|}{\na} & \multicolumn{1}{c|}{\na} \\ \cline{3-7} 
         & \multicolumn{1}{r|}{Lack of group awareness} & \multicolumn{1}{c|}{\y} & \multicolumn{1}{c|}{\n} & \multicolumn{1}{c|}{\na} & \multicolumn{1}{c|}{\y} & \multicolumn{1}{c|}{\y} \\ \cline{2-7} 
         & \multicolumn{1}{r|}{Educational disparities} & \multicolumn{1}{c|}{\na} & \multicolumn{1}{c|}{\na} & \multicolumn{1}{c|}{\na} & \multicolumn{1}{c|}{\na} & \multicolumn{1}{c|}{\na} \\ \cline{3-7} 
         & \multicolumn{1}{r|}{Methodology detachment} & \multicolumn{1}{c|}{\y} & \multicolumn{1}{c|}{\na} & \multicolumn{1}{c|}{\n} & \multicolumn{1}{c|}{\na} & \multicolumn{1}{c|}{\y} \\ \cline{3-7} 
         & \multicolumn{1}{r|}{Different artifacts} & \multicolumn{1}{c|}{\y} & \multicolumn{1}{c|}{\na} & \multicolumn{1}{c|}{\n} & \multicolumn{1}{c|}{\na} & \multicolumn{1}{c|}{\na} \\ \cline{2-7} 
         & \multicolumn{1}{r|}{UX de-emphasized} & \multicolumn{1}{c|}{\na} & \multicolumn{1}{c|}{\na} & \multicolumn{1}{c|}{\na} & \multicolumn{1}{c|}{\na} & \multicolumn{1}{c|}{\na} \\ \cline{3-7} 
         & \multicolumn{1}{r|}{Absence of trust} & \multicolumn{1}{c|}{\na} & \multicolumn{1}{c|}{\na} & \multicolumn{1}{c|}{\na} & \multicolumn{1}{c|}{\na} & \multicolumn{1}{c|}{\na} \\ \cline{3-7} 
         & \multicolumn{1}{r|}{Mismatch of expectations} & \multicolumn{1}{c|}{\na} & \multicolumn{1}{c|}{\y} & \multicolumn{1}{c|}{\na} & \multicolumn{1}{c|}{\na} & \multicolumn{1}{c|}{\na} \\ 
         \Xhline{2\arrayrulewidth}
    \end{tabular}
    \label{tab:tool_mapping}
\end{table}

Next, we provide a condensed overview of the collaboration-related comments made by practitioners concerning these five tools/platforms. 

 \textbf{Figma} is the most popular tool among SDEs and UXDs that has been mentioned in seven out of eight posts for its highly effective \emph{cross-team collaboration features}. 
It is recommended for designer-developer collaboration for its version control functionality and the ``lock and branch'' feature~[P\ref{UXR4}], which can effectively specify designs to developers, enabling them to understand their assigned tasks clearly and increase group awareness, and allows designers to continue working on their designs without affecting the work handed to developers.
Other posts noted that Figma involves the whole team in the project with different roles~[P\ref{UXR12}, P\ref{UXR15}, P\ref{UXR33}, P\ref{UXR35}, P\ref{UXR36}]. 
From the developers' perspective, Figma links design and development artifacts by allowing them to access and inspect code on top of the design.
It also offers style codes and colour variables, facilitating handoff and enhancing overall development efficiency~[P\ref{UXR15}, P\ref{UXR35}, P\ref{UXR36}].
Moreover, its artboard feature enables developers and stakeholders to comment on their designs, simplifying the feedback-gathering process~[P\ref{UXR35}]. 
However, some developers complained that Figma's user interface is perplexing and has a steep learning curve~[P\ref{UXR8}]. 

 \textbf{Adobe XD} significantly simplifies the design presentation process by allowing designers to generate prototypes with a simple button click, allowing the collaboration team to quickly understand the expectation~[P\ref{UXR12}]. However, the poor document syncing and low parallelization make it hard for UXDs and SDEs to quickly be aware of what has been done by the other group.
\textbf{Axure} is recommended for its interaction design tools, making communication and interaction between SDEs and UXDs more accessible~[P\ref{UXR22}].
Similarly, \textbf{Protopie} is nominated for its ability to facilitate communication and design handoff; it enhances collaboration and mutual understanding between two groups by operating within a design logic that makes designers' work easier to understand~[P\ref{UXR22}].
Some designers struggle to find a single software that meets their diverse daily requirements, so they employ a combination of collaboration tools~[P\ref{UXR8}]. For example, \textbf{Miro}, though not good for prototyping or designer-developer handoff, is used for problem-solving and decision-making as a complement to Figma~[P\ref{UXR8}]. 
This will, to some extent, make their work more convenient.
However, when it comes to collaborating with developers, this complex situation can pose challenges for developers, making it difficult for them to dive in and work on top of designers' design.

\subsubsection{\textbf{Understanding the Interaction Between UXDs and SDEs via Issue Discussion From The \emph{VS Code} Project}} \label{vscode}
\begin{figure*}[h]
\tiny
\centering
    \noindent\includegraphics[width=0.4\textwidth]{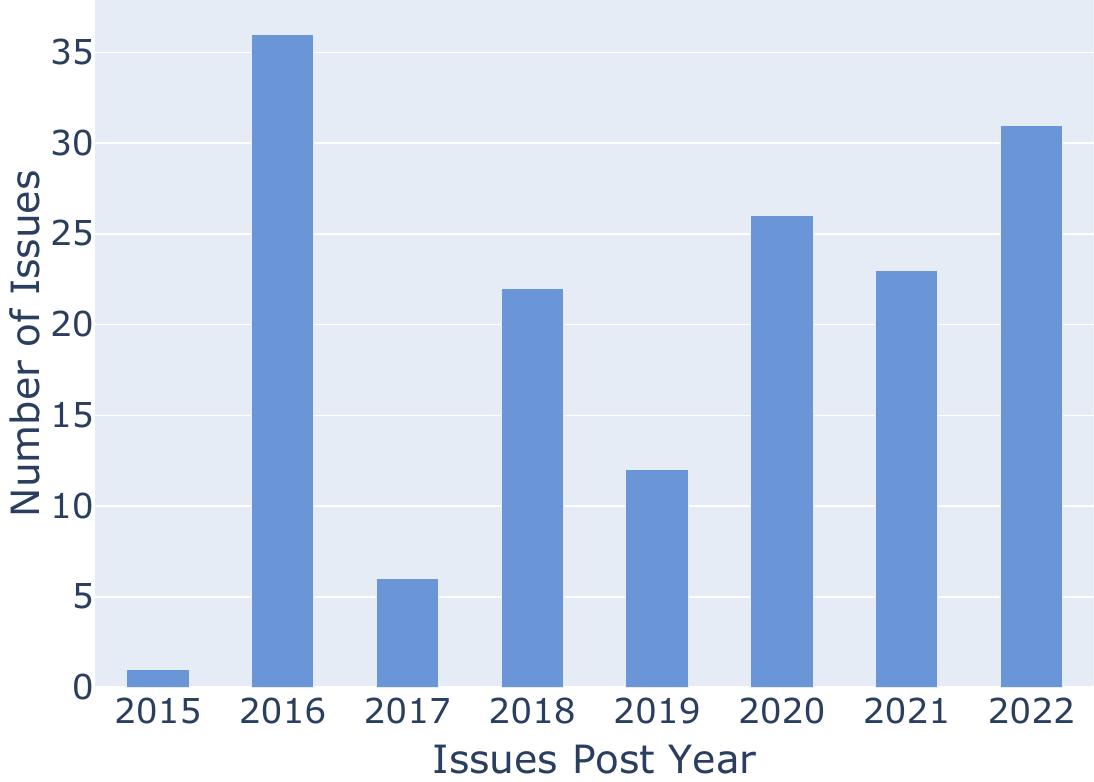}
    \caption{Number of UX-label issues per year since late 2015 in the VS Code project.}
    \label{fig:issue_popularity}
\end{figure*}

\hspace*{\fill} \\
\emph{VS Code} is a source code editor owned by Microsoft, and it is an open-source project hosted on GitHub. 
Though the development team might have internal communication channels and meetings, we can still trace designer-developer collaboration by reviewing GitHub issues as documented within the project's Wiki page.\footnote{\url{https://github.com/microsoft/vscode/wiki/UX/127f1567d5155bcd4dab20dee8f170e82f4bc63a}} 
Figure~\ref{fig:issue_popularity}  illustrates the yearly occurrence of 
\uxlabel~issues recorded since the project's creation in September 2015.

\looseness = -1
By reviewing the discussion in each collected issue, we compile the following collaboration scenarios. Some of these scenarios conform to the best practices outlined in Section~\ref{ref:RQ2} (RQ2) and/or could be regarded as best practices capable of mitigating the challenges delineated in Section~\ref{ref:RQ1} (RQ1). 
First, designers and developers collaborate closely throughout the issue time span, which aligns well with the best practice of \emph{enhance collaboration between UXD and SDE throughout the process}.  
Specifically, we observe that designers and developers initiate their collaboration during the ideation stage of an issue by clearly outlining ideas and design proposals~[I\ref{issue_156179}, I\ref
{issue_98614}].
For example, one designer stated, ``\textit{From a design perspective, I think we can establish some principles for how the interactions should work'' and ``Things that are unclear so far}''~[I\ref{issue_156179}].
Developers then actively engage by asking for clarification and providing feedback on further discussions~[I\ref{issue_130611}, I\ref{issue_98614}, I\ref{issue_63152}, I\ref{issue_100347}].
For example, one developer asked, ``\emph{All sounds good to me except the `Go to running cell'. Maybe I'm misunderstanding that one though. Is that only available when actually running?}''~[I\ref{issue_98614}]
During the implementation phases, designers monitor the project progress and address inquiries or concerns from developers [I\ref{issue_147903}]. 
Ultimately, designers verify that the final implementation aligns with their original design by adding the ``verified'' label, ensuring consistency with their intended vision [I\ref{issue_131641}, I\ref{issue_144198}].

Another scenario we observed is the respect between SDEs and UXDs.
In terms of communication, this is represented by the cordial relationship between two groups of experts.
Through our findings, we have discovered the mutual recognition and assistance demonstrated by both groups.
This is done by either direct follow-up comments such as ``\emph{Love that we are tackling this problem!}'' 
or ``\emph{I like where we are heading towards!}'' to show merits or positive reaction emojis such as  \thumbsup~ under the original thread [I\ref{issue_156179}, I\ref{issue_144324}, I\ref{issue_146806}]. 
By observing the dynamics of the interaction, it becomes evident that team members \emph{exhibit a shared sense of respect for each other}, which could directly eliminate the challenge of \emph{absence of trust} summarized in Section~\ref{ref:RQ1} (RQ1).

In terms of cooperation, when SDEs intend to modify design works, they first notify and discuss designers rather than make changes without acknowledging them [I\ref{issue_133622}], showing that \emph{UX is promoted in the development team} and the team members are exerting themselves to \emph{ensure group awareness}. 
For example, one developer suggested that ``\textit{I am not familiar with the details of the interactive window and how its editor is created but this should be possible}'' and used ``\textit{to verify}'' to start a discussion with designers before making changes~[I\ref{issue_133622}].
Another notable finding is that both groups clearly understand each other's field and background.
For instance, developers can showcase their ideas by creating simple prototypes, indicating their capacity to convey design concepts [I\ref{issue_144823}]. 
Designers have basic technical knowledge and understanding of the codebase, enabling them to locate UX bugs or even make coding contributions [I\ref{issue_101375}, I\ref{issue_89533}, I\ref{issue_115808}]. 
This mutual understanding and proficiency across disciplines mitigate the challenge of \emph{Educational disparities}, contributing to smoother collaboration and effective communication between the two groups of experts. 

However, we also identified several collaboration challenges from those issues. For instance, although design artifacts are included in the issue posts, most of the time when developers want to comment on the design artifacts, they are not able to add annotations freely but can only react with emojis, leave text comments or regenerate new visual artifacts under the original thread [I\ref{issue_130611}], indicating that the current platform is inadequate for facilitating efficient communication involving non-textual artifacts and may lead to misunderstandings. Moreover, geographical separation causes further inefficiency in communication. Unlike face-to-face discussions or meetings where team members can provide instant reactions and feedback, sometimes designers do not participate in further discussion [I\ref{issue_119776}]. For the VScode project, we noticed that most developers are based in Zurich Switzerland, whereas two major designers are based in Seattle, USA. The time difference between them causes a delay in responses~[I\ref{issue_130516}].

\section{Discussion}
\begin{figure} [th]
\centering
\includegraphics[width=1.0\linewidth]{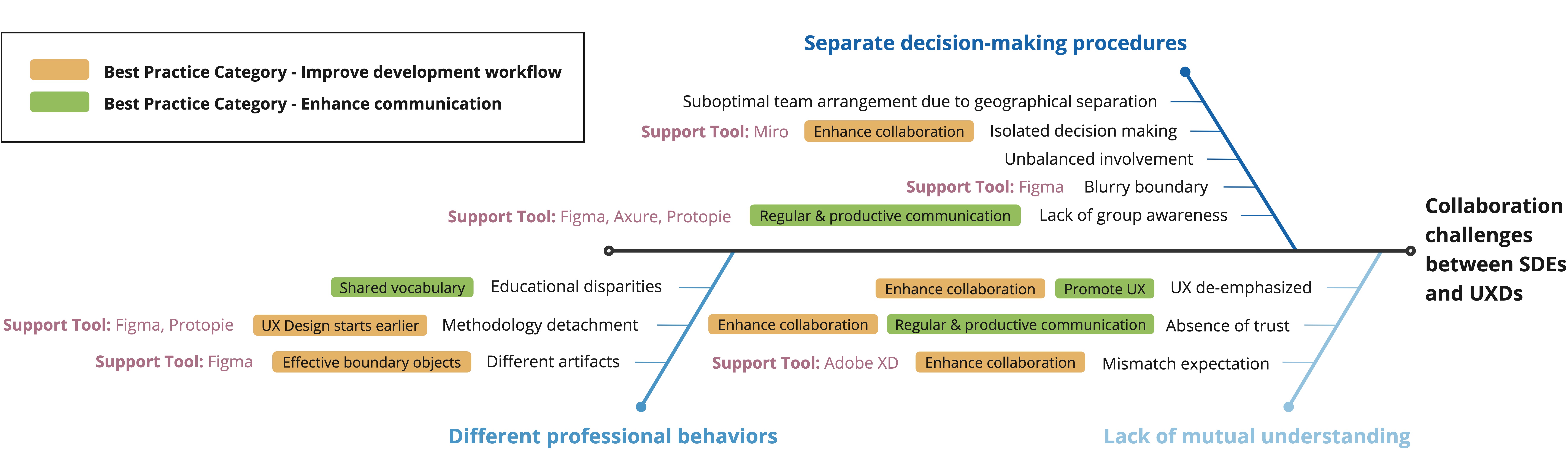} 
\caption{Fishbone diagram of collaboration challenges, corresponding best practices, and supporting tools.}
\label{fig:fishbone}
\end{figure}

Through our comprehensive examination of the existing academic literature and non-academic sources, we have pinpointed (1) three significant challenges faced in collaborative efforts between SDEs and UXDs, (2) two main categories of best practices that hold the potential to alleviate these challenges, and (3) how current collaboration tools and practices reflect these challenges and best practices. Figure~\ref{fig:fishbone} provides an overview of these findings, representing their connections. 
Below, we discuss our contributions to advancing previous SLRs on similar topics, as well as the implications of our findings on tooling support, education, and future research.

\subsection{Addressing Research Gaps and Advancing Prior SLR Studies}
Compared to previous SLR studies, our work has several unique contributions. 
First, most previous SLR studies on UXD-SDE collaboration have examined the integration of Agile development methodologies and user-centered design, emphasizing process and management aspects~\cite{2010_integration_slr,2011_integration_slr,2016_slr_integration,2014_Integration_slr,2022_slr_management}. 
In contrast, our research broadens the focus to include software development as a whole, without limiting it to specific development models. 
Although the Agile model appears promising, there is no evidence to suggest it has been universally adopted across all software development contexts, including the crucial and influential realm of open-source software development~\cite{harzl2017can,muller2018agile}. 
Agile literature and frameworks emphasize the necessity for teams to be small, stable, and whole, but it is unrealistic to achieve these goals in many software development environments~\cite{staahl2023dynamic}. Our expanded focus on UXD-SDE collaboration has contributed to identifying challenges and best practices that are important for both agile teams and beyond. For example, we identified that the challenge of \textit{UX de-emphasized} is rarely discussed in papers focused on agile teams but more frequently investigated in open-source software projects~\cite{paper_94}. At the same time, we also found that papers touching on agile, open-source, and traditional software development models have discussed many common issues regarding UXD-SDE collaboration. These results highlighted the importance and prevalence of collaboration issues between UXDs and SDEs. Regarding this commonality, our review also included papers that are closely related but were published in venues not considered in earlier research, such as the Journal of System Software and Information and Software Technology. This expanded collection has contributed to identifying new challenges, such as blurry boundary~\cite{paper_46} and mismatch of expectations~\cite{paper_103}. 

\looseness = -1
Moreover, our review encompasses a broader scope of resource collection, covering literature from important venues in both HCI and SE and triangulating it with real-world discussions in forums and open-source issue trackers. When it comes to the topic of UXD-SDE collaboration, the boundary between the fields of HCI and SE is blurry, as is the line between practice and research in this area. We found that papers published on this topic in HCI and SE venues discussed similar problems. The differences in their perspectives are much less prominent than the similarities in the problems these two fields targeted and the findings the studies revealed. The previous SLRs have never adopted this broad angle, encompassing both academic literature in multiple fields and real-world scenarios. As a result, our study has identified topics that earlier SLR studies overlooked, particularly related to human factors that tackle specific challenges, such as a lack of mutual understanding, absence of trust, and mismatched expectations, as well as the corresponding best practices, including improved communication, promotion of UX in development-driven teams, and the establishment of a shared vocabulary. By analyzing online discussions and open-source project repositories, we also used insights from practitioners with expertise in UX and SE fields to triangulate previously identified challenges and understand their practical implications.

\subsection{Better Tooling Support to Integrate Both SE and UX Artifacts.}
Based on our literature review, a considerable number of identified challenges are related to the isolation between SE and UX, especially their processes and artifacts. While boundary artifacts are often defined between SDEs and UXDs, the iteration of the artifacts is often done in the corresponding field, with little input from the counterpart. When exploring open-source platforms for information on collaborations between UXDs and SDEs, we also discovered a scarcity of projects that actively emphasized discussions about UX-related issues. 
Likewise, conversations among UX designers on Reddit revealed that these designers often do not feel comfortable participating in code reviews on GitHub because of the barrier faced by those who are not developers~\cite{reddit-uxd-github}.
This observation implies the underutilization of GitHub among UX designers or a lack of consideration for UX-related quality attributes during the open-source development process, echoing findings from previous work~\cite{Rajanen2015, Hellman2021, sanei2023characterizing,terry2010perceptions}.
In future research, it would be valuable to explore tool support that lowers the barrier for experts from both fields, integrates SE and UX artifacts, and facilitates collaboration between UXDs and SDEs. This integration could be achieved by considering the two potential scenarios:

\looseness=-1
\textbf{Scenario 1: Decentralized Platforms With Seamless Integration.} Improved tooling support is needed to seamlessly integrate platforms used by software developers (e.g., GitHub, GitLab) with those used by UX designers (as described in Section~\ref{ref:GLR}). 
Since UX artifacts are often visual and not supported in software development platforms, UX designers resort to suboptimal workarounds, such as sharing and commenting on screenshots, to convey their ideas~\cite{sanei2023characterizing, Agrawal2022}. 
Despite efforts to integrate these two types of platforms (e.g., GitLab Design Management\footnote{\url{https://docs.gitlab.com/ee/user/project/issues/design_management.html}}), there remains significant room for improvement. 
For instance, Sketchboard allows for the embedding of designs into markdown files in GitHub repositories~\cite{sketchboard} but is not primarily designed for enhancing collaboration purposes. 
Figma also introduced FigJam~\cite{github_figma}, which advocates for the integration of collaborative software development and collaborative UX/UI design. 
Since the feature has been recently introduced, it remains uncertain whether it adequately addresses the needs and resolves the collaboration challenges. 

\textbf{Scenario 2: Centralized Platforms.} Alternatively, an ideal platform should accommodate both UX designers and software developers within a shared workspace, despite their distinct objectives, by facilitating the seamless exchange of boundary objects~\cite{akkerman2011boundary}, thereby eliminating the need to switch between disparate websites and incompatible artifacts. For example, the current software development tools such as GitHub and Jira include sufficient features to support various SE roles, such as developers, quality assurance experts, and project managers. New features in those tools can be considered incorporating the tasks typically performed by UX roles, such as user researchers and UX designers. Several discussions indicate that using multiple platforms for collaboration is sub-optimal. For instance, a UX designer inquired on Stack Overflow about the process of uploading Figma files directly to GitHub to enhance team collaboration; however, subsequent feedback revealed that this approach inadvertently obstructed the timely editing of Figma files when needed, proving counterproductive~\cite{SOUploadFile}. Similarly, inquiries on Reddit underscore the ongoing challenge of streamlining the integration between design and development processes, revealing a lack of universally accepted solutions to address this issue~\cite{reddit-figma-github}. These problems might be addressed with a centralized platform that puts SE and UX tasks and artifacts together.

\subsection{Educators From HCI, CSCW, and SE Fields Need to Work Together to Design an Integrated and Systematic Training Process.} 
Out of the 45 papers gathered, only a single paper examined the collaboration between  UXDs and SDEs in a classroom setting~\cite {paper_3}. This suggests that there may be a necessity to focus on developing more effective training programs to raise awareness of better collaboration.
Moreover, many challenges we identified in the literature review, such as \textit{different professional behaviors} and \textit{lack of mutual understanding}, are rooted in discrepancies in the mindsets of SDEs and UXDs. These mindsets, in turn, originated from the education of SDE and UXD practitioners. SE education typically focuses on software development best practices, such as software design and the agile model, which offer numerous advantages, but they frequently neglect concerns associated with UX design~\cite{paper_3}. For example, while the proposed curriculum~\cite{Shaw2005SoftwareEF} for software engineering programs has acknowledged collaboration with stakeholders, the documentation does not explicitly address the decision-making process and potential trade-offs that can arise during such collaborations. Moreover, even though an increasing number of software engineering courses started to incorporate UX into software development, the short sprints limited by course length make fitting UX into software development challenging~\cite{paper_3}. 

On the other hand, UX and HCI education often addresses user-centered design-related topics, such as user inquiry, prototyping, and usability testing, omitting subjects like software process and development concerns. In the HCI curriculum defined by the ACM SIGCHI~\cite{acm_hci_curriculum}, software developers (SDEs) were merely mentioned, let alone the aspect of collaboration between these two expert groups. Our results indicate that the fundamental solution for the SDE-UXD collaboration issue might be to incorporate the counterparts in the education of both fields. We believe that such an integration in training would ripple out and result in synchronization in practice. 
Developers and designers should collaborate in creating enhanced training programs for upcoming generations, aiming to improve the efficiency of collaboration within interdisciplinary teams.

\subsection{Future Research Needs to Investigate Better Strategies to Align UX and SE Lifecycles and Activities.} 
Earlier studies have put forth the idea of integrated software development lifecycles~\cite{paper_1, paper_4, paper_14, paper_16, paper_54, paper_s1, 2011_integration_slr}, where UX design could be conducted from Sprint 0 before the development starts and also has frequent collaboration points to communicate to ensure the group awareness and avoid misunderstanding. \citet{paper_3} summarized a case study regarding the lesson learned on integrating UX lean into the agile development model via a software engineering course. Similarly, \citet{paper_58} also reported a case study with a similar setting with the focus on the defects decrease. 
However, anecdotal evidence still indicates the persistence of inefficient collaboration and challenges outlined in RQ1 (Section~\ref{ref:RQ1}). 
Subsequent research may necessitate a reexamination of lifecycle designs and the execution of field studies in diverse real-world settings encompassing various organizational structures. These efforts can shed light on trade-offs and support deliberate decision-making in the collaboration between UXDs and SDEs.

\emph{The challenges intensify significantly when creating software tailored to specific domains.} 
As the challenges of collaboration between UXDs and SDEs remain unresolved, the situation becomes increasingly dire and complex when additional stakeholders become involved, making resolution more challenging. 
With the rising popularity of artificial intelligence (AI) and its integration into numerous software systems, researchers in the SE communities have explored the collaboration difficulties associated with constructing AI-enabled software projects~\cite{2022_collaboration_ml}. 
Similarly, in the HCI communities, researchers also investigated how to improve the collaboration efficiency between UX designers and AI experts~\cite{yang2018investigating, liao2023designerly}.
Nevertheless, we did not find papers discussing the three groups of experts working together. We aspire that this research can serve as a catalyst for fostering collaboration between and beyond SDEs and UXDs, facilitating the refinement of best practices and the development of efficient collaboration processes. The same idea would be applicable to any domain-specific software, such as improving the usability of scientific software~\cite{Queiroz2017}, where collaboration among UXDs, SDEs, and research scientists is essential.

\section{Limitations}
All research design comes with limitations. 
Therefore, readers should exercise caution when generalizing findings beyond the boundaries defined by the methods employed. Firstly, despite conducting extensive searches across five academic databases,  we may have overlooked some pertinent papers. 
To address this, two researchers independently retrieved literature data to ensure the most comprehensive data collection. Additionally, forward and backward snowball sampling were employed to thoroughly review moderation research. 
Secondly, our research primarily concentrated on the collaboration between UXDs and SDEs, excluding studies from other domains. Subsequent research endeavours could explore the collaboration between UXDs and SDEs in the development of domain-specific software, such as AI-based software or scientific software. This broader investigation is warranted as collaboration in these contexts often involves domain experts.
Another limitation is that our case study on open-source project issues only focuses on the VS Code project as an initial exploratory case. As a result, the findings may not be applicable to a broader range of projects. Future research should consider examining additional projects to expand the scope and validate the generalizability of the results.

\section{Conclusion}
\looseness = -1
We presented an SLR about collaboration challenges between SDEs and UXDs, followed by a post and issue search to examine the current state of collaboration. 
By analyzing selected papers, we identified three key collaboration challenges and six best practices.
We also complement the SLR with an analysis of online forum posts and issue discussions to examine the current state of designer-developer collaboration.
Our findings can be applied beyond the partnership between SDEs and UXDs, such as collaboration with product managers and machine learning engineers. Those good practices and interventions suggested also act as a reference for several future research directions.

\section{Data Availability}
All papers, posts, and issues we reviewed are available in this repository:  \url{https://github.com/FORCOLAB-UofToronto/CSCW2025-WhoIsToBlame}.

\begin{acks}
This work is partially supported by the Natural Sciences and Engineering Research Council of Canada (NSERC), RGPIN2021-03538 and RGPIN2018-04470.
\end{acks}

\newpage
\bibliographystyle{ACM-Reference-Format}
\bibliography{ref}


\begin{thebibliography}{115}


\ifx \showCODEN    \undefined \def \showCODEN     #1{\unskip}     \fi
\ifx \showISBNx    \undefined \def \showISBNx     #1{\unskip}     \fi
\ifx \showISBNxiii \undefined \def \showISBNxiii  #1{\unskip}     \fi
\ifx \showISSN     \undefined \def \showISSN      #1{\unskip}     \fi
\ifx \showLCCN     \undefined \def \showLCCN      #1{\unskip}     \fi
\ifx \shownote     \undefined \def \shownote      #1{#1}          \fi
\ifx \showarticletitle \undefined \def \showarticletitle #1{#1}   \fi
\ifx \showURL      \undefined \def \showURL       {\relax}        \fi
\providecommand\bibfield[2]{#2}
\providecommand\bibinfo[2]{#2}
\providecommand\natexlab[1]{#1}
\providecommand\showeprint[2][]{arXiv:#2}

\bibitem[Adobe(2023)]%
        {adobe_xd}
\bibfield{author}{\bibinfo{person}{Adobe}.} \bibinfo{year}{2023}\natexlab{}.
\newblock \bibinfo{title}{Adobe XD Platform}.
\newblock
\urldef\tempurl%
\url{https://adobexdplatform.com/}
\showURL{%
\tempurl}


\bibitem[Agrawal et~al\mbox{.}(2022)]%
        {Agrawal2022}
\bibfield{author}{\bibinfo{person}{Vishakha Agrawal}, \bibinfo{person}{Yong-Han
  Lin}, {and} \bibinfo{person}{Jinghui Cheng}.}
  \bibinfo{year}{2022}\natexlab{}.
\newblock \showarticletitle{Understanding the Characteristics of Visual
  Contents in Open Source Issue Discussions: A Case Study of Jupyter Notebook}.
  In \bibinfo{booktitle}{\emph{Proceedings of the 26th International Conference
  on Evaluation and Assessment in Software Engineering}}
  \emph{(\bibinfo{series}{EASE '22})}. \bibinfo{publisher}{Association for
  Computing Machinery}, \bibinfo{address}{New York, NY, USA},
  \bibinfo{pages}{249–254}.
\newblock
\showISBNx{9781450396134}
\href{https://doi.org/10.1145/3530019.3534082}{doi:\nolinkurl{10.1145/3530019.3534082}}


\bibitem[Akkerman and Bakker(2011)]%
        {akkerman2011boundary}
\bibfield{author}{\bibinfo{person}{Sanne~F Akkerman} {and}
  \bibinfo{person}{Arthur Bakker}.} \bibinfo{year}{2011}\natexlab{}.
\newblock \showarticletitle{Boundary Crossing and Boundary Objects}.
\newblock \bibinfo{journal}{\emph{Review of educational research}}
  (\bibinfo{year}{2011}), \bibinfo{pages}{132--169}.
\newblock
\href{https://doi.org/10.3102/0034654311404435}{doi:\nolinkurl{10.3102/0034654311404435}}


\bibitem[Alhammad and Moreno(2022)]%
        {paper_3}
\bibfield{author}{\bibinfo{person}{Manal~M. Alhammad} {and}
  \bibinfo{person}{Ana~M. Moreno}.} \bibinfo{year}{2022}\natexlab{}.
\newblock \showarticletitle{Integrating User Experience into Agile: an
  Experience Report on Lean UX and Scrum}. In
  \bibinfo{booktitle}{\emph{Proceedings of the ACM/IEEE 44th International
  Conference on Software Engineering: Software Engineering Education and
  Training}} \emph{(\bibinfo{series}{ICSE-SEET '22})}.
  \bibinfo{publisher}{Association for Computing Machinery},
  \bibinfo{address}{New York, NY, USA}, \bibinfo{pages}{146–157}.
\newblock
\showISBNx{9781450392259}
\href{https://doi.org/10.1145/3510456.3514156}{doi:\nolinkurl{10.1145/3510456.3514156}}


\bibitem[Almughram and Alyahya(2017)]%
        {paper_7}
\bibfield{author}{\bibinfo{person}{Ohoud Almughram} {and}
  \bibinfo{person}{Sultan Alyahya}.} \bibinfo{year}{2017}\natexlab{}.
\newblock \showarticletitle{Coordination Support for Integrating User Centered
  Design in Distributed Agile Projects}. In \bibinfo{booktitle}{\emph{2017 IEEE
  15th International Conference on Software Engineering Research, Management
  and Applications (SERA)}}. \bibinfo{pages}{229--238}.
\newblock
\href{https://doi.org/10.1109/SERA.2017.7965732}{doi:\nolinkurl{10.1109/SERA.2017.7965732}}


\bibitem[Andreasen et~al\mbox{.}(2006)]%
        {paper_94}
\bibfield{author}{\bibinfo{person}{Morten~Sieker Andreasen},
  \bibinfo{person}{Henrik~Villemann Nielsen}, \bibinfo{person}{Simon~Ormholt
  Schrøder}, {and} \bibinfo{person}{Jan Stage}.}
  \bibinfo{year}{2006}\natexlab{}.
\newblock \showarticletitle{Usability in Open Source Software Development:
  Opinions and Practice}.
\newblock \bibinfo{journal}{\emph{Information Technology And Control}}
  \bibinfo{volume}{35}, \bibinfo{number}{3} (\bibinfo{year}{2006}).
\newblock
\showISSN{1392-124X}
\href{https://doi.org/10.5755/j01.itc.35.3.11776}{doi:\nolinkurl{10.5755/j01.itc.35.3.11776}}


\bibitem[Anslow(2014)]%
        {anslow2014reflections}
\bibfield{author}{\bibinfo{person}{Craig Anslow}.}
  \bibinfo{year}{2014}\natexlab{}.
\newblock \showarticletitle{Reflections on Collaborative Software Visualization
  in Co-located Environments}. In \bibinfo{booktitle}{\emph{2014 IEEE
  International Conference on Software Maintenance and Evolution}}.
  \bibinfo{pages}{645--650}.
\newblock
\href{https://doi.org/10.1109/ICSME.2014.115}{doi:\nolinkurl{10.1109/ICSME.2014.115}}


\bibitem[Axure(2023)]%
        {axure}
\bibfield{author}{\bibinfo{person}{Axure}.} \bibinfo{year}{2023}\natexlab{}.
\newblock \bibinfo{title}{Axure RP - UX prototypes, specifications, and
  diagrams in one tool}.
\newblock
\urldef\tempurl%
\url{https://www.axure.com/}
\showURL{%
\tempurl}


\bibitem[Bach et~al\mbox{.}(2009)]%
        {paper_45}
\bibfield{author}{\bibinfo{person}{Paula~M. Bach}, \bibinfo{person}{Robert
  DeLine}, {and} \bibinfo{person}{John~M. Carroll}.}
  \bibinfo{year}{2009}\natexlab{}.
\newblock \showarticletitle{Designers Wanted: Participation and the User
  Experience in Open Source Software Development}. In
  \bibinfo{booktitle}{\emph{Proceedings of the SIGCHI Conference on Human
  Factors in Computing Systems}} \emph{(\bibinfo{series}{CHI '09})}.
  \bibinfo{publisher}{Association for Computing Machinery},
  \bibinfo{address}{New York, NY, USA}, \bibinfo{pages}{985–994}.
\newblock
\showISBNx{9781605582467}
\href{https://doi.org/10.1145/1518701.1518852}{doi:\nolinkurl{10.1145/1518701.1518852}}


\bibitem[Bang et~al\mbox{.}(2017)]%
        {paper_54}
\bibfield{author}{\bibinfo{person}{Kristine Bang}, \bibinfo{person}{Martin~Akto
  Kanstrup}, \bibinfo{person}{Adam Kjems}, {and} \bibinfo{person}{Jan Stage}.}
  \bibinfo{year}{2017}\natexlab{}.
\newblock \showarticletitle{Adoption of UX Evaluation in Practice: An Action
  Research Study in a Software Organization}. In
  \bibinfo{booktitle}{\emph{Human-Computer Interaction - INTERACT 2017}}.
  \bibinfo{publisher}{Springer-Verlag}, \bibinfo{address}{Berlin, Heidelberg},
  \bibinfo{pages}{169–188}.
\newblock
\showISBNx{9783319680583}
\href{https://doi.org/10.1007/978-3-319-68059-0_11}{doi:\nolinkurl{10.1007/978-3-319-68059-0_11}}


\bibitem[Banker et~al\mbox{.}(2006)]%
        {2006_collaboration_tool}
\bibfield{author}{\bibinfo{person}{Rajiv~D. Banker}, \bibinfo{person}{Indranil
  Bardhan}, {and} \bibinfo{person}{Ozer Asdemir}.}
  \bibinfo{year}{2006}\natexlab{}.
\newblock \showarticletitle{Understanding the Impact of Collaboration Software
  on Product Design and Development}.
\newblock \bibinfo{journal}{\emph{Information Systems Research}}
  \bibinfo{volume}{17}, \bibinfo{number}{4} (\bibinfo{year}{2006}),
  \bibinfo{pages}{352–373}.
\newblock
\showISSN{1526-5536}
\href{https://doi.org/10.1287/isre.1060.0104}{doi:\nolinkurl{10.1287/isre.1060.0104}}


\bibitem[Bassil(2012)]%
        {2012_waterfall_lifecycle}
\bibfield{author}{\bibinfo{person}{Youssef Bassil}.}
  \bibinfo{year}{2012}\natexlab{}.
\newblock \bibinfo{title}{A Simulation Model for the Waterfall Software
  Development Life Cycle}.
\newblock
\showeprint{1205.6904}


\bibitem[Beck et~al\mbox{.}(2001)]%
        {2001_agilemanifesto}
\bibfield{author}{\bibinfo{person}{K. Beck}, \bibinfo{person}{M. Beedle},
  \bibinfo{person}{A. van Bennekum}, \bibinfo{person}{A. Cockburn},
  \bibinfo{person}{W. Cunningham}, \bibinfo{person}{M. Fowler},
  \bibinfo{person}{...}, {and} \bibinfo{person}{J. Kern}.}
  \bibinfo{year}{2001}\natexlab{}.
\newblock \bibinfo{title}{Manifesto for Agile Software Development}.
\newblock \bibinfo{howpublished}{Agile Alliance}.
\newblock
\urldef\tempurl%
\url{http://www.agilemanifesto.org/}
\showURL{%
\tempurl}


\bibitem[Bjarnason et~al\mbox{.}(2022)]%
        {bjarnason2022inter}
\bibfield{author}{\bibinfo{person}{Elizabeth Bjarnason},
  \bibinfo{person}{Baldvin Gislason~Bern}, {and} \bibinfo{person}{Linda
  Svedberg}.} \bibinfo{year}{2022}\natexlab{}.
\newblock \showarticletitle{Inter-Team Communication in Large-Scale Co-located
  Software Engineering: A Case Study}.
\newblock \bibinfo{journal}{\emph{Empirical Software Engineering}}
  \bibinfo{volume}{27}, \bibinfo{number}{2} (\bibinfo{year}{2022}).
\newblock
\showISSN{1573-7616}
\href{https://doi.org/10.1007/s10664-021-10027-z}{doi:\nolinkurl{10.1007/s10664-021-10027-z}}


\bibitem[Boehm(1988)]%
        {Boehm_waterfall}
\bibfield{author}{\bibinfo{person}{B.~W. Boehm}.}
  \bibinfo{year}{1988}\natexlab{}.
\newblock \showarticletitle{A Spiral Model of Software Development and
  Enhancement}.
\newblock \bibinfo{journal}{\emph{Computer}} \bibinfo{volume}{21},
  \bibinfo{number}{5} (\bibinfo{year}{1988}), \bibinfo{pages}{61--72}.
\newblock
\href{https://doi.org/10.1109/2.59}{doi:\nolinkurl{10.1109/2.59}}


\bibitem[Boland and Fitzgerald(2004)]%
        {boland2004transitioning}
\bibfield{author}{\bibinfo{person}{David Boland} {and} \bibinfo{person}{Brian
  Fitzgerald}.} \bibinfo{year}{2004}\natexlab{}.
\newblock \showarticletitle{Transitioning from a Co-located to a
  Globally-distributed Software Development Team: A Case Study at Analog
  Devices Inc}. In \bibinfo{booktitle}{\emph{Third International Workshop on
  Global Software Development (GSD 2004) in 26th International Conference on
  Software Engineering}}. \bibinfo{publisher}{IEEE}.
\newblock
\href{https://doi.org/10.1049/ic:20040303}{doi:\nolinkurl{10.1049/ic:20040303}}


\bibitem[Brown et~al\mbox{.}(2008)]%
        {paper_19}
\bibfield{author}{\bibinfo{person}{Judith Brown}, \bibinfo{person}{Gitte
  Lindgaard}, {and} \bibinfo{person}{Robert Biddle}.}
  \bibinfo{year}{2008}\natexlab{}.
\newblock \showarticletitle{Stories, Sketches, and Lists: Developers and
  Interaction Designers Interacting Through Artefacts}. In
  \bibinfo{booktitle}{\emph{Agile 2008 Conference}}. \bibinfo{pages}{39--50}.
\newblock
\href{https://doi.org/10.1109/Agile.2008.54}{doi:\nolinkurl{10.1109/Agile.2008.54}}


\bibitem[Bruun(2010)]%
        {2010_slr_training}
\bibfield{author}{\bibinfo{person}{Anders Bruun}.}
  \bibinfo{year}{2010}\natexlab{}.
\newblock \showarticletitle{Training Software Developers in Usability
  Engineering: A Literature Review}. In \bibinfo{booktitle}{\emph{Proceedings
  of the 6th Nordic Conference on Human-Computer Interaction: Extending
  Boundaries}} \emph{(\bibinfo{series}{NordiCHI '10})}.
  \bibinfo{publisher}{Association for Computing Machinery},
  \bibinfo{address}{New York, NY, USA}, \bibinfo{pages}{82–91}.
\newblock
\showISBNx{9781605589343}
\href{https://doi.org/10.1145/1868914.1868928}{doi:\nolinkurl{10.1145/1868914.1868928}}


\bibitem[Caballero et~al\mbox{.}(2016)]%
        {2016_slr_integration_3}
\bibfield{author}{\bibinfo{person}{Leydi Caballero}, \bibinfo{person}{Ana~M.
  Moreno}, {and} \bibinfo{person}{Ahmed Seffah}.}
  \bibinfo{year}{2016}\natexlab{}.
\newblock \showarticletitle{How Agile Developers Integrate User-Centered Design
  Into Their Processes: A Literature Review}.
\newblock \bibinfo{journal}{\emph{International Journal of Software Engineering
  and Knowledge Engineering}} \bibinfo{volume}{26}, \bibinfo{number}{08}
  (\bibinfo{year}{2016}), \bibinfo{pages}{1175–1201}.
\newblock
\showISSN{1793-6403}
\href{https://doi.org/10.1142/s0218194016500418}{doi:\nolinkurl{10.1142/s0218194016500418}}


\bibitem[Cajander et~al\mbox{.}(2022)]%
        {new_3}
\bibfield{author}{\bibinfo{person}{Åsa Cajander}, \bibinfo{person}{Marta
  Larusdottir}, {and} \bibinfo{person}{Johannes~L. Geiser}.}
  \bibinfo{year}{2022}\natexlab{}.
\newblock \showarticletitle{UX Professionals’ Learning and Usage of UX
  Methods in Agile}.
\newblock \bibinfo{journal}{\emph{Information and Software Technology}}
  \bibinfo{volume}{151} (\bibinfo{year}{2022}), \bibinfo{pages}{107005}.
\newblock
\showISSN{0950-5849}
\href{https://doi.org/10.1016/j.infsof.2022.107005}{doi:\nolinkurl{10.1016/j.infsof.2022.107005}}


\bibitem[Cassivi et~al\mbox{.}(2004)]%
        {2004_collaboration_tool}
\bibfield{author}{\bibinfo{person}{Luc Cassivi}, \bibinfo{person}{Élisabeth
  Lefebvre}, \bibinfo{person}{Louis~A. Lefebvre}, {and}
  \bibinfo{person}{Pierre‐ Majorique~Léger}.}
  \bibinfo{year}{2004}\natexlab{}.
\newblock \showarticletitle{The Impact of E‐collaboration Tools on Firms’
  Performance}.
\newblock \bibinfo{journal}{\emph{The International Journal of Logistics
  Management}} \bibinfo{volume}{15}, \bibinfo{number}{1}
  (\bibinfo{year}{2004}), \bibinfo{pages}{91–110}.
\newblock
\showISSN{0957-4093}
\href{https://doi.org/10.1108/09574090410700257}{doi:\nolinkurl{10.1108/09574090410700257}}


\bibitem[Chamberlain et~al\mbox{.}(2006)]%
        {paper_s9}
\bibfield{author}{\bibinfo{person}{Stephanie Chamberlain},
  \bibinfo{person}{Helen Sharp}, {and} \bibinfo{person}{Neil Maiden}.}
  \bibinfo{year}{2006}\natexlab{}.
\newblock \showarticletitle{Towards a Framework for Integrating Agile
  Development and User-Centred Design}. In
  \bibinfo{booktitle}{\emph{Proceedings of the 7th International Conference on
  Extreme Programming and Agile Processes in Software Engineering}}
  \emph{(\bibinfo{series}{XP'06})}. \bibinfo{publisher}{Springer-Verlag},
  \bibinfo{address}{Berlin, Heidelberg}, \bibinfo{pages}{143–153}.
\newblock
\showISBNx{3540350942}
\href{https://doi.org/10.1007/11774129_15}{doi:\nolinkurl{10.1007/11774129_15}}


\bibitem[Dabbish et~al\mbox{.}(2012)]%
        {dabbish2012social}
\bibfield{author}{\bibinfo{person}{Laura Dabbish}, \bibinfo{person}{Colleen
  Stuart}, \bibinfo{person}{Jason Tsay}, {and} \bibinfo{person}{Jim Herbsleb}.}
  \bibinfo{year}{2012}\natexlab{}.
\newblock \showarticletitle{Social Coding in Github: Transparency and
  Collaboration in an Open Software Repository}. In
  \bibinfo{booktitle}{\emph{Proceedings of the ACM 2012 Conference on Computer
  Supported Cooperative Work}} \emph{(\bibinfo{series}{CSCW '12})}.
  \bibinfo{publisher}{Association for Computing Machinery},
  \bibinfo{address}{New York, NY, USA}, \bibinfo{pages}{1277–1286}.
\newblock
\showISBNx{9781450310864}
\href{https://doi.org/10.1145/2145204.2145396}{doi:\nolinkurl{10.1145/2145204.2145396}}


\bibitem[Dam and Siang(2022)]%
        {affinity_diagrams}
\bibfield{author}{\bibinfo{person}{Rikke~Friis Dam} {and}
  \bibinfo{person}{Teo~Yu Siang}.} \bibinfo{year}{2022}\natexlab{}.
\newblock \bibinfo{title}{Affinity Diagrams: How to Cluster Your Ideas and
  Reveal Insights}.
\newblock
\urldef\tempurl%
\url{https://www.interaction-design.org/literature/article/affinity-diagrams-learn-how-to-cluster-and-bundle-ideas-and-facts}
\showURL{%
\tempurl}


\bibitem[Dekel(2005)]%
        {dekel2005supporting}
\bibfield{author}{\bibinfo{person}{Uri Dekel}.}
  \bibinfo{year}{2005}\natexlab{}.
\newblock \showarticletitle{Supporting Distributed Software Design Meetings:
  What Can We Learn from Co-located Meetings?}
\newblock \bibinfo{journal}{\emph{ACM SIGSOFT Software Engineering Notes}}
  \bibinfo{volume}{30}, \bibinfo{number}{4} (\bibinfo{year}{2005}),
  \bibinfo{pages}{1–7}.
\newblock
\showISSN{0163-5948}
\href{https://doi.org/10.1145/1082983.1083109}{doi:\nolinkurl{10.1145/1082983.1083109}}


\bibitem[Dhandapani(2016)]%
        {paper_16}
\bibfield{author}{\bibinfo{person}{Sowmya Dhandapani}.}
  \bibinfo{year}{2016}\natexlab{}.
\newblock \showarticletitle{Integration of User Centered Design and Software
  Development Process}. In \bibinfo{booktitle}{\emph{2016 IEEE 7th Annual
  Information Technology, Electronics and Mobile Communication Conference
  (IEMCON)}}. \bibinfo{pages}{1--5}.
\newblock
\href{https://doi.org/10.1109/IEMCON.2016.7746075}{doi:\nolinkurl{10.1109/IEMCON.2016.7746075}}


\bibitem[Dourish and Bellotti(1992)]%
        {dourish1992awareness}
\bibfield{author}{\bibinfo{person}{Paul Dourish} {and}
  \bibinfo{person}{Victoria Bellotti}.} \bibinfo{year}{1992}\natexlab{}.
\newblock \showarticletitle{Awareness and Coordination in Shared Workspaces}.
  In \bibinfo{booktitle}{\emph{Proceedings of the 1992 ACM Conference on
  Computer-Supported Cooperative Work}} (Toronto, Ontario, Canada)
  \emph{(\bibinfo{series}{CSCW '92})}. \bibinfo{publisher}{Association for
  Computing Machinery}, \bibinfo{address}{New York, NY, USA},
  \bibinfo{pages}{107–114}.
\newblock
\showISBNx{0897915429}
\href{https://doi.org/10.1145/143457.143468}{doi:\nolinkurl{10.1145/143457.143468}}


\bibitem[Exchange(2023a)]%
        {stack_overflow}
\bibfield{author}{\bibinfo{person}{Stack Exchange}.}
  \bibinfo{year}{2023}\natexlab{a}.
\newblock \bibinfo{title}{Stack Overflow}.
\newblock \bibinfo{howpublished}{\url{https://stackoverflow.com/}}.
\newblock


\bibitem[Exchange(2023b)]%
        {UX_stackexchange}
\bibfield{author}{\bibinfo{person}{Stack Exchange}.}
  \bibinfo{year}{2023}\natexlab{b}.
\newblock \bibinfo{title}{User Experience Stack Exchange}.
\newblock \bibinfo{howpublished}{\url{https://ux.stackexchange.com/}}.
\newblock


\bibitem[Feng et~al\mbox{.}(2023)]%
        {feng2023understanding}
\bibfield{author}{\bibinfo{person}{K.~J.~Kevin Feng}, \bibinfo{person}{Tony~W
  Li}, {and} \bibinfo{person}{Amy~X. Zhang}.} \bibinfo{year}{2023}\natexlab{}.
\newblock \showarticletitle{Understanding Collaborative Practices and Tools of
  Professional UX Practitioners in Software Organizations}. In
  \bibinfo{booktitle}{\emph{Proceedings of the 2023 CHI Conference on Human
  Factors in Computing Systems}} \emph{(\bibinfo{series}{CHI '23})}.
  \bibinfo{publisher}{Association for Computing Machinery},
  \bibinfo{address}{New York, NY, USA}, Article \bibinfo{articleno}{764},
  \bibinfo{numpages}{20}~pages.
\newblock
\showISBNx{9781450394215}
\href{https://doi.org/10.1145/3544548.3581273}{doi:\nolinkurl{10.1145/3544548.3581273}}


\bibitem[Feng and Zhang(2022)]%
        {feng2022handoffs}
\bibfield{author}{\bibinfo{person}{KJ~Kevin Feng} {and} \bibinfo{person}{Amyx
  Zhang}.} \bibinfo{year}{2022}\natexlab{}.
\newblock \showarticletitle{From Handoffs to Co-creation: Deepening
  Collaboration Between Designers, Developers, and Data Science Workers in UX
  Design}.
\newblock \bibinfo{journal}{\emph{Proceedings of the InContext: Futuring
  User-Experience Design Tools Workshop at CHI Conference on Human Factors in
  Computing Systems}} (\bibinfo{year}{2022}).
\newblock


\bibitem[Ferreira et~al\mbox{.}(2023)]%
        {new_2}
\bibfield{author}{\bibinfo{person}{Bruna Ferreira}, \bibinfo{person}{Silvio
  Marques}, \bibinfo{person}{Marcos Kalinowski}, \bibinfo{person}{Hélio
  Lopes}, {and} \bibinfo{person}{Simone~D.J. Barbosa}.}
  \bibinfo{year}{2023}\natexlab{}.
\newblock \showarticletitle{Lessons Learned to Improve the UX Practices in
  Agile Projects Involving Data Science and Process Automation}.
\newblock \bibinfo{journal}{\emph{Information and Software Technology}}
  \bibinfo{volume}{155} (\bibinfo{year}{2023}), \bibinfo{pages}{107106}.
\newblock
\showISSN{0950-5849}
\href{https://doi.org/10.1016/j.infsof.2022.107106}{doi:\nolinkurl{10.1016/j.infsof.2022.107106}}


\bibitem[Ferreira et~al\mbox{.}(2007a)]%
        {paper_22}
\bibfield{author}{\bibinfo{person}{Jennifer Ferreira}, \bibinfo{person}{James
  Noble}, {and} \bibinfo{person}{Robert Biddle}.}
  \bibinfo{year}{2007}\natexlab{a}.
\newblock \showarticletitle{Agile Development Iterations and UI Design}. In
  \bibinfo{booktitle}{\emph{Agile 2007 (AGILE 2007)}}. \bibinfo{pages}{50--58}.
\newblock
\href{https://doi.org/10.1109/AGILE.2007.8}{doi:\nolinkurl{10.1109/AGILE.2007.8}}


\bibitem[Ferreira et~al\mbox{.}(2007b)]%
        {paper_95}
\bibfield{author}{\bibinfo{person}{Jennifer Ferreira}, \bibinfo{person}{James
  Noble}, {and} \bibinfo{person}{Robert Biddle}.}
  \bibinfo{year}{2007}\natexlab{b}.
\newblock \showarticletitle{Interaction Designers on eXtreme Programming Teams:
  Two Case Studies From the Real World}. In
  \bibinfo{booktitle}{\emph{Proceedings of the Fifth New Zealand Computer
  Science Research Student Conference (NZCSRSC2007)}},
  Vol.~\bibinfo{volume}{55}. \bibinfo{pages}{93--94}.
\newblock


\bibitem[Ferreira et~al\mbox{.}(2011)]%
        {paper_46}
\bibfield{author}{\bibinfo{person}{Jennifer Ferreira}, \bibinfo{person}{Helen
  Sharp}, {and} \bibinfo{person}{Hugh Robinson}.}
  \bibinfo{year}{2011}\natexlab{}.
\newblock \showarticletitle{User Experience Design and Agile Development:
  Managing Cooperation Through Articulation Work}.
\newblock \bibinfo{journal}{\emph{Software: Practice and Experience}}
  \bibinfo{volume}{41}, \bibinfo{number}{9} (\bibinfo{year}{2011}),
  \bibinfo{pages}{963–974}.
\newblock
\showISSN{1097-024X}
\href{https://doi.org/10.1002/spe.1012}{doi:\nolinkurl{10.1002/spe.1012}}


\bibitem[Ferreira et~al\mbox{.}(2012)]%
        {paper_11}
\bibfield{author}{\bibinfo{person}{Jennifer Ferreira}, \bibinfo{person}{Helen
  Sharp}, {and} \bibinfo{person}{Hugh Robinson}.}
  \bibinfo{year}{2012}\natexlab{}.
\newblock \showarticletitle{Agile Development and User Experience Design
  Integration as an Ongoing Achievement in Practice}. In
  \bibinfo{booktitle}{\emph{2012 Agile Conference}}. \bibinfo{pages}{11--20}.
\newblock
\href{https://doi.org/10.1109/Agile.2012.33}{doi:\nolinkurl{10.1109/Agile.2012.33}}


\bibitem[Figma(2023a)]%
        {figma}
\bibfield{author}{\bibinfo{person}{Figma}.} \bibinfo{year}{2023}\natexlab{a}.
\newblock \bibinfo{title}{Figma}.
\newblock \bibinfo{howpublished}{\url{https://www.figma.com/}}.
\newblock


\bibitem[Figma(2023b)]%
        {github_figma}
\bibfield{author}{\bibinfo{person}{Figma}.} \bibinfo{year}{2023}\natexlab{b}.
\newblock \bibinfo{title}{Github and Figma—Better Together}.
\newblock
  \bibinfo{howpublished}{\url{https://www.figma.com/github-for-figjam/}}.
\newblock


\bibitem[Fox et~al\mbox{.}(2008)]%
        {paper_14}
\bibfield{author}{\bibinfo{person}{David Fox}, \bibinfo{person}{Jonathan
  Sillito}, {and} \bibinfo{person}{Frank Maurer}.}
  \bibinfo{year}{2008}\natexlab{}.
\newblock \showarticletitle{Agile Methods and User-Centered Design: How These
  Two Methodologies are Being Successfully Integrated in Industry}. In
  \bibinfo{booktitle}{\emph{Agile 2008 Conference}}. \bibinfo{pages}{63--72}.
\newblock
\href{https://doi.org/10.1109/Agile.2008.78}{doi:\nolinkurl{10.1109/Agile.2008.78}}


\bibitem[Garcia et~al\mbox{.}(2017)]%
        {2017_slr_artifacts}
\bibfield{author}{\bibinfo{person}{Andrei Garcia}, \bibinfo{person}{Tiago
  Da~Silva}, {and} \bibinfo{person}{Milene Silveira}.}
  \bibinfo{year}{2017}\natexlab{}.
\newblock \showarticletitle{Artifacts for Agile User-Centered Design: A
  Systematic Mapping}. In \bibinfo{booktitle}{\emph{Hawaii International
  Conference on System Sciences}}. \bibinfo{pages}{5859--5868}.
\newblock


\bibitem[Green et~al\mbox{.}(2010)]%
        {green2010understanding}
\bibfield{author}{\bibinfo{person}{R Green}, \bibinfo{person}{THOMAS Mazzuchi},
  {and} \bibinfo{person}{SHAHRAM Sarkani}.} \bibinfo{year}{2010}\natexlab{}.
\newblock \showarticletitle{Understanding the Role of Synchronous \&
  Asynchronous Communication in Agile Software Development and Its Effects on
  Quality}.
\newblock \bibinfo{journal}{\emph{Journal of Information Technology
  Management}} \bibinfo{volume}{21}, \bibinfo{number}{2}
  (\bibinfo{year}{2010}), \bibinfo{pages}{8--23}.
\newblock


\bibitem[Gutwin and Greenberg(2004)]%
        {gutwin2004importance}
\bibfield{author}{\bibinfo{person}{Carl Gutwin} {and} \bibinfo{person}{Saul
  Greenberg}.} \bibinfo{year}{2004}\natexlab{}.
\newblock \showarticletitle{The Importance of Awareness for Team Cognition in
  Distributed Collaboration}. In \bibinfo{booktitle}{\emph{Team cognition:
  Understanding the factors that drive process and performance.}}
  \bibinfo{publisher}{American Psychological Association},
  \bibinfo{pages}{177–201}.
\newblock
\showISBNx{1591471036}
\href{https://doi.org/10.1037/10690-009}{doi:\nolinkurl{10.1037/10690-009}}


\bibitem[Gutwin et~al\mbox{.}(2004)]%
        {gutwin2004group}
\bibfield{author}{\bibinfo{person}{Carl Gutwin}, \bibinfo{person}{Reagan
  Penner}, {and} \bibinfo{person}{Kevin Schneider}.}
  \bibinfo{year}{2004}\natexlab{}.
\newblock \showarticletitle{Group Awareness in Distributed Software
  Development}. In \bibinfo{booktitle}{\emph{Proceedings of the 2004 ACM
  Conference on Computer Supported Cooperative Work}}
  \emph{(\bibinfo{series}{CSCW '04})}. \bibinfo{publisher}{Association for
  Computing Machinery}, \bibinfo{address}{New York, NY, USA},
  \bibinfo{pages}{72–81}.
\newblock
\showISBNx{1581138105}
\href{https://doi.org/10.1145/1031607.1031621}{doi:\nolinkurl{10.1145/1031607.1031621}}


\bibitem[Göransson et~al\mbox{.}(2004)]%
        {2004_new}
\bibfield{author}{\bibinfo{person}{Bengt Göransson}, \bibinfo{person}{Jan
  Gulliksen}, {and} \bibinfo{person}{Inger Boivie}.}
  \bibinfo{year}{2004}\natexlab{}.
\newblock \showarticletitle{The usability design process - Integrating
  user-centered systems design in the software development process}.
\newblock \bibinfo{journal}{\emph{Software Process: Improvement and Practice}}
  \bibinfo{volume}{8} (\bibinfo{date}{04} \bibinfo{year}{2004}),
  \bibinfo{pages}{111--131}.
\newblock
\href{https://doi.org/10.1002/spip.174}{doi:\nolinkurl{10.1002/spip.174}}


\bibitem[Hartson and Pyla(2019)]%
        {2019_UXbook}
\bibfield{author}{\bibinfo{person}{Rex Hartson} {and} \bibinfo{person}{Pardha~S
  Pyla}.} \bibinfo{year}{2019}\natexlab{}.
\newblock \bibinfo{booktitle}{\emph{The {UX} book} (\bibinfo{edition}{2} ed.)}.
\newblock \bibinfo{publisher}{Morgan Kaufmann}, \bibinfo{address}{Oxford,
  England}.
\newblock


\bibitem[Harzl(2017)]%
        {harzl2017can}
\bibfield{author}{\bibinfo{person}{Annemarie Harzl}.}
  \bibinfo{year}{2017}\natexlab{}.
\newblock \showarticletitle{Can Foss Projects Benefit From Integrating Kanban:
  A Case Study}.
\newblock \bibinfo{journal}{\emph{Journal of Internet Services and
  Applications}} \bibinfo{volume}{8}, \bibinfo{number}{1}
  (\bibinfo{year}{2017}), \bibinfo{pages}{7}.
\newblock
\href{https://doi.org/10.1186/s13174-017-0058-z}{doi:\nolinkurl{10.1186/s13174-017-0058-z}}


\bibitem[Hassenzahl and Tractinsky(2006)]%
        {2006_uxlifecycle}
\bibfield{author}{\bibinfo{person}{Marc Hassenzahl} {and} \bibinfo{person}{Noam
  Tractinsky}.} \bibinfo{year}{2006}\natexlab{}.
\newblock \showarticletitle{User Experience - A Research Agenda}.
\newblock \bibinfo{journal}{\emph{Behaviour \& Information Technology}}
  \bibinfo{volume}{25}, \bibinfo{number}{2} (\bibinfo{year}{2006}),
  \bibinfo{pages}{91–97}.
\newblock
\showISSN{1362-3001}
\href{https://doi.org/10.1080/01449290500330331}{doi:\nolinkurl{10.1080/01449290500330331}}


\bibitem[Hattori(2010)]%
        {hattori2010enhancing}
\bibfield{author}{\bibinfo{person}{Lile Hattori}.}
  \bibinfo{year}{2010}\natexlab{}.
\newblock \showarticletitle{Enhancing Collaboration of Multi-Developer Projects
  With Synchronous Changes}. In \bibinfo{booktitle}{\emph{Proceedings of the
  32nd ACM/IEEE International Conference on Software Engineering - Volume 2}}
  \emph{(\bibinfo{series}{ICSE '10})}. \bibinfo{publisher}{Association for
  Computing Machinery}, \bibinfo{address}{New York, NY, USA},
  \bibinfo{pages}{377–380}.
\newblock
\showISBNx{9781605587196}
\href{https://doi.org/10.1145/1810295.1810397}{doi:\nolinkurl{10.1145/1810295.1810397}}


\bibitem[Hellman et~al\mbox{.}(2021)]%
        {Hellman2021}
\bibfield{author}{\bibinfo{person}{Jazlyn Hellman}, \bibinfo{person}{Jinghui
  Cheng}, {and} \bibinfo{person}{Jin~L.C. Guo}.}
  \bibinfo{year}{2021}\natexlab{}.
\newblock \showarticletitle{Facilitating Asynchronous Participatory Design of
  Open Source Software: Bringing End Users into the Loop}. In
  \bibinfo{booktitle}{\emph{Extended Abstracts of the 2021 CHI Conference on
  Human Factors in Computing Systems}} \emph{(\bibinfo{series}{CHI EA '21})}.
  \bibinfo{publisher}{Association for Computing Machinery},
  \bibinfo{address}{New York, NY, USA}, Article \bibinfo{articleno}{438},
  \bibinfo{numpages}{7}~pages.
\newblock
\showISBNx{9781450380959}
\href{https://doi.org/10.1145/3411763.3451643}{doi:\nolinkurl{10.1145/3411763.3451643}}


\bibitem[Herbsleb(2007)]%
        {herbsleb2007global}
\bibfield{author}{\bibinfo{person}{James~D. Herbsleb}.}
  \bibinfo{year}{2007}\natexlab{}.
\newblock \showarticletitle{Global Software Engineering: The Future of
  Socio-Technical Coordination}. In \bibinfo{booktitle}{\emph{Future of
  Software Engineering (FOSE '07)}}. \bibinfo{pages}{188--198}.
\newblock
\href{https://doi.org/10.1109/FOSE.2007.11}{doi:\nolinkurl{10.1109/FOSE.2007.11}}


\bibitem[Hewett et~al\mbox{.}(1992)]%
        {acm_hci_curriculum}
\bibfield{author}{\bibinfo{person}{Thomas Hewett}, \bibinfo{person}{Ronald
  Baecker}, \bibinfo{person}{Stuart Card}, \bibinfo{person}{Tom Carey},
  \bibinfo{person}{Jean Gasen}, \bibinfo{person}{Marilyn Mantei},
  \bibinfo{person}{Gary Perlman}, \bibinfo{person}{Gary Strong}, {and}
  \bibinfo{person}{William Verplank}.} \bibinfo{year}{1992}\natexlab{}.
\newblock \bibinfo{booktitle}{\emph{ACM SIGCHI Curricula for Human-Computer
  Interaction}}.
\newblock \bibinfo{publisher}{Association for Computing Machinery}.
\newblock
\showISBNx{0897914740}
\href{https://doi.org/10.1145/2594128}{doi:\nolinkurl{10.1145/2594128}}


\bibitem[Hinderks et~al\mbox{.}(2022)]%
        {2022_slr_management}
\bibfield{author}{\bibinfo{person}{Andreas Hinderks},
  \bibinfo{person}{Francisco~José Domínguez~Mayo}, \bibinfo{person}{J\"{o}rg
  Thomaschewski}, {and} \bibinfo{person}{María~José Escalona}.}
  \bibinfo{year}{2022}\natexlab{}.
\newblock \showarticletitle{Approaches to Manage the User Experience Process in
  Agile Software Development: A Systematic Literature Review}.
\newblock \bibinfo{journal}{\emph{Information and Software Technology}}
  \bibinfo{volume}{150} (\bibinfo{year}{2022}), \bibinfo{pages}{106957}.
\newblock
\showISSN{0950-5849}
\href{https://doi.org/10.1016/j.infsof.2022.106957}{doi:\nolinkurl{10.1016/j.infsof.2022.106957}}


\bibitem[Hodgetts(2005)]%
        {paper_17}
\bibfield{author}{\bibinfo{person}{P. Hodgetts}.}
  \bibinfo{year}{2005}\natexlab{}.
\newblock \showarticletitle{Experiences Integrating Sophisticated User
  Experience Design Practices Into Agile Processes}. In
  \bibinfo{booktitle}{\emph{Agile Development Conference (ADC'05)}}.
  \bibinfo{pages}{235--242}.
\newblock
\href{https://doi.org/10.1109/ADC.2005.24}{doi:\nolinkurl{10.1109/ADC.2005.24}}


\bibitem[Iivari(2005)]%
        {paper_57}
\bibfield{author}{\bibinfo{person}{Netta Iivari}.}
  \bibinfo{year}{2005}\natexlab{}.
\newblock \showarticletitle{Usability Specialists -- `A Mommy Mob', `Realistic
  Humanists' or `Staid Researchers'? An Analysis of Usability Work in the
  Software Product Development}. In \bibinfo{booktitle}{\emph{Human-Computer
  Interaction - INTERACT 2005}}. \bibinfo{publisher}{Springer-Verlag},
  \bibinfo{address}{Berlin, Heidelberg}, \bibinfo{pages}{418–430}.
\newblock
\showISBNx{3540289437}
\href{https://doi.org/10.1007/11555261_35}{doi:\nolinkurl{10.1007/11555261_35}}


\bibitem[Isomursu et~al\mbox{.}(2012)]%
        {paper_59}
\bibfield{author}{\bibinfo{person}{Minna Isomursu}, \bibinfo{person}{Andrey
  Sirotkin}, \bibinfo{person}{Petri Voltti}, {and} \bibinfo{person}{Markku
  Halonen}.} \bibinfo{year}{2012}\natexlab{}.
\newblock \showarticletitle{User Experience Design Goes Agile in Lean
  Transformation -- A Case Study}. In \bibinfo{booktitle}{\emph{2012 Agile
  Conference}}. \bibinfo{pages}{1--10}.
\newblock
\href{https://doi.org/10.1109/Agile.2012.10}{doi:\nolinkurl{10.1109/Agile.2012.10}}


\bibitem[Jolak et~al\mbox{.}(2020)]%
        {jolak2020design}
\bibfield{author}{\bibinfo{person}{Rodi Jolak}, \bibinfo{person}{Andreas
  Wortmann}, \bibinfo{person}{Grischa Liebel}, \bibinfo{person}{Eric Umuhoza},
  {and} \bibinfo{person}{Michel R.~V. Chaudron}.}
  \bibinfo{year}{2020}\natexlab{}.
\newblock \showarticletitle{The Design Thinking of Co-located vs. Distributed
  Software Developers: Distance Strikes Again!}. In
  \bibinfo{booktitle}{\emph{Proceedings of the 15th International Conference on
  Global Software Engineering}} \emph{(\bibinfo{series}{ICGSE '20})}.
  \bibinfo{publisher}{Association for Computing Machinery},
  \bibinfo{address}{New York, NY, USA}, \bibinfo{pages}{106–116}.
\newblock
\showISBNx{9781450370936}
\href{https://doi.org/10.1145/3372787.3390438}{doi:\nolinkurl{10.1145/3372787.3390438}}


\bibitem[Joshi et~al\mbox{.}(2010)]%
        {2010_integration}
\bibfield{author}{\bibinfo{person}{Anirudha Joshi}, \bibinfo{person}{N.L.
  Sarda}, {and} \bibinfo{person}{Sanjay Tripathi}.}
  \bibinfo{year}{2010}\natexlab{}.
\newblock \showarticletitle{Supporting Multidisciplinary Collaboration:
  Requirements From Novel HCI Education}.
\newblock \bibinfo{journal}{\emph{Journal of Systems and Software}}
  \bibinfo{volume}{83}, \bibinfo{number}{11} (\bibinfo{year}{2010}),
  \bibinfo{pages}{2045–2058}.
\newblock
\showISSN{0164-1212}
\href{https://doi.org/10.1016/j.jss.2010.03.078}{doi:\nolinkurl{10.1016/j.jss.2010.03.078}}


\bibitem[Jurca et~al\mbox{.}(2014)]%
        {2014_Integration_slr}
\bibfield{author}{\bibinfo{person}{Gabriela Jurca},
  \bibinfo{person}{Theodore~D. Hellmann}, {and} \bibinfo{person}{Frank
  Maurer}.} \bibinfo{year}{2014}\natexlab{}.
\newblock \showarticletitle{Integrating Agile and User-Centered Design: A
  Systematic Mapping and Review of Evaluation and Validation Studies of
  Agile-UX}. In \bibinfo{booktitle}{\emph{2014 Agile Conference}}.
  \bibinfo{pages}{24--32}.
\newblock
\href{https://doi.org/10.1109/AGILE.2014.17}{doi:\nolinkurl{10.1109/AGILE.2014.17}}


\bibitem[Juristo et~al\mbox{.}(2007)]%
        {paper_50}
\bibfield{author}{\bibinfo{person}{Natalia Juristo}, \bibinfo{person}{Ana
  Moreno}, \bibinfo{person}{Maria-Isabel Sanchez-Segura}, {and}
  \bibinfo{person}{Maria Cecília~Calani Baranauskas}.}
  \bibinfo{year}{2007}\natexlab{}.
\newblock \showarticletitle{A Glass Box Design: Making the Impact of Usability
  on Software Development Visible}. In \bibinfo{booktitle}{\emph{Human-Computer
  Interaction – INTERACT 2007}}. \bibinfo{publisher}{Springer Berlin
  Heidelberg}, \bibinfo{pages}{541–554}.
\newblock
\showISBNx{9783540748007}
\showISSN{1611-3349}
\href{https://doi.org/10.1007/978-3-540-74800-7_49}{doi:\nolinkurl{10.1007/978-3-540-74800-7_49}}


\bibitem[Kashfi et~al\mbox{.}(2019)]%
        {2019_integration}
\bibfield{author}{\bibinfo{person}{Pariya Kashfi}, \bibinfo{person}{Robert
  Feldt}, {and} \bibinfo{person}{Agneta Nilsson}.}
  \bibinfo{year}{2019}\natexlab{}.
\newblock \showarticletitle{Integrating UX Principles and Practices Into
  Software Development Organizations: A Case Study of Influencing Events}.
\newblock \bibinfo{journal}{\emph{Journal of Systems and Software}}
  \bibinfo{volume}{154} (\bibinfo{year}{2019}), \bibinfo{pages}{37–58}.
\newblock
\showISSN{0164-1212}
\href{https://doi.org/10.1016/j.jss.2019.03.066}{doi:\nolinkurl{10.1016/j.jss.2019.03.066}}


\bibitem[Kashfi et~al\mbox{.}(2017)]%
        {paper_105}
\bibfield{author}{\bibinfo{person}{Pariya Kashfi}, \bibinfo{person}{Agneta
  Nilsson}, {and} \bibinfo{person}{Robert Feldt}.}
  \bibinfo{year}{2017}\natexlab{}.
\newblock \showarticletitle{Integrating User eXperience Practices Into Software
  Development Processes: Implications of the UX Characteristics}.
\newblock \bibinfo{journal}{\emph{PeerJ Computer Science}}  \bibinfo{volume}{3}
  (\bibinfo{year}{2017}), \bibinfo{pages}{e130}.
\newblock
\showISSN{2376-5992}
\href{https://doi.org/10.7717/peerj-cs.130}{doi:\nolinkurl{10.7717/peerj-cs.130}}


\bibitem[Kitchenham and Brereton(2013)]%
        {kitchenham2013systematic}
\bibfield{author}{\bibinfo{person}{Barbara Kitchenham} {and}
  \bibinfo{person}{Pearl Brereton}.} \bibinfo{year}{2013}\natexlab{}.
\newblock \showarticletitle{A Systematic Review of Systematic Review Process
  Research in Software Engineering}.
\newblock \bibinfo{journal}{\emph{Information and Software Technology}}
  \bibinfo{volume}{55}, \bibinfo{number}{12} (\bibinfo{year}{2013}),
  \bibinfo{pages}{2049–2075}.
\newblock
\showISSN{0950-5849}
\href{https://doi.org/10.1016/j.infsof.2013.07.010}{doi:\nolinkurl{10.1016/j.infsof.2013.07.010}}


\bibitem[Kitchenham and Charters(2007)]%
        {2007_SLRguide}
\bibfield{author}{\bibinfo{person}{Barbara Kitchenham} {and}
  \bibinfo{person}{Stuart Charters}.} \bibinfo{year}{2007}\natexlab{}.
\newblock \showarticletitle{Guidelines for Performing Systematic Literature
  Reviews in Software Engineering}.
\newblock \bibinfo{journal}{\emph{EBSE Technical Report}}
  (\bibinfo{year}{2007}).
\newblock


\bibitem[Klein et~al\mbox{.}(2012)]%
        {klein2012scaling}
\bibfield{author}{\bibinfo{person}{Harald Klein}, \bibinfo{person}{Eric
  Knauss}, {and} \bibinfo{person}{Andreas Rausch}.}
  \bibinfo{year}{2012}\natexlab{}.
\newblock \showarticletitle{Scaling Software Development Methods from
  Co-located to Distributed}. In \bibinfo{booktitle}{\emph{Software Quality.
  Process Automation in Software Development}},
  \bibfield{editor}{\bibinfo{person}{Stefan Biffl}, \bibinfo{person}{Dietmar
  Winkler}, {and} \bibinfo{person}{Johannes Bergsmann}} (Eds.).
  \bibinfo{publisher}{Springer Berlin Heidelberg}, \bibinfo{address}{Berlin,
  Heidelberg}, \bibinfo{pages}{71--83}.
\newblock
\showISBNx{978-3-642-27213-4}


\bibitem[Kollmann et~al\mbox{.}(2009)]%
        {paper_s12}
\bibfield{author}{\bibinfo{person}{Johanna Kollmann}, \bibinfo{person}{Helen
  Sharp}, {and} \bibinfo{person}{Ann Blandford}.}
  \bibinfo{year}{2009}\natexlab{}.
\newblock \showarticletitle{The Importance of Identity and Vision to User
  Experience Designers on Agile Projects}. In \bibinfo{booktitle}{\emph{2009
  Agile Conference}}. \bibinfo{pages}{11--18}.
\newblock
\href{https://doi.org/10.1109/AGILE.2009.58}{doi:\nolinkurl{10.1109/AGILE.2009.58}}


\bibitem[Kuusinen(2015)]%
        {paper_48}
\bibfield{author}{\bibinfo{person}{Kati Kuusinen}.}
  \bibinfo{year}{2015}\natexlab{}.
\newblock \showarticletitle{Task Allocation Between UX Specialists and
  Developers in Agile Software Development Projects}. In
  \bibinfo{booktitle}{\emph{Human-Computer Interaction – INTERACT 2015}}.
  \bibinfo{publisher}{Springer-Verlag}, \bibinfo{address}{Berlin, Heidelberg},
  \bibinfo{pages}{27–44}.
\newblock
\showISBNx{978-3-319-22697-2}
\href{https://doi.org/10.1007/978-3-319-22698-9_3}{doi:\nolinkurl{10.1007/978-3-319-22698-9_3}}


\bibitem[Larusdottir et~al\mbox{.}(2017)]%
        {paper_103}
\bibfield{author}{\bibinfo{person}{Marta Larusdottir}, \bibinfo{person}{Jan
  Gulliksen}, {and} \bibinfo{person}{Åsa Cajander}.}
  \bibinfo{year}{2017}\natexlab{}.
\newblock \showarticletitle{A License to Kill – Improving UCSD in Agile
  Development}.
\newblock \bibinfo{journal}{\emph{Journal of Systems and Software}}
  \bibinfo{volume}{123} (\bibinfo{year}{2017}), \bibinfo{pages}{214–222}.
\newblock
\showISSN{0164-1212}
\href{https://doi.org/10.1016/j.jss.2016.01.024}{doi:\nolinkurl{10.1016/j.jss.2016.01.024}}


\bibitem[Leiva et~al\mbox{.}(2019)]%
        {paper_41}
\bibfield{author}{\bibinfo{person}{Germ\'{a}n Leiva}, \bibinfo{person}{Nolwenn
  Maudet}, \bibinfo{person}{Wendy Mackay}, {and} \bibinfo{person}{Michel
  Beaudouin-Lafon}.} \bibinfo{year}{2019}\natexlab{}.
\newblock \showarticletitle{Enact: Reducing Designer–Developer Breakdowns
  When Prototyping Custom Interactions}.
\newblock \bibinfo{journal}{\emph{ACM Transactions on Computer-Human
  Interaction}} \bibinfo{volume}{26}, \bibinfo{number}{3}, Article
  \bibinfo{articleno}{19} (\bibinfo{year}{2019}), \bibinfo{numpages}{48}~pages.
\newblock
\showISSN{1073-0516}
\href{https://doi.org/10.1145/3310276}{doi:\nolinkurl{10.1145/3310276}}


\bibitem[Liao et~al\mbox{.}(2023)]%
        {liao2023designerly}
\bibfield{author}{\bibinfo{person}{Q.~Vera Liao}, \bibinfo{person}{Hariharan
  Subramonyam}, \bibinfo{person}{Jennifer Wang}, {and}
  \bibinfo{person}{Jennifer Wortman~Vaughan}.} \bibinfo{year}{2023}\natexlab{}.
\newblock \showarticletitle{Designerly Understanding: Information Needs for
  Model Transparency to Support Design Ideation for AI-Powered User
  Experience}. In \bibinfo{booktitle}{\emph{Proceedings of the 2023 CHI
  Conference on Human Factors in Computing Systems}}
  \emph{(\bibinfo{series}{CHI '23})}. \bibinfo{publisher}{Association for
  Computing Machinery}, \bibinfo{address}{New York, NY, USA}, Article
  \bibinfo{articleno}{9}, \bibinfo{numpages}{21}~pages.
\newblock
\showISBNx{9781450394215}
\href{https://doi.org/10.1145/3544548.3580652}{doi:\nolinkurl{10.1145/3544548.3580652}}


\bibitem[Magües et~al\mbox{.}(2016)]%
        {2016_slr_integration}
\bibfield{author}{\bibinfo{person}{Daniel~A. Magües}, \bibinfo{person}{John~W.
  Castro}, {and} \bibinfo{person}{Silvia~T. Acuña}.}
  \bibinfo{year}{2016}\natexlab{}.
\newblock \showarticletitle{Usability in Agile Development: A Systematic
  Mapping Study}. In \bibinfo{booktitle}{\emph{2016 XLII Latin American
  Computing Conference (CLEI)}}. \bibinfo{pages}{1--8}.
\newblock
\href{https://doi.org/10.1109/CLEI.2016.7833347}{doi:\nolinkurl{10.1109/CLEI.2016.7833347}}


\bibitem[Mailach and Siegmund(2023)]%
        {mailach2023socio}
\bibfield{author}{\bibinfo{person}{Alina Mailach} {and}
  \bibinfo{person}{Norbert Siegmund}.} \bibinfo{year}{2023}\natexlab{}.
\newblock \showarticletitle{Socio-Technical Anti-patterns in Building
  ML-Enabled Software: Insights From Leaders on the Forefront}. In
  \bibinfo{booktitle}{\emph{Proceedings of the 45th International Conference on
  Software Engineering}} \emph{(\bibinfo{series}{ICSE '23})}.
  \bibinfo{publisher}{IEEE Press}, \bibinfo{pages}{690–702}.
\newblock
\showISBNx{9781665457019}
\href{https://doi.org/10.1109/ICSE48619.2023.00067}{doi:\nolinkurl{10.1109/ICSE48619.2023.00067}}


\bibitem[Maudet et~al\mbox{.}(2017)]%
        {paper_42}
\bibfield{author}{\bibinfo{person}{Nolwenn Maudet}, \bibinfo{person}{Germ\'{a}n
  Leiva}, \bibinfo{person}{Michel Beaudouin-Lafon}, {and}
  \bibinfo{person}{Wendy Mackay}.} \bibinfo{year}{2017}\natexlab{}.
\newblock \showarticletitle{Design Breakdowns: Designer-Developer Gaps in
  Representing and Interpreting Interactive Systems}. In
  \bibinfo{booktitle}{\emph{Proceedings of the 2017 ACM Conference on Computer
  Supported Cooperative Work and Social Computing}}
  \emph{(\bibinfo{series}{CSCW '17})}. \bibinfo{publisher}{Association for
  Computing Machinery}, \bibinfo{address}{New York, NY, USA},
  \bibinfo{pages}{630–641}.
\newblock
\showISBNx{9781450343350}
\href{https://doi.org/10.1145/2998181.2998190}{doi:\nolinkurl{10.1145/2998181.2998190}}


\bibitem[Memmel et~al\mbox{.}(2007)]%
        {2007_gather_requirements}
\bibfield{author}{\bibinfo{person}{Thomas Memmel}, \bibinfo{person}{Fredrik
  Gundelsweiler}, {and} \bibinfo{person}{Harald Reiterer}.}
  \bibinfo{year}{2007}\natexlab{}.
\newblock \showarticletitle{Agile Human-Centered Software Engineering}. In
  \bibinfo{booktitle}{\emph{Proceedings of the 21st British HCI Group Annual
  Conference on People and Computers - Volume 1}}
  \emph{(\bibinfo{series}{BCS-HCI '07})}. \bibinfo{publisher}{BCS Learning \&
  Development Ltd.}, \bibinfo{address}{Swindon, GBR},
  \bibinfo{pages}{167–175}.
\newblock
\showISBNx{9781902505947}


\bibitem[Meszaros and Aston(2006)]%
        {paper_79}
\bibfield{author}{\bibinfo{person}{Gerard Meszaros} {and}
  \bibinfo{person}{Janice Aston}.} \bibinfo{year}{2006}\natexlab{}.
\newblock \showarticletitle{Adding Usability Testing to an Agile Project}. In
  \bibinfo{booktitle}{\emph{Conference on AGILE 2006}}
  \emph{(\bibinfo{series}{AGILE '06})}. \bibinfo{publisher}{IEEE Computer
  Society}, \bibinfo{address}{USA}, \bibinfo{pages}{289–294}.
\newblock
\showISBNx{0769525628}
\href{https://doi.org/10.1109/AGILE.2006.5}{doi:\nolinkurl{10.1109/AGILE.2006.5}}


\bibitem[Miro(2023)]%
        {miro}
\bibfield{author}{\bibinfo{person}{Miro}.} \bibinfo{year}{2023}\natexlab{}.
\newblock \bibinfo{title}{Miro}.
\newblock \bibinfo{howpublished}{\url{https://miro.com/}}.
\newblock


\bibitem[Moher et~al\mbox{.}(2009)]%
        {moher2009preferred}
\bibfield{author}{\bibinfo{person}{D. Moher}, \bibinfo{person}{A. Liberati},
  \bibinfo{person}{J. Tetzlaff}, {and} \bibinfo{person}{D.~G Altman}.}
  \bibinfo{year}{2009}\natexlab{}.
\newblock \showarticletitle{Preferred Reporting Items for Systematic Reviews
  and Meta-Analyses: The Prisma Statement}.
\newblock \bibinfo{journal}{\emph{BMJ}} \bibinfo{volume}{339},
  \bibinfo{number}{jul21 1} (\bibinfo{year}{2009}), \bibinfo{pages}{b2535}.
\newblock
\showISSN{1468-5833}
\href{https://doi.org/10.1136/bmj.b2535}{doi:\nolinkurl{10.1136/bmj.b2535}}


\bibitem[M{\"u}ller(2018)]%
        {muller2018agile}
\bibfield{author}{\bibinfo{person}{Matthias M{\"u}ller}.}
  \bibinfo{year}{2018}\natexlab{}.
\newblock \showarticletitle{Agile Challenges and Chances for Open Source:
  Lessons Learned From Managing a Floss Project}. In
  \bibinfo{booktitle}{\emph{2018 IEEE Conference on Open Systems (ICOS)}}.
  \bibinfo{publisher}{IEEE}, \bibinfo{pages}{1--6}.
\newblock
\href{https://doi.org/10.1109/ICOS.2018.8632819}{doi:\nolinkurl{10.1109/ICOS.2018.8632819}}


\bibitem[Nahar et~al\mbox{.}(2022)]%
        {2022_collaboration_ml}
\bibfield{author}{\bibinfo{person}{Nadia Nahar}, \bibinfo{person}{Shurui Zhou},
  \bibinfo{person}{Grace Lewis}, {and} \bibinfo{person}{Christian
  K\"{a}stner}.} \bibinfo{year}{2022}\natexlab{}.
\newblock \showarticletitle{Collaboration Challenges in Building ML-Enabled
  Systems: Communication, Documentation, Engineering, and Process}. In
  \bibinfo{booktitle}{\emph{Proceedings of the 44th International Conference on
  Software Engineering}} \emph{(\bibinfo{series}{ICSE '22})}.
  \bibinfo{publisher}{Association for Computing Machinery},
  \bibinfo{address}{New York, NY, USA}, \bibinfo{pages}{413–425}.
\newblock
\showISBNx{9781450392211}
\href{https://doi.org/10.1145/3510003.3510209}{doi:\nolinkurl{10.1145/3510003.3510209}}


\bibitem[Najafi and Toyoshiba(2008)]%
        {paper_4}
\bibfield{author}{\bibinfo{person}{Maryam Najafi} {and} \bibinfo{person}{Len
  Toyoshiba}.} \bibinfo{year}{2008}\natexlab{}.
\newblock \showarticletitle{Two Case Studies of User Experience Design and
  Agile Development}. In \bibinfo{booktitle}{\emph{Proceedings of the Agile
  2008 Conference}} \emph{(\bibinfo{series}{AGILE '08})}.
  \bibinfo{publisher}{IEEE Computer Society}, \bibinfo{address}{USA},
  \bibinfo{pages}{531–536}.
\newblock
\showISBNx{9780769533216}
\href{https://doi.org/10.1109/Agile.2008.67}{doi:\nolinkurl{10.1109/Agile.2008.67}}


\bibitem[Noll et~al\mbox{.}(2011)]%
        {noll2011global}
\bibfield{author}{\bibinfo{person}{John Noll}, \bibinfo{person}{Sarah Beecham},
  {and} \bibinfo{person}{Ita Richardson}.} \bibinfo{year}{2011}\natexlab{}.
\newblock \showarticletitle{Global Software Development and Collaboration:
  Barriers and Solutions}.
\newblock \bibinfo{journal}{\emph{ACM Inroads}} \bibinfo{volume}{1},
  \bibinfo{number}{3} (\bibinfo{year}{2011}), \bibinfo{pages}{66–78}.
\newblock
\showISSN{2153-2184}
\href{https://doi.org/10.1145/1835428.1835445}{doi:\nolinkurl{10.1145/1835428.1835445}}


\bibitem[Nørgaard and Hornbæk(2010)]%
        {paper_99}
\bibfield{author}{\bibinfo{person}{Mie Nørgaard} {and} \bibinfo{person}{Kasper
  Hornbæk}.} \bibinfo{year}{2010}\natexlab{}.
\newblock \showarticletitle{Working Together to Improve Usability: Exploring
  Challenges and Successful Practices}.
\newblock \bibinfo{journal}{\emph{International Journal of Technology and Human
  Interaction}} \bibinfo{volume}{6}, \bibinfo{number}{1}
  (\bibinfo{year}{2010}), \bibinfo{pages}{33–53}.
\newblock
\showISSN{1548-3916}
\href{https://doi.org/10.4018/jthi.2010091703}{doi:\nolinkurl{10.4018/jthi.2010091703}}


\bibitem[Overflow(2023)]%
        {SOUploadFile}
\bibfield{author}{\bibinfo{person}{Stack Overflow}.}
  \bibinfo{year}{2023}\natexlab{}.
\newblock \bibinfo{title}{How to Upload Figma File(Wireframe) on Github?}
\newblock
  \bibinfo{howpublished}{\url{https://stackoverflow.com/questions/70094767/how-to-upload-figma-filewireframe-on-github}}.
\newblock


\bibitem[Paz and Pow-Sang(2015)]%
        {2015_slr_evaluation}
\bibfield{author}{\bibinfo{person}{Freddy Paz} {and}
  \bibinfo{person}{José~Antonio Pow-Sang}.} \bibinfo{year}{2015}\natexlab{}.
\newblock \showarticletitle{Usability Evaluation Methods for Software
  Development: A Systematic Mapping Review}. In \bibinfo{booktitle}{\emph{2015
  8th International Conference on Advanced Software Engineering \& Its
  Applications (ASEA)}}. \bibinfo{pages}{1--4}.
\newblock
\href{https://doi.org/10.1109/ASEA.2015.8}{doi:\nolinkurl{10.1109/ASEA.2015.8}}


\bibitem[Pereira et~al\mbox{.}(2022)]%
        {paper_37}
\bibfield{author}{\bibinfo{person}{Anathan Pereira}, \bibinfo{person}{Abner
  Cleto~Filho}, \bibinfo{person}{Eduardo Guerra}, {and}
  \bibinfo{person}{Luciana Zaina}.} \bibinfo{year}{2022}\natexlab{}.
\newblock \showarticletitle{Towards a Pattern Language to Embed UX Information
  in Agile Software Requirements}. In \bibinfo{booktitle}{\emph{Proceedings of
  the 26th European Conference on Pattern Languages of Programs}}
  \emph{(\bibinfo{series}{EuroPLoP '21})}. \bibinfo{publisher}{Association for
  Computing Machinery}, \bibinfo{address}{New York, NY, USA}, Article
  \bibinfo{articleno}{18}, \bibinfo{numpages}{8}~pages.
\newblock
\showISBNx{9781450389976}
\href{https://doi.org/10.1145/3489449.3489991}{doi:\nolinkurl{10.1145/3489449.3489991}}


\bibitem[Pereira and Russo(2018)]%
        {2018_integration_slr}
\bibfield{author}{\bibinfo{person}{Julio~Cesar Pereira} {and}
  \bibinfo{person}{Rosaria de~F.S.M. Russo}.} \bibinfo{year}{2018}\natexlab{}.
\newblock \showarticletitle{Design Thinking Integrated in Agile Software
  Development: A Systematic Literature Review}.
\newblock \bibinfo{journal}{\emph{Procedia Computer Science}}
  \bibinfo{volume}{138} (\bibinfo{year}{2018}), \bibinfo{pages}{775–782}.
\newblock
\showISSN{1877-0509}
\href{https://doi.org/10.1016/j.procs.2018.10.101}{doi:\nolinkurl{10.1016/j.procs.2018.10.101}}


\bibitem[Pillay and Wing(2019)]%
        {paper_1}
\bibfield{author}{\bibinfo{person}{Narendren Pillay} {and}
  \bibinfo{person}{Jeanette Wing}.} \bibinfo{year}{2019}\natexlab{}.
\newblock \showarticletitle{Agile UX: Integrating good UX development practices
  in Agile}. In \bibinfo{booktitle}{\emph{2019 Conference on Information
  Communications Technology and Society (ICTAS)}}. \bibinfo{pages}{1--6}.
\newblock
\href{https://doi.org/10.1109/ICTAS.2019.8703607}{doi:\nolinkurl{10.1109/ICTAS.2019.8703607}}


\bibitem[ProtoPie(2023)]%
        {protopie}
\bibfield{author}{\bibinfo{person}{ProtoPie}.} \bibinfo{year}{2023}\natexlab{}.
\newblock \bibinfo{title}{ProtoPie}.
\newblock \bibinfo{howpublished}{\url{https://www.protopie.io/}}.
\newblock


\bibitem[Queiroz et~al\mbox{.}(2017)]%
        {Queiroz2017}
\bibfield{author}{\bibinfo{person}{Francisco Queiroz}, \bibinfo{person}{Raniere
  Silva}, \bibinfo{person}{Jonah Miller}, \bibinfo{person}{Sandor Brockhauser},
  {and} \bibinfo{person}{Hans Fangohr}.} \bibinfo{year}{2017}\natexlab{}.
\newblock \showarticletitle{Good Usability Practices in Scientific Software
  Development}. In \bibinfo{booktitle}{\emph{Proceedings of WSSSPE5.1}}.
\newblock


\bibitem[Rajanen and Iivari(2015)]%
        {Rajanen2015}
\bibfield{author}{\bibinfo{person}{Mikko Rajanen} {and} \bibinfo{person}{Netta
  Iivari}.} \bibinfo{year}{2015}\natexlab{}.
\newblock \showarticletitle{Power, Empowerment and Open Source Usability}. In
  \bibinfo{booktitle}{\emph{Proceedings of the 33rd Annual ACM Conference on
  Human Factors in Computing Systems}} \emph{(\bibinfo{series}{CHI '15})}.
  \bibinfo{publisher}{Association for Computing Machinery},
  \bibinfo{address}{New York, NY, USA}, \bibinfo{pages}{3413–3422}.
\newblock
\showISBNx{9781450331456}
\href{https://doi.org/10.1145/2702123.2702441}{doi:\nolinkurl{10.1145/2702123.2702441}}


\bibitem[Rajanen et~al\mbox{.}(2015)]%
        {paper_53}
\bibfield{author}{\bibinfo{person}{Mikko Rajanen}, \bibinfo{person}{Netta
  Iivari}, {and} \bibinfo{person}{Arto Lanam\"{a}ki}.}
  \bibinfo{year}{2015}\natexlab{}.
\newblock \showarticletitle{Non-response, Social Exclusion, and False
  Acceptance: Gatekeeping Tactics and Usability Work in Free-Libre Open Source
  Software Development}. In \bibinfo{booktitle}{\emph{Human-Computer
  Interaction – INTERACT 2015}}. \bibinfo{publisher}{Springer-Verlag},
  \bibinfo{address}{Berlin, Heidelberg}, \bibinfo{pages}{9–26}.
\newblock
\showISBNx{978-3-319-22697-2}
\href{https://doi.org/10.1007/978-3-319-22698-9_2}{doi:\nolinkurl{10.1007/978-3-319-22698-9_2}}


\bibitem[Reddit(2023a)]%
        {reddit-uxd-github}
\bibfield{author}{\bibinfo{person}{Reddit}.} \bibinfo{year}{2023}\natexlab{a}.
\newblock \bibinfo{title}{Designers - Do You Go in Github?}
\newblock
  \bibinfo{howpublished}{\url{https://www.reddit.com/r/DesignSystems/comments/1bjj95r/how_do_you_manage_release_versions_with_figma/}}.
\newblock


\bibitem[Reddit(2023b)]%
        {reddit-figma-github}
\bibfield{author}{\bibinfo{person}{Reddit}.} \bibinfo{year}{2023}\natexlab{b}.
\newblock \bibinfo{title}{Is There a Way to Streamline What We Create in Figma
  and What the Developers Have Created on Their End?}
\newblock
  \bibinfo{howpublished}{\url{https://www.reddit.com/r/FigmaDesign/comments/1278o7z/is_there_a_way_to_streamline_what_we_create_in/}}.
\newblock


\bibitem[Reddit(2023c)]%
        {Reddit_SE}
\bibfield{author}{\bibinfo{person}{Reddit}.} \bibinfo{year}{2023}\natexlab{c}.
\newblock \bibinfo{title}{Reddit Software Engineering}.
\newblock
  \bibinfo{howpublished}{\url{https://www.reddit.com/r/SoftwareEngineering/}}.
\newblock


\bibitem[Reddit(2023d)]%
        {Reddit_UX}
\bibfield{author}{\bibinfo{person}{Reddit}.} \bibinfo{year}{2023}\natexlab{d}.
\newblock \bibinfo{title}{Reddit UXDesign: For Experienced and Veteran UX
  Practitioners}.
\newblock \bibinfo{howpublished}{\url{https://www.reddit.com/r/UXDesign/}}.
\newblock


\bibitem[Salah et~al\mbox{.}(2014)]%
        {2014_slr_integration}
\bibfield{author}{\bibinfo{person}{Dina Salah}, \bibinfo{person}{Richard~F.
  Paige}, {and} \bibinfo{person}{Paul Cairns}.}
  \bibinfo{year}{2014}\natexlab{}.
\newblock \showarticletitle{A Systematic Literature Review for Agile
  Development Processes and User Centred Design Integration}. In
  \bibinfo{booktitle}{\emph{Proceedings of the 18th International Conference on
  Evaluation and Assessment in Software Engineering}}
  \emph{(\bibinfo{series}{EASE '14})}. \bibinfo{publisher}{Association for
  Computing Machinery}, \bibinfo{address}{New York, NY, USA}, Article
  \bibinfo{articleno}{5}, \bibinfo{numpages}{10}~pages.
\newblock
\showISBNx{9781450324762}
\href{https://doi.org/10.1145/2601248.2601276}{doi:\nolinkurl{10.1145/2601248.2601276}}


\bibitem[Sanei and Cheng(2024)]%
        {sanei2023characterizing}
\bibfield{author}{\bibinfo{person}{Arghavan Sanei} {and}
  \bibinfo{person}{Jinghui Cheng}.} \bibinfo{year}{2024}\natexlab{}.
\newblock \showarticletitle{Characterizing Usability Issue Discussions in Open
  Source Software Projects}.
\newblock \bibinfo{journal}{\emph{Proceedings of the ACM on Human-Computer
  Interaction}} \bibinfo{volume}{8}, \bibinfo{number}{CSCW1}, Article
  \bibinfo{articleno}{30} (\bibinfo{year}{2024}), \bibinfo{numpages}{26}~pages.
\newblock
\href{https://doi.org/10.1145/3637307}{doi:\nolinkurl{10.1145/3637307}}


\bibitem[Sangwan et~al\mbox{.}(2020)]%
        {sangwan2020asynchronous}
\bibfield{author}{\bibinfo{person}{Raghvinder~S. Sangwan},
  \bibinfo{person}{Kathryn~W. Jablokow}, {and} \bibinfo{person}{Joanna~F.
  DeFranco}.} \bibinfo{year}{2020}\natexlab{}.
\newblock \showarticletitle{Asynchronous Collaboration: Bridging the Cognitive
  Distance in Global Software Development Projects}.
\newblock \bibinfo{journal}{\emph{IEEE Transactions on Professional
  Communication}} \bibinfo{volume}{63}, \bibinfo{number}{4}
  (\bibinfo{year}{2020}), \bibinfo{pages}{361--371}.
\newblock
\href{https://doi.org/10.1109/TPC.2020.3029674}{doi:\nolinkurl{10.1109/TPC.2020.3029674}}


\bibitem[Sarma et~al\mbox{.}(2012)]%
        {sarma2011palantir}
\bibfield{author}{\bibinfo{person}{Anita Sarma}, \bibinfo{person}{David~F.
  Redmiles}, {and} \bibinfo{person}{André van~der Hoek}.}
  \bibinfo{year}{2012}\natexlab{}.
\newblock \showarticletitle{Palantir: Early Detection of Development Conflicts
  Arising from Parallel Code Changes}.
\newblock \bibinfo{journal}{\emph{IEEE Transactions on Software Engineering}}
  \bibinfo{volume}{38}, \bibinfo{number}{4} (\bibinfo{year}{2012}),
  \bibinfo{pages}{889--908}.
\newblock
\href{https://doi.org/10.1109/TSE.2011.64}{doi:\nolinkurl{10.1109/TSE.2011.64}}


\bibitem[Sfetsos et~al\mbox{.}(2016)]%
        {paper_58}
\bibfield{author}{\bibinfo{person}{Panagiotis Sfetsos},
  \bibinfo{person}{Lefteris Angelis}, \bibinfo{person}{Ioannis Stamelos}, {and}
  \bibinfo{person}{Pashalis Raptis}.} \bibinfo{year}{2016}\natexlab{}.
\newblock \showarticletitle{Integrating User-Centered Design Practices Into
  Agile Web Development: A Case Study}. In \bibinfo{booktitle}{\emph{2016 7th
  International Conference on Information, Intelligence, Systems \&
  Applications (IISA)}}. \bibinfo{pages}{1--6}.
\newblock
\href{https://doi.org/10.1109/IISA.2016.7785424}{doi:\nolinkurl{10.1109/IISA.2016.7785424}}


\bibitem[Shaw(2005)]%
        {Shaw2005SoftwareEF}
\bibfield{author}{\bibinfo{person}{Mary Shaw}.}
  \bibinfo{year}{2005}\natexlab{}.
\newblock \showarticletitle{Software Engineering for the 21st Century: A Basis
  for Rethinking the Curriculum}. In \bibinfo{booktitle}{\emph{School of
  Computer Science Carnegie Mellon University}}.
\newblock


\bibitem[Silva~da Silva et~al\mbox{.}(2011)]%
        {2011_integration_slr}
\bibfield{author}{\bibinfo{person}{Tiago Silva~da Silva},
  \bibinfo{person}{Angela Martin}, \bibinfo{person}{Frank Maurer}, {and}
  \bibinfo{person}{Milene Silveira}.} \bibinfo{year}{2011}\natexlab{}.
\newblock \showarticletitle{User-Centered Design and Agile Methods: A
  Systematic Review}. In \bibinfo{booktitle}{\emph{2011 Agile Conference}}.
  \bibinfo{pages}{77--86}.
\newblock
\href{https://doi.org/10.1109/AGILE.2011.24}{doi:\nolinkurl{10.1109/AGILE.2011.24}}


\bibitem[Sketchboard(2014)]%
        {sketchboard}
\bibfield{author}{\bibinfo{person}{Sketchboard}.}
  \bibinfo{year}{2014}\natexlab{}.
\newblock \bibinfo{title}{Online Collaborative Whiteboard}.
\newblock
  \bibinfo{howpublished}{\url{https://sketchboard.io/blog/2014/03/06/github-sketchboard}}.
\newblock


\bibitem[Sohaib and Khan(2010)]%
        {2010_integration_slr}
\bibfield{author}{\bibinfo{person}{Osama Sohaib} {and} \bibinfo{person}{Khalid
  Khan}.} \bibinfo{year}{2010}\natexlab{}.
\newblock \showarticletitle{Integrating Usability Engineering and Agile
  Software Development: A Literature Review}. In \bibinfo{booktitle}{\emph{2010
  International Conference On Computer Design and Applications}},
  Vol.~\bibinfo{volume}{2}. \bibinfo{pages}{V2--32--V2--38}.
\newblock
\href{https://doi.org/10.1109/ICCDA.2010.5540916}{doi:\nolinkurl{10.1109/ICCDA.2010.5540916}}


\bibitem[St{\aa}hl(2023)]%
        {staahl2023dynamic}
\bibfield{author}{\bibinfo{person}{Daniel St{\aa}hl}.}
  \bibinfo{year}{2023}\natexlab{}.
\newblock \showarticletitle{The Dynamic Versus the Stable Team: The Unspoken
  Question in Large-Scale Agile Development}.
\newblock \bibinfo{journal}{\emph{Journal of Software: Evolution and Process}}
  \bibinfo{volume}{35}, \bibinfo{number}{12} (\bibinfo{year}{2023}),
  \bibinfo{pages}{e2589}.
\newblock
\href{https://doi.org/10.1002/smr.2589}{doi:\nolinkurl{10.1002/smr.2589}}


\bibitem[Strode et~al\mbox{.}(2012)]%
        {strode2012coordination}
\bibfield{author}{\bibinfo{person}{Diane~E. Strode}, \bibinfo{person}{Sid~L.
  Huff}, \bibinfo{person}{Beverley Hope}, {and} \bibinfo{person}{Sebastian
  Link}.} \bibinfo{year}{2012}\natexlab{}.
\newblock \showarticletitle{Coordination in Co-located Agile Software
  Development Projects}.
\newblock \bibinfo{journal}{\emph{Journal of Systems and Software}}
  \bibinfo{volume}{85}, \bibinfo{number}{6} (\bibinfo{year}{2012}),
  \bibinfo{pages}{1222–1238}.
\newblock
\showISSN{0164-1212}
\href{https://doi.org/10.1016/j.jss.2012.02.017}{doi:\nolinkurl{10.1016/j.jss.2012.02.017}}


\bibitem[Sy(2007)]%
        {paper_s1}
\bibfield{author}{\bibinfo{person}{Desir\'{e}e Sy}.}
  \bibinfo{year}{2007}\natexlab{}.
\newblock \showarticletitle{Adapting Usability Investigations for Agile
  User-Centered Design}.
\newblock \bibinfo{journal}{\emph{Journal of Usability Studies}}
  \bibinfo{volume}{2}, \bibinfo{number}{3} (\bibinfo{year}{2007}),
  \bibinfo{pages}{112–132}.
\newblock


\bibitem[Terry et~al\mbox{.}(2010)]%
        {terry2010perceptions}
\bibfield{author}{\bibinfo{person}{Michael Terry}, \bibinfo{person}{Matthew
  Kay}, {and} \bibinfo{person}{Ben Lafreniere}.}
  \bibinfo{year}{2010}\natexlab{}.
\newblock \showarticletitle{Perceptions and Practices of Usability in the
  Free/Open Source Software (FoSS) Community}. In
  \bibinfo{booktitle}{\emph{Proceedings of the SIGCHI Conference on Human
  Factors in Computing Systems}} \emph{(\bibinfo{series}{CHI '10})}.
  \bibinfo{publisher}{Association for Computing Machinery},
  \bibinfo{address}{New York, NY, USA}, \bibinfo{pages}{999–1008}.
\newblock
\showISBNx{9781605589299}
\href{https://doi.org/10.1145/1753326.1753476}{doi:\nolinkurl{10.1145/1753326.1753476}}


\bibitem[Treude and Storey(2010)]%
        {treude2010awareness}
\bibfield{author}{\bibinfo{person}{Christoph Treude} {and}
  \bibinfo{person}{Margaret-Anne Storey}.} \bibinfo{year}{2010}\natexlab{}.
\newblock \showarticletitle{Awareness 2.0: Staying Aware of Projects,
  Developers and Tasks Using Dashboards and Feeds}. In
  \bibinfo{booktitle}{\emph{Proceedings of the ACM/IEEE 32nd International
  Conference on Software Engineering}}, Vol.~\bibinfo{volume}{1}.
  \bibinfo{pages}{365--374}.
\newblock
\href{https://doi.org/10.1145/1806799.1806854}{doi:\nolinkurl{10.1145/1806799.1806854}}


\bibitem[Ungar(2008)]%
        {paper_26}
\bibfield{author}{\bibinfo{person}{Jim Ungar}.}
  \bibinfo{year}{2008}\natexlab{}.
\newblock \showarticletitle{The Design Studio: Interface Design for Agile
  Teams}. In \bibinfo{booktitle}{\emph{Agile 2008 Conference}}.
  \bibinfo{pages}{519--524}.
\newblock
\href{https://doi.org/10.1109/Agile.2008.51}{doi:\nolinkurl{10.1109/Agile.2008.51}}


\bibitem[Wale-Kolade and Nielsen(2015)]%
        {paper_104}
\bibfield{author}{\bibinfo{person}{Adeola Wale-Kolade} {and}
  \bibinfo{person}{Peter~Axel Nielsen}.} \bibinfo{year}{2015}\natexlab{}.
\newblock \showarticletitle{Apathy Towards the Integration of Usability Work: A
  Case of System Justification}.
\newblock \bibinfo{journal}{\emph{Interacting with Computers}}
  \bibinfo{volume}{28}, \bibinfo{number}{4} (\bibinfo{year}{2015}),
  \bibinfo{pages}{437–450}.
\newblock
\showISSN{1873-7951}
\href{https://doi.org/10.1093/iwc/iwv016}{doi:\nolinkurl{10.1093/iwc/iwv016}}


\bibitem[Wale-Kolade(2015)]%
        {paper_102}
\bibfield{author}{\bibinfo{person}{Adeola~Yetunde Wale-Kolade}.}
  \bibinfo{year}{2015}\natexlab{}.
\newblock \showarticletitle{Integrating Usability Work Into a Large
  Inter-Organisational Agile Development Project: Tactics Developed by
  Usability Designers}.
\newblock \bibinfo{journal}{\emph{Journal of Systems and Software}}
  \bibinfo{volume}{100} (\bibinfo{year}{2015}), \bibinfo{pages}{54–66}.
\newblock
\showISSN{0164-1212}
\href{https://doi.org/10.1016/j.jss.2014.10.036}{doi:\nolinkurl{10.1016/j.jss.2014.10.036}}


\bibitem[Wohlin(2014)]%
        {2014_snowballing}
\bibfield{author}{\bibinfo{person}{Claes Wohlin}.}
  \bibinfo{year}{2014}\natexlab{}.
\newblock \showarticletitle{Guidelines for Snowballing in Systematic Literature
  Studies and a Replication in Software Engineering}. In
  \bibinfo{booktitle}{\emph{Proceedings of the 18th International Conference on
  Evaluation and Assessment in Software Engineering}}
  \emph{(\bibinfo{series}{EASE '14})}. \bibinfo{publisher}{Association for
  Computing Machinery}, \bibinfo{address}{New York, NY, USA}, Article
  \bibinfo{articleno}{38}, \bibinfo{numpages}{10}~pages.
\newblock
\showISBNx{9781450324762}
\href{https://doi.org/10.1145/2601248.2601268}{doi:\nolinkurl{10.1145/2601248.2601268}}


\bibitem[Yang et~al\mbox{.}(2018)]%
        {yang2018investigating}
\bibfield{author}{\bibinfo{person}{Qian Yang}, \bibinfo{person}{Alex Scuito},
  \bibinfo{person}{John Zimmerman}, \bibinfo{person}{Jodi Forlizzi}, {and}
  \bibinfo{person}{Aaron Steinfeld}.} \bibinfo{year}{2018}\natexlab{}.
\newblock \showarticletitle{Investigating How Experienced UX Designers
  Effectively Work with Machine Learning}. In
  \bibinfo{booktitle}{\emph{Proceedings of the 2018 Designing Interactive
  Systems Conference}} \emph{(\bibinfo{series}{DIS '18})}.
  \bibinfo{publisher}{Association for Computing Machinery},
  \bibinfo{address}{New York, NY, USA}, \bibinfo{pages}{585–596}.
\newblock
\showISBNx{9781450351980}
\href{https://doi.org/10.1145/3196709.3196730}{doi:\nolinkurl{10.1145/3196709.3196730}}


\bibitem[Zdanowska and Taylor(2022)]%
        {zdanowska2022study}
\bibfield{author}{\bibinfo{person}{Sabah Zdanowska} {and}
  \bibinfo{person}{Alex~S Taylor}.} \bibinfo{year}{2022}\natexlab{}.
\newblock \showarticletitle{A study of UX practitioners roles in designing
  real-world, enterprise ML systems}. In \bibinfo{booktitle}{\emph{Proceedings
  of the 2022 CHI Conference on Human Factors in Computing Systems}}.
  \bibinfo{pages}{1--15}.
\newblock


\bibitem[Çetin and Göktürk(2008)]%
        {paper_77}
\bibfield{author}{\bibinfo{person}{Görkem Çetin} {and}
  \bibinfo{person}{Mehmet Göktürk}.} \bibinfo{year}{2008}\natexlab{}.
\newblock \showarticletitle{A Measurement Based Framework for Assessment of
  Usability-Centricness of Open Source Software Projects}. In
  \bibinfo{booktitle}{\emph{2008 IEEE International Conference on Signal Image
  Technology and Internet Based Systems}}. \bibinfo{pages}{585--592}.
\newblock
\href{https://doi.org/10.1109/SITIS.2008.106}{doi:\nolinkurl{10.1109/SITIS.2008.106}}


\end{thebibliography}
\newpage
\appendix
\section{Appendix}
\subsection{Post links}
This section presents a list of all the referenced post links in the main text.
\begin{myblock}
[P1] \url{https://ux.stackexchange.com/questions/144725/}\label{UX1}
\end{myblock}

\begin{myblock}
[P2] \url{https://ux.stackexchange.com/questions/125859/}\label{UX5}
\end{myblock}

\begin{myblock}
[P3] \url{https://ux.stackexchange.com/questions/138215/}\label{UX13}
\end{myblock}

\begin{myblock}
[P4] \url{https://www.reddit.com/r/UXDesign/comments/lrpngo/}\label{UXR8}
\end{myblock}

\begin{myblock}
[P5] \url{https://ux.stackexchange.com/questions/140837/}\label{UX2}
\end{myblock}

\begin{myblock}
[P6] \url{https://www.reddit.com/r/UXDesign/comments/uhn981/}\label{UXR4}
\end{myblock}

\begin{myblock}
[P7] \url{https://www.reddit.com/r/UXDesign/comments/v0sh4c/}\label{UXR17}
\end{myblock}

\begin{myblock}
[P8] \url{https://www.reddit.com/r/UXDesign/comments/11xupv0/}\label{UXR46}
\end{myblock}

\begin{myblock}
[P9] \url{https://www.reddit.com/r/SoftwareEngineering/comments/sytmb2/}\label{SER3}
\end{myblock}

\begin{myblock}
[P10] \url{https://ux.stackexchange.com/questions/132945/}\label{UX15}
\end{myblock}

\begin{myblock}
[P11] \url{https://www.reddit.com/r/UXDesign/comments/pp15aj/}\label{UXR5}
\end{myblock}

\begin{myblock}
[P12] \url{https://ux.stackexchange.com/questions/135804/}\label{UX3}
\end{myblock}

\begin{myblock}
[P13] \url{https://www.reddit.com/r/UXDesign/comments/z2nis5/}\label{UXR18}
\end{myblock}

\begin{myblock}
[P14] \url{https://ux.stackexchange.com/questions/131656/}\label{UX4}
\end{myblock}

\begin{myblock}
[P15] \url{https://ux.stackexchange.com/questions/133644/}\label{UX10}
\end{myblock}

\begin{myblock}
[P16] \url{https://www.reddit.com/r/UXDesign/comments/129rdsk/}\label{UXR44}
\end{myblock}

\begin{myblock}
[P17] \url{https://www.reddit.com/r/SoftwareEngineering/comments/107jqtl/}\label{SER2}
\end{myblock}

\begin{myblock}
[P18] \url{https://ux.stackexchange.com/questions/136765/}\label{UX14}
\end{myblock}

\begin{myblock}
[P19] \url{https://ux.stackexchange.com/questions/131927/}\label{UX11}
\end{myblock}

\begin{myblock}
[P20] \url{https://www.reddit.com/r/UXDesign/comments/gcm2o0/}\label{UXR12}
\end{myblock}

\begin{myblock}
[P21] \url{https://www.reddit.com/r/UXDesign/comments/e25iuu/}\label{UXR15}
\end{myblock}

\begin{myblock}
[P22] \url{https://www.reddit.com/r/UXDesign/comments/z3foqp/}\label{UXR33}
\end{myblock}

\begin{myblock}
[P23] \url{https://www.reddit.com/r/UXDesign/comments/rlorbm/}\label{UXR35}
\end{myblock}

\begin{myblock}
[P24] \url{https://www.reddit.com/r/UXDesign/comments/pklqo7/}\label{UXR36}
\end{myblock}

\begin{myblock}
[P25] \url{https://www.reddit.com/r/UXDesign/comments/10kz2zw/}\label{UXR22}
\end{myblock}

\subsection{Issue Links}
This section presents a list of all the referenced issue links in the main text.
\begin{issue}
[I1] \url{https://github.com/microsoft/vscode/issues/156179}\label{issue_156179}
\end{issue}

\begin{issue}
[I2] \url{https://github.com/microsoft/vscode/issues/98614}\label{issue_98614}
\end{issue}

\begin{issue}
[I3] \url{https://github.com/microsoft/vscode/issues/130611}\label{issue_130611}
\end{issue}

\begin{issue}
[I4] \url{https://github.com/microsoft/vscode/issues/63152}\label{issue_63152}
\end{issue}

\begin{issue}
[I5] \url{https://github.com/microsoft/vscode/issues/147903}\label{issue_147903}
\end{issue}

\begin{issue}
[I6] \url{https://github.com/microsoft/vscode/issues/131641}\label{issue_131641}
\end{issue}

\begin{issue}
[I7] \url{https://github.com/microsoft/vscode/issues/133622}\label{issue_133622}
\end{issue}

\begin{issue}
[I8] \url{https://github.com/microsoft/vscode/issues/144324}\label{issue_144324}
\end{issue}

\begin{issue}
[I9] \url{https://github.com/microsoft/vscode/issues/146806}\label{issue_146806}
\end{issue}

\begin{issue}
[I10] \url{https://github.com/microsoft/vscode/issues/100347}\label{issue_100347}
\end{issue}

\begin{issue}
[I11] \url{https://github.com/microsoft/vscode/issues/101375}\label{issue_101375}
\end{issue}

\begin{issue}
[I12] \url{https://github.com/microsoft/vscode/issues/89533}\label{issue_89533}
\end{issue}

\begin{issue}
[I13] \url{https://github.com/microsoft/vscode/issues/144198}\label{issue_144198}
\end{issue}

\begin{issue}
[I14] \url{https://github.com/microsoft/vscode/issues/115808}\label{issue_115808}
\end{issue}

\begin{issue}
[I15] \url{https://github.com/microsoft/vscode/issues/144823}\label{issue_144823}
\end{issue}

\begin{issue}
[I16] \url{https://github.com/microsoft/vscode/issues/130516}\label{issue_130516}
\end{issue}

\begin{issue}
[I17] \url{https://github.com/microsoft/vscode/issues/119776}\label{issue_119776}
\end{issue}

\subsection{List of Keywords}
\label{apd:keywords}
In the data processing process, we utilized specific keywords in both the paper titles and publication titles to exclude studies in other fields that are not related to UX and SE, or those presented at minor or regional conferences. \\
\textbf{Small and local conferences}: symposium, workshop, seminar, congress, phd school association, country, jubilee, Asia-Pacific, latin, north american, asia, australasian, arabic, euromicro, brazilian, malaysian \\
\textbf{Other fields}: avionics, cyber security, security, distributed computing, image processing, blockchain, wearable, api, java, reality, augmented, virtual, 3d, hardware, image, visual, algorithm, query, machine, cloud, network, companion, extended abstracts, smart contract, crypto, dataset, computing, data, graph, machine learning, operating system, computational, securing, health, secure, power, deep learning, memory, prediction, optimization, optimize, xml, sql, clone, generator, generation, calculation, calculate, auto, toolkit \\
In this step, we filter out 824 papers. All filtered papers are reviewed by the first author to ensure there is no false negative.

\end{document}